\documentclass[twocolumn,showpacs,superscriptaddress,eqsecnum,longbibliography,nofootinbib]{revtex4-2}
\usepackage{mathrsfs,amsmath,amsthm,latexsym,amssymb,amsfonts,epsfig,cancel,enumerate,graphicx,txfonts,subfigure}
\usepackage{multirow}
\usepackage{xcolor}
\usepackage[colorlinks,linkcolor=blue,citecolor=blue,urlcolor=blue,plainpages=false,pdfstartview=FitH]{hyperref}
\usepackage{appendix}
\allowdisplaybreaks
\newcommand{\GZU}{School of Physics, Guizhou University, Guiyang 550025, China}
\begin{document}

\title{Optical appearance of Schwarzschild black holes with optically thin and thick accretion disks at various inclination angles}

\author{Jiawei Chen}
\email{gs.chenjw23@gzu.edu.cn}
\affiliation{\GZU}

\author{Jinsong Yang}
\thanks{Corresponding author}
\email{jsyang@gzu.edu.cn}
\affiliation{\GZU}

\begin{abstract}
In this paper, we systematically investigate the optical appearance of a Schwarzschild black hole illuminated by three geometrically thin accretion disk models under varying observational inclination angles. Based on the geometric relationship between the black hole and observer, we first divide the accretion disk into co-side and counter-side semi-disks. We then analyze light ray trajectories, and calculate the total number of orbits and transfer functions for both semi-disks. The results reveal distinct inclination-dependence of lensed regions on different semi-disks: as inclination increases, the lensed region contracts for the counter-side semi-disk while expanding for the co-side one. Furthermore, through explicit specification of the emission profiles of the three models, we present optical images for both optically thin and thick disk scenarios at different inclinations. The results demonstrate that: (i) the bright rings in all three models become progressively compressed and deviate from circularity as inclination increases; (ii) for thick disks, partial rings are obscured and the overall intensity is lower than thin disks. These results may advance our understanding of general black hole imaging processes and provide a new approach to test gravitational theories through optical morphology studies.
\end{abstract}

\maketitle

\section{Introduction}

As one of the most influential theories of gravity, general relativity (GR) has profoundly influenced our understanding of the nature of gravity. In weak-field regions, the validity of GR has been confirmed by numerous classical experiments~\cite{Will:2014kxa}. Moreover, in strong-field regions, the successful detection of gravitational waves (GWs) from a binary black hole (BH) merger~\cite{LIGOScientific:2016aoc} and the release of the first image of the supermassive BH~\cite{EventHorizonTelescope:2019dse} provided further confirmation of GR's predictions. The released BH image of M87* reveals a distinct bright ring encircling a central dark region. This luminous structure arises from photons that are deflected by the BH's extreme gravity and redirected toward the observer, whereas the dark shadow area corresponds to photons that have fallen into the BH. The shadow image of a BH reflects the properties of its surrounding spacetime geometry. By analyzing the shadow, one can extract key parameters of the BH~\cite{Kumar:2018ple,Hioki:2022mdg}, which has drawn significant attention to BH shadow research. Synge was the first to theoretically analyze the formation of shadows for Schwarzschild BH~\cite{Synge:1966okc}. Later, Bardeen investigated the shadow of Kerr BH, and found that the Kerr BH's shadow deviates from the circular shape of Schwarzschild's case, with the distortion increasing as the spin parameter grows~\cite{Bardeen:1972fi}. Since then, extensive theoretical studies have been conducted on BH shadows in various gravitational theories, including matter couplings~\cite{Claudel:2000yi,Bambi:2008jg,Wei:2013kza,Younsi:2016azx,Cunha:2016wzk,Stuchlik:2018qyz,Wang:2018prk,Jusufi:2019nrn,Liu:2020ola,Hou:2021okc,Ban:2024qsa}.

Furthermore, considering astrophysical BHs in the Universe, their surroundings inevitably contain various matter components. Under the influence of the BH's strong gravitational field, this matter is accreted by the BH, forming different types of accretion structures around the BH~\cite{Abramowicz:2011xu}. Based on their geometric thickness and optical properties, accretion disks around BHs can be classified into thin and thick disks. The geometric thickness is typically quantified by the ratio of the disk thickness $D$ to its radius $R$~\cite{Abramowicz:2011xu,Chen:2025doc}, where a disk is considered geometrically thin if $D/R \ll 1$. In contrast, the optical thickness relates to the disk's transparency; specifically, a disk is optically thick if it prevents the transmission of photons. The distinct imaging features produced by these different accretion disk models make it possible to test various gravitational theories through the optical appearances of BH accretion systems. Previous studies have extensively investigated both geometrically thick and thin disks~\cite{Abramowicz:2011xu,Zhang:2024lsf}. The first simulated image of a dynamically geometrically thin and optically thick accretion disk surrounding a Schwarzschild BH was produced by Luminet~\cite{Luminet:1979nyg}. This pioneering work incorporated the observer's inclination angle relative to the accretion disk, established angular relationships through spherical geometry, and ultimately generated a contour map on the observer's plane displaying both the primary and secondary images of the BH. Subsequently, the study of BH imaging with such accretion disk models as emission sources, particularly under inclined viewing angles, has garnered widespread attention across various gravitational theories~\cite{,Chen:2025doc,Gyulchev:2019tvk,Guo:2022rql,Huang:2023ilm,Guo:2023grt,Asukula:2023akj,You:2024jeu,Cui:2024wvz,You:2024uql}.

For optically thick accretion disks, the propagation of light is obstructed by the disk, which may lead to degradation of additional bright rings. Consequently, later studies adopted simpler geometrically thin and optically thin disk models. When investigating their imaging properties, Gralla et al. found that the observed image features are determined by the location and nature of emitting matter near the BH~\cite{Gralla:2019xty}. They further derived the optical appearance of three types of static, geometrically thin and optically thin accretion disks viewed face-on. Subsequent works have extensively investigated the luminous features of BHs illuminated by three static thin-disk models across various gravitational theories~\cite{Peng:2020wun,Zeng:2020vsj,Zeng:2020dco,Cardoso:2021sip,Gan:2021xdl,Uniyal:2022vdu,Wang:2022yvi,Zeng:2021dlj,Guerrero:2021ues,Hou:2022eev,Zeng:2022pvb,Rosa:2022tfv,Yang:2022btw,Zhang:2023okw,Wang:2023vcv,Sui:2023tje,Gao:2023mjb,Meng:2024puu,Darvishi:2024ndu,Zare:2024dtf,Chen:2025ifv,Rani:2025esb,Luo:2025xjb,Li:2025ftb}.

In realistic astrophysical scenarios, observers rarely view BHs perfectly face-on to the equatorial plane, but substantial inclination angles exist. For example, the line of sight from Earth to M87* is not face-on its equatorial plane but maintains a specific angle~\cite{CraigWalker:2018vam,EventHorizonTelescope:2019dse}. However, in current studies of these three static accretion disk models, systematic investigations of their BH images at varying inclination angles remain lacking. This motivates our focused investigation into the optical appearance of these three disk models under varying inclination conditions.

In this study, we examine three geometrically thin accretion disk models under both optically thin and optically thick conditions, and investigate the optical appearance characteristics of a Schwarzschild BH illuminated by these distinct types of accretion disks. Specifically, we divide the accretion disk into co-side and counter-side semi-disks based on their positions relative to the observer and BH. Based on the structure of two semi-disks, we analyze the photon orbit number distributions and light-ray intersection patterns for both semi-disks at various inclinations. Then we introduce corresponding transfer functions to establish the positional relationship between the accretion disk of the BH and observer's plane, which are then used to calculate the received intensities. Finally, we obtain optical images for both optically thin and thick disk scenarios under the three static accretion disk models.

The structure of this paper is organized as follows. In Sec.~\ref{section2}, we briefly recall photon trajectories around a Schwarzschild BH. In Sec.~\ref{section3}, based on the relative positions between the BH and observer, we divide the accretion disk into two semi-disks for separate analysis, examining photon trajectory intersections with both semi-disks and introducing their respective transfer functions. In Sec.~\ref{section4}, we obtain the optical appearance of a Schwarzschild BH illuminated by three types of optically thin accretion disk models at various observational inclination angles. Sec.~\ref{section5} presents the corresponding optical images for the optically thick disk scenarios of all three models, with systematic comparisons to their thin-disk counterparts. Finally, a summary is provided in Sec.~\ref{section6}. Throughout this paper, we adopt geometric units with $G=c=1$ and set the BH mass $M=1$ for computational purposes.

\section{A Brief Recall of Photon Trajectories in Schwarzschild Black Hole Spacetime}\label{section2}

\begin{figure}[htbp]
	\centering
	\includegraphics[width=7cm]{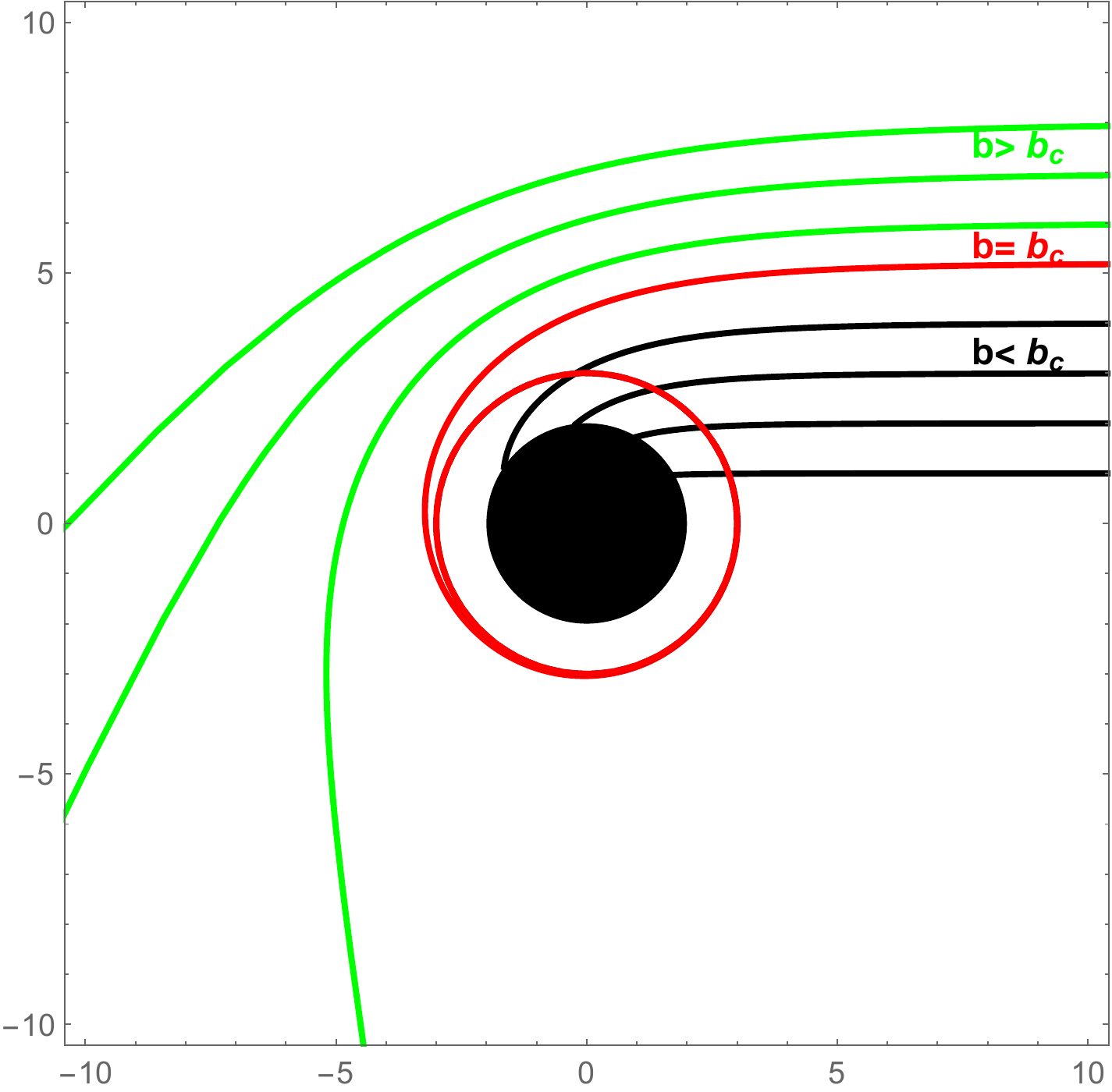}
	\caption{The photon trajectories around a Schwarzschild BH. The black curves represent trajectories with impact parameters $b<b_c$, the red curve represents the photon trajectory corresponding to the critical impact parameter $b_c$, and the green curves depict trajectories with $b>b_c$. The black disk represents the BH.}
	\label{fig_photon}
\end{figure}

The optical appearance of a Schwarzschild BH arises from photons near the BH being deflected and subsequently reaching the observer's screen, forming an image. This implies that we need study the trajectories of photons around the BH to analyze its optical appearance. In Schwarzschild coordinates, the spacetime line element can be expressed as:
\begin{equation}
	{\rm d} s^{2}=-f(r) {\rm d} t^{2}+ \frac{1}{f(r)} {\rm d} r^{2}+r^2\left({\rm d} \theta^{2}+\sin ^{2} \theta {\rm d} \varphi^{2}\right),\label{xianyuan}
\end{equation}
where
\begin{equation}
	\begin{split}
		f(r)=1-\frac{2 M}{r}.
	\end{split}
\end{equation}
Here $M$ denotes the mass of the BH. Due to spherical symmetry of spacetime, we restrict our analysis to photon trajectories confined to the equatorial plane ($\theta = \pi/2$). The tangent vector field $\left(\partial /\partial \tau \right)^a$ of the null geodesic satisfy~\cite{Wald:1984rg,Liang:2023ahd}
\begin{equation}
 \begin{split}
 0:=g_{ab}\left(\frac{\partial}{\partial \tau}\right)^a\left(\frac{\partial}{\partial \tau}\right)^b, \label{keq}
 \end{split}
\end{equation}
where $\tau$ is the affine parameter along the null geodesic. In the Schwarzschild coordinate system, Eq.~\eqref{keq} can be written as
\begin{equation}
 \begin{split}
 0&=g_{ab}\left(\frac{\partial}{\partial \tau}\right)^a\left(\frac{\partial}{\partial \tau}\right)^b\\
 &=-f(r)\left(\frac{{\rm d} t}{{\rm d} \tau}\right)^2+\frac{1}{f(r)}\left(\frac{{\rm d} r}{{\rm d} \tau}\right)^2+r^2\left(\frac{{\rm d} \varphi}{{\rm d} \tau}\right)^2.\label{keq1}
 \end{split}
\end{equation}
Note that in the Schwarzschild spacetime, there exist two Killing vector fields, $\left(\partial/\partial t\right)^{a}$ and $\left(\partial/\partial \varphi\right)^{a}$, which determine two conserved quantities, the energy $E$ and angular momentum $L$, by~\cite{Wald:1984rg,Liang:2023ahd}
\begin{equation}
 \begin{split}
 &E:=-g_{a b}\left(\frac{\partial}{\partial t}\right)^{a}\left(\frac{\partial}{\partial \tau}\right)^{b}=f(r) \frac{{\rm d} t}{{\rm d} \tau},\\
 &L:=g_{a b}\left(\frac{\partial}{\partial \varphi}\right)^{a}\left(\frac{\partial}{\partial \tau}\right)^{b}=r^2 \frac{{\rm d} \varphi}{{\rm d} \tau}.\label{EL1}
 \end{split}
\end{equation}
Inserting Eq.~\eqref{EL1} into Eq.~\eqref{keq1} yields the photon trajectory equation, namely,
\begin{equation}
 \begin{split}
 \left(\frac{{\rm d} r}{{\rm d} \varphi}\right)^2 =r^4 \left[\frac{1}{b^2}- \frac{f(r)}{r^2}\right].\label{ef}
 \end{split}
\end{equation}
Here, $b \equiv L/E$ represents the photon's impact parameter. For subsequent discussion, we reformulate the above expression using the substitution \( u = 1/r \), and get
\begin{equation}
	\begin{split}
		\left(\frac{{\rm d} u}{{\rm d} \varphi}\right)^2 = \frac{1}{b^2}- f\left(\frac{1}{u}\right)u^2.\label{guiji}
	\end{split}
\end{equation}
Hence, given a specific impact parameter $b$ associated to a photon, the trajectory of the photon can be determined through Eq.~\eqref{ef} [or Eq.~\eqref{guiji}]. Figure~\ref{fig_photon} schematically displays the photon trajectories around the BH (represented by the black disk) for different impact parameters. Specifically, the red curve corresponds to photons with $b = b_c$ forming an unstable circular orbit around the BH due to gravitational deflection, where $b_c$ is referred to as the critical impact parameter. The black curves depict trajectories of photons with $b<b_c$ that ultimately fall into the event horizon, while the green curves represent photons with $b>b_c$ that are deflected by the BH and escape back to infinity.

Notably, due to spherical symmetry, it turns out that any one photon trajectory around a Schwarzschild BH can always be confined to a plane passing through the center of the BH. In such a plane, the aforementioned unstable circular orbits exist. These unstable circular orbits collectively form an unstable photon sphere around the BH, where perturbed photons either fall into the BH or escape to infinity. Hence, when a BH is illuminated from behind by a large, distant planar screen that emits isotropiclly with uniform brightness, a perfectly circular BH shadow silhouette is formed in the observer's view at all inclination angles~\cite{Gralla:2019xty}. However, when considering luminous accretion disks, the optical appearance of the BH becomes dependent on both the disk type and observer's inclination angle. In what follows, we will examine the optical appearance of BHs surrounded by accretion disks at various inclination angles.

\section{Trajectories of photons emitted from the thin accretion disk}\label{section3}

\subsection{Geometry of the black hole-observer system}

\begin{figure}[htbp]
	\centering
	\includegraphics[width=8.8cm]{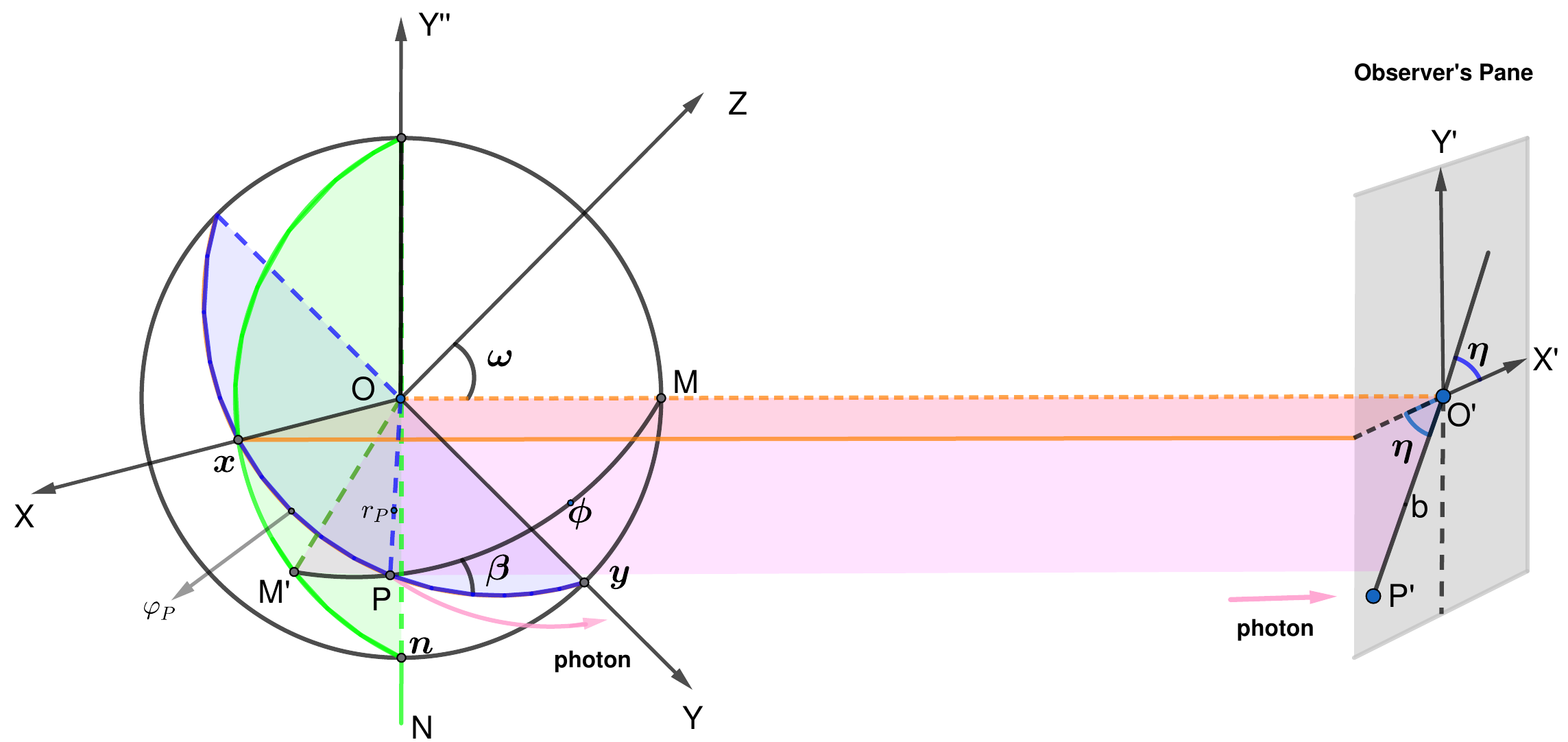}
	\caption{The schematic diagram of the observer and BH coordinate systems. The BH center is located at point $O$, and the observer is positioned at point $O'$. The viewing inclination is denoted by $\omega$. A photon emitted from point $P$ on the accretion disk arrives at point $P'$ on the observer's plane, undergoing a total azimuthal angle change $\angle POM \equiv \phi$, with $\eta$ representing the polar angle in the observer's plane.}
	\label{fig_yanshinew}
\end{figure}

For the Schwarzschild BH, we consider the presence of an optically and geometrically thin disk as the light source surrounding it. The disk emits radiation isotropically and is observed by distant observers at various inclination angles. To study the appearance of a Schwarzschild BH surrounded by the disk, it is convenient to introduce a (local) Cartesian coordinate system $\{X,Y,Z\}$ based on the center of the BH in Fig.~\ref{fig_yanshinew}~\cite{Luminet:1979nyg,Cui:2024wvz,Igata:2025glk}. Assuming that the BH center and a distant observer are located at point $O$ and $O'$ respectively, and the accretion disk lies strictly in the equatorial plane ($\theta=\pi/2$) of the BH. We specify the disk's symmetry axis as the $OZ$-axis, and the intersection between the equatorial plane (the blue plane in Fig.~\ref{fig_yanshinew}) and a plane passing through the $OZ$-axis and the radial line $OO'$ as the $OY$-axis. The plane through the $OZ$-axis and $OO'$ can be obtained by rotating the equatorial plane about $O$. Thus, the $OX$-axis in the equatorial plane is determined by requiring that the coordinate system $\{X,Y,Z\}$ is a right-handed one, in the other words, the $OX$-axis is obtained by rotating counterclockwise about the $OZ$-axis through $\pi/2$. Once the coordinate system $\{X,Y,Z\}$ is specified, the spherical coordinate system $\{r,\theta,\varphi\}$ adapted to the coordinate system $\{X,Y,Z\}$ is determined in the conventional sense. Hence, any one point in space can then be described by spherical coordinates $(r,\theta,\varphi)$, for instance, $O'$ has coordinates $(r_o,\omega,\pi/2)$ with $r_o \gg M$. The angle $\omega$ is called the inclination angle. Next, we introduce the observer's plane which passes through the observer's position $O'$ and orthogonal (measured by the Schwarzschild metric) to the radial line $OO'$. Furthermore, in this plane, we construct a coordinate system $\{X',Y'\}$ based on $O'$, called as the observer's coordinate system. The $O'Y'$-axis is specified as the intersection of the observer's plane with the plane formed by the $OZ$-axis and the radial line $OO'$, oriented in the direction of decreasing $\theta$. The $O'X'$-axis is aligned with the intersection of the observer's plane and the plane through the $OX$-axis and the radial line $OO'$, pointing in the direction of increasing $\varphi$. Moreover, we perform a counterclockwise rotation of the $\{X,Y,Z\}$ frame by $\pi/2-\omega$ about the $OX$-axis, which brings the $OZ$-axis into alignment with $OY''$. Thus, the plane $XOY''$ maintains a $\pi/2$ angle relative to $OO'$.

Consider a point $P$ with coordinates $(r_P, \pi/2, \varphi_P)$ in the accretion disk plane and a spherical surface with a radius $r_P$ centered at $O$. Denote the intersections of the surface with the coordinate axes $OX$ and $OY$ by points $x$ and $y$, respectively. The trajectory of a photon emitted from $P$ is confined to a plane through $O$ (the pink plane in Fig.~\ref{fig_yanshinew}), referred to as a trajectory plane. The intersections of these planes with the spherical surface are marked in Fig.~\ref{fig_yanshinew}. The photon ultimately reaches a point $P'$ on the observer's plane, forming a polar angle $\eta$ with the $O'X'$-axis while accumulating a total azimuthal angle change $\phi$ during propagation. By applying the spherical sine theorem to the spherical triangles, one can derive the fundamental relation among $\phi$, $\omega$, and $\eta$ as~\cite{Luminet:1979nyg}
\begin{equation}
	\cos{\phi}=\frac{\sin{\eta}}{\sqrt{\sin^2{\eta}+\cot^2{\omega}}}.\label{jiaodu}
\end{equation}
Hence, we obtain
\begin{equation}
	\phi=\arctan\left(\frac{1}{\tan{\omega} \sin{\eta}}\right).\label{jiaodu1}
\end{equation}
\begin{figure}[htbp]
	\centering
	\includegraphics[width=9cm]{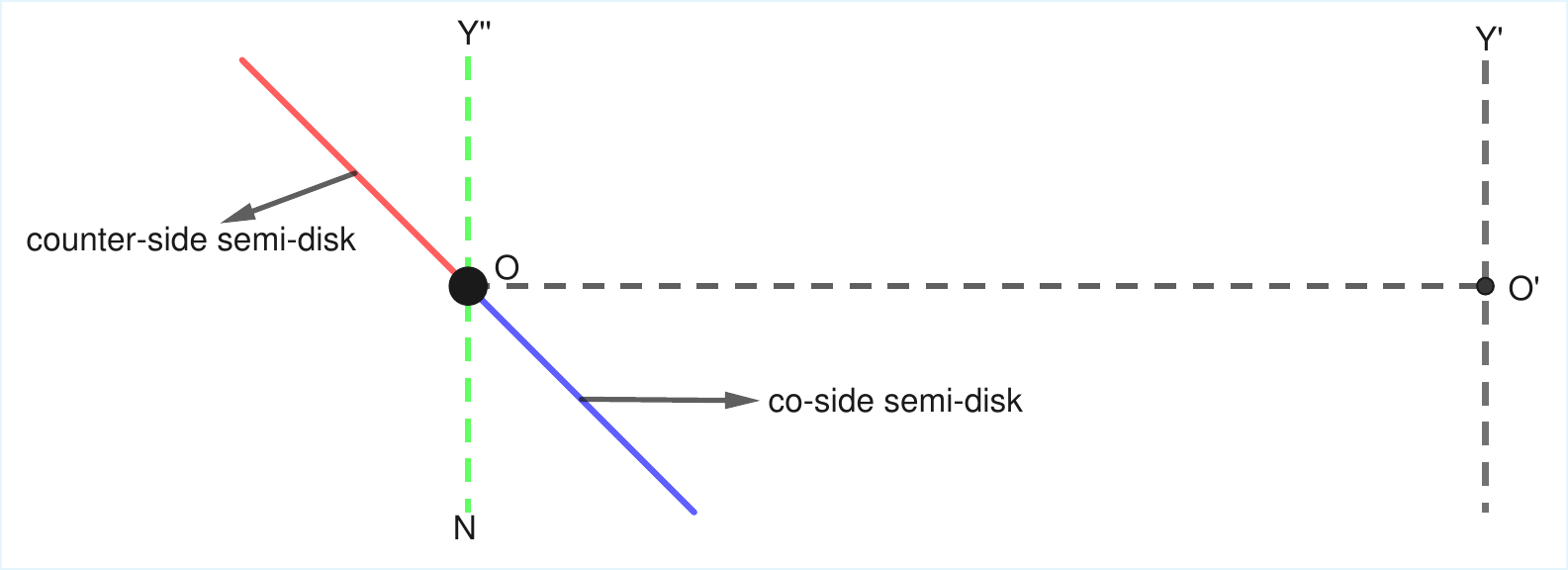}
	\caption{Schematic diagram of the co-side and counter-side semi-disks, where $O$ and $O'$ denote the positions of the BH and the observer, respectively. The dashed lines $NY''$ and $O'Y'$ represent the intersections of the plane $OO'Y'$ with the planes $XOY''$ and $X'O'Y'$ from Fig.~\ref{fig_yanshinew}, respectively.}
	\label{fig_semidisk}
\end{figure}

To establish a one-to-one correspondence between all points on the accretion disk and those on the observer's plane, it is convenient to partition the BH's accretion disk along the $OX$-axis by the $XOY''$ plane (the green plane in Fig.~\ref{fig_yanshinew}) into two distinct semi-disks: the co-side semi-disk (located on the same side as the observer, to the right of the $XOY''$ plane) and the counter-side semi-disk (on the opposite side, to the left of the $XOY''$ plane). Figure~\ref{fig_semidisk} shows a schematic diagram of the cross-section formed by intersecting the accretion disk with a plane (a dash line) passing through both the BH and the observer, clearly illustrating the relative positions of the co-side and counter-side semi-disks.

\subsection{Classification of photon trajectories}

Photons emitted from the accretion disk and reaching the observer can equivalently be treated as backward-traced light rays from the observer intersecting with the disk near the BH. Each intersection corresponds to brightness contribution from the disk. Since light rays may intersect the accretion disk multiple times in the case of optically thin, we follow the conventional classification scheme of photon trajectories proposed in~\cite{Gralla:2019xty}, and define direct emission for rays intersecting the disk once, lensed emission for twice, and photon ring emission for three or more times by considering the orbit number $n =\phi/{(2\pi)}$, where $\phi$ represents the total change in orbital plane azimuthal angle outside the horizon. Combining with Eq.~\eqref{jiaodu1}, we have
\begin{equation}
	n =\frac{\phi}{2\pi}=\frac{1}{{2\pi}}\arctan\left(\frac{1}{\tan{\omega} \sin{\eta}}\right).\label{n1}
\end{equation}

\begin{figure}[htbp]
	\centering
	\includegraphics[width=9cm]{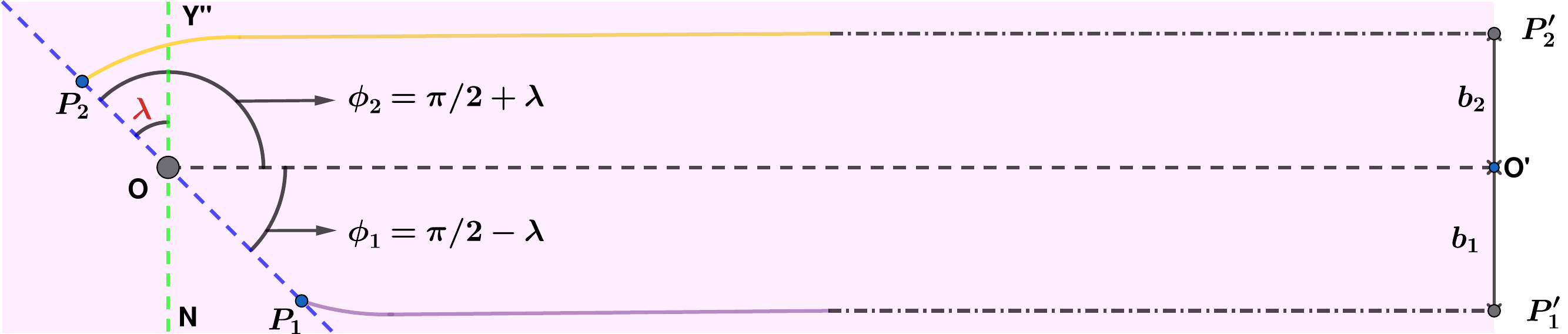}
	\caption{Schematic of geometric relations in the photon trajectory plane, where $O$ is the BH position and $O'$ corresponds to the center of the observer’s plane. The blue dashed line represents the accretion disk, while the green dashed line indicates the plane $XOY''$.}
	\label{fig_yanshi11}
\end{figure}

When photons arrive at the plane $XOY''$ from a distant observer, they undergo an azimuthal angle change of $\pi/2$. In Fig.~\ref{fig_yanshi11}, we single out the trajectory plane (the pink plane) in Fig.~\ref{fig_yanshinew} and take the case where photons intersect each semi-disk once as an example, in order to show the dependence of the total changes $\phi$ on these two semi-disks. The two points $P_{2}$ and $P_{1}$ represent the intersection points of photons with the counter-side and co-side semi-disks, respectively. It is easy to see that from Fig.~\ref{fig_yanshi11} one only needs to calculate the angle $\lambda$ between the plane $XOY''$ and the accretion disk within this trajectory plane to determine the total azimuthal angle of photons reaching the two semi-disks, and thus obtain the total change $\phi=\pi/2\pm\lambda$, where $+$ for the counter-side semi-disk while $-$ for the co-side semi-disk. Based on the geometric relations in Fig.~\ref{fig_yanshinew}, we get
\begin{align}
 \lambda=\arctan{(\tan{\omega} \sin{\eta})}.
\end{align}
Through further generalization, we find that for a light ray emitted backward from the observer, which initially intersects the counter-side semi-disk and subsequently experiences multiple intersections with the accretion disk, the azimuthal angle accumulated at each intersection can be expressed as
\begin{align}
	\phi=\frac{2m-1}{2}\pi+\arctan{({\tan{\omega} \sin{\eta}})}, \; m=1,2,3,\cdots, \label{counterside}
\end{align}
where $m$ denotes the $m$-th intersection with the accretion disk. Analogously, for the co-side semi-disk, the corresponding azimuthal angle reads
\begin{align}
	\phi=\frac{2m-1}{2}\pi-\arctan{({\tan{\omega} \sin{\eta}})}, \; m=1,2,3,\cdots.\label{coside}
\end{align}
Hence, for a given $m$-th intersection, the the azimuthal angles of the two distinct panels differ. This difference serves as a more significant rationale for partitioning the accretion disk into two panels.

\begin{figure*}[htbp]
	\centering
	 \begin{minipage}{0.45\textwidth}
		\includegraphics[width=3in,height=5in,keepaspectratio]{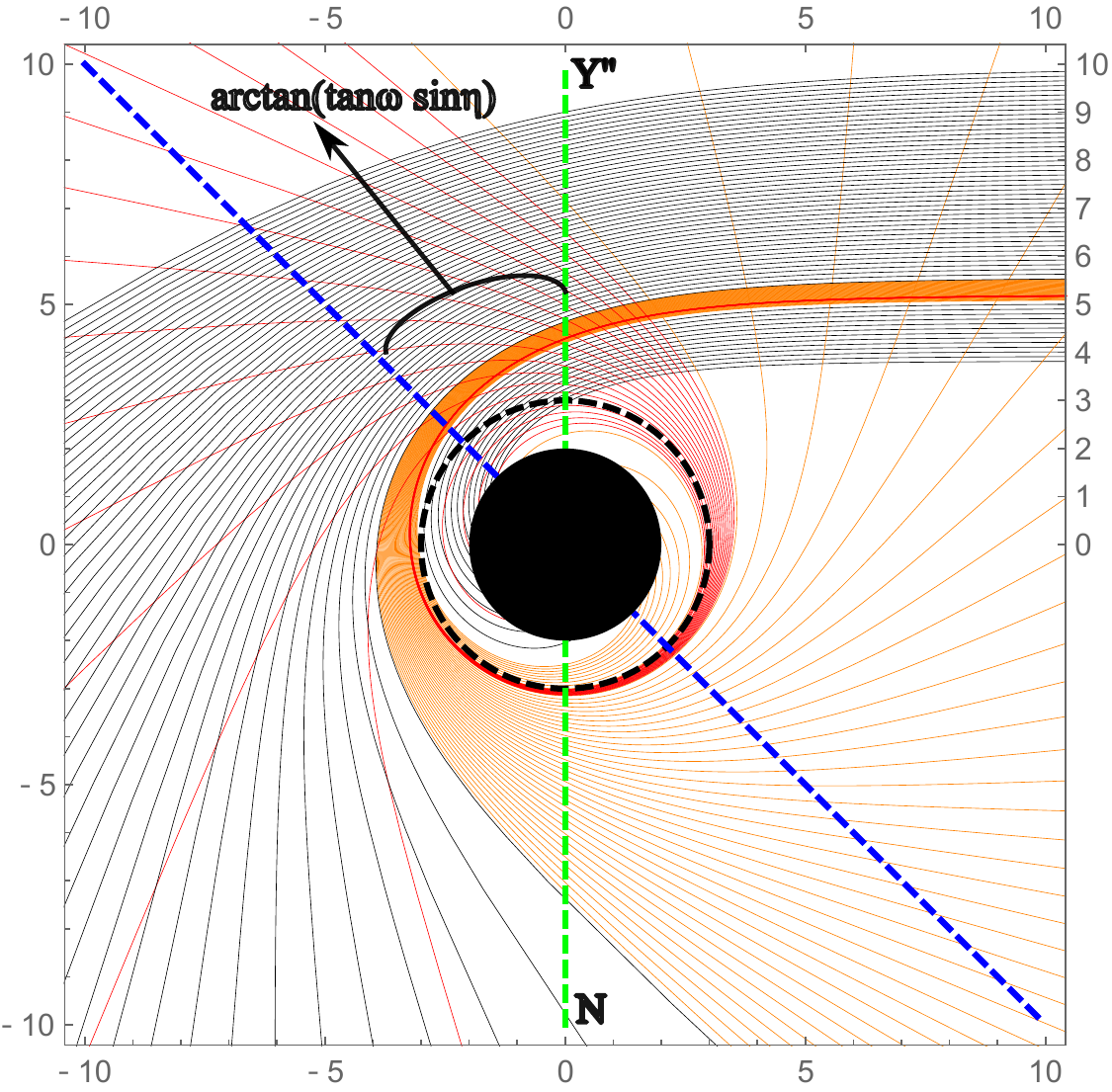}
	\end{minipage}
	\hfill
	 \begin{minipage}{0.45\textwidth}
		\includegraphics[width=3in,height=5in,keepaspectratio]{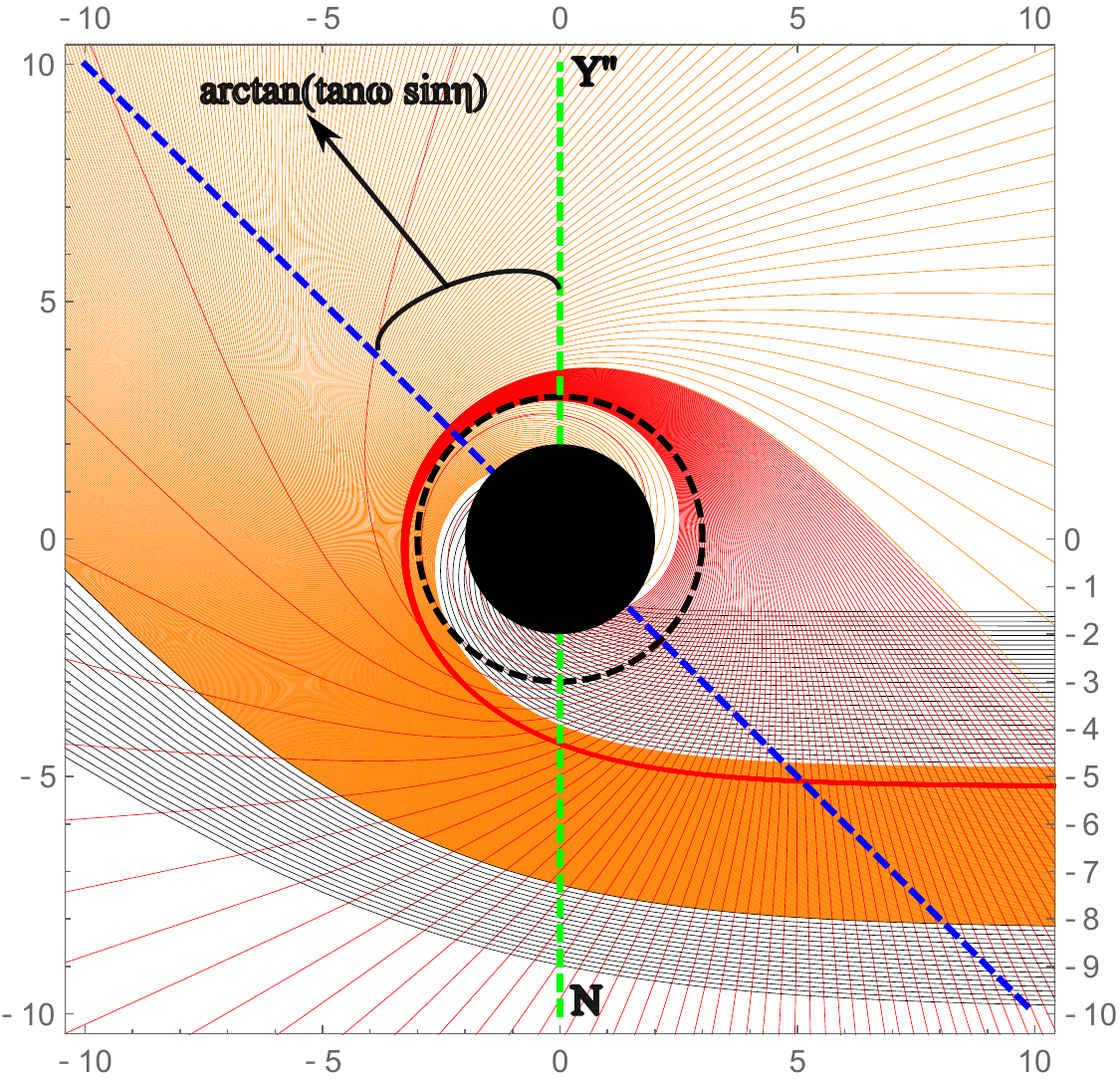}
	\end{minipage}

	\caption{Light trajectories from a distant right-side observer with inclination $\omega$, intersecting the counter-side semi-disk (left) and co-side semi-disk (right) of a Schwarzschild BH. The black, orange, and red curves correspond to direct, lensed, and photon ring trajectories, respectively. The blue dashed curve depicts the accretion disk, while the green dashed line represents the $XOY''$ plane.}
	\label{fig_photon_new}
\end{figure*}

In Fig.~\ref{fig_photon_new}, we demonstrate the intersection patterns between light rays (emitted from a distant observer at infinity on the right side) and the accretion disk around a Schwarzschild BH at non-zero inclination angles, specifically showing the counter-side semi-disk (left panel) and co-side semi-disk (right panel). In Fig.~\ref{fig_photon_new}, the green dashed line $NY''$ represents the $XOY''$ plane in Fig.~\ref{fig_yanshinew}, while the blue dashed line indicates the equatorial plane of the accretion disk. Black, orange, and red curves in Fig.~\ref{fig_photon_new} denote direct, lensed, and photon ring trajectories, respectively. The distinct intersection patterns between the counter-side semi-disk and co-side semi-disk motivates us to take separate treatments of the orbit numbers for the two semi-disks. In what follows, we denote the orbit numbers for the counter-side and the co-side semi-disks by $n_1$ and $n_2$, respectively.

\begin{figure*}[htbp]
	\centering
	\begin{minipage}{0.45\textwidth}
		\includegraphics[width=3in,height=5in,keepaspectratio]{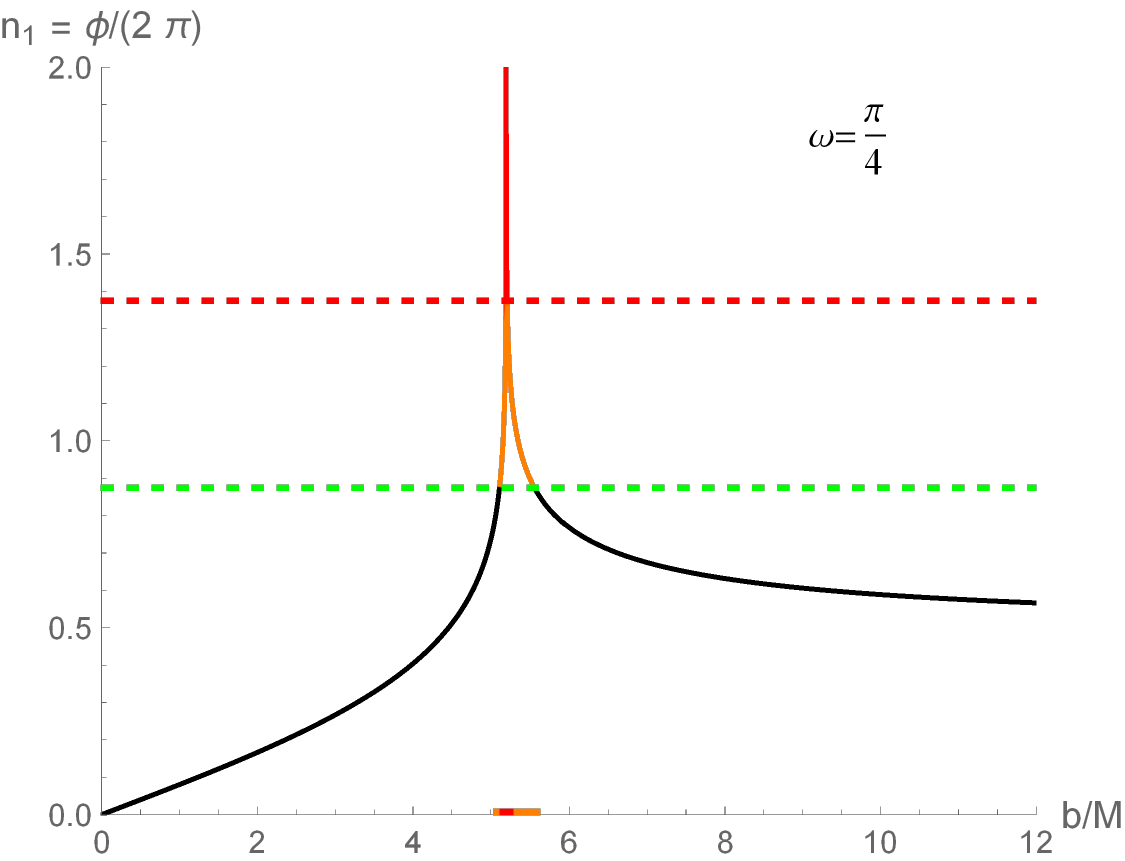}
	\end{minipage}
	\hfill
	\begin{minipage}{0.45\textwidth}
		\includegraphics[width=3in,height=5in,keepaspectratio]{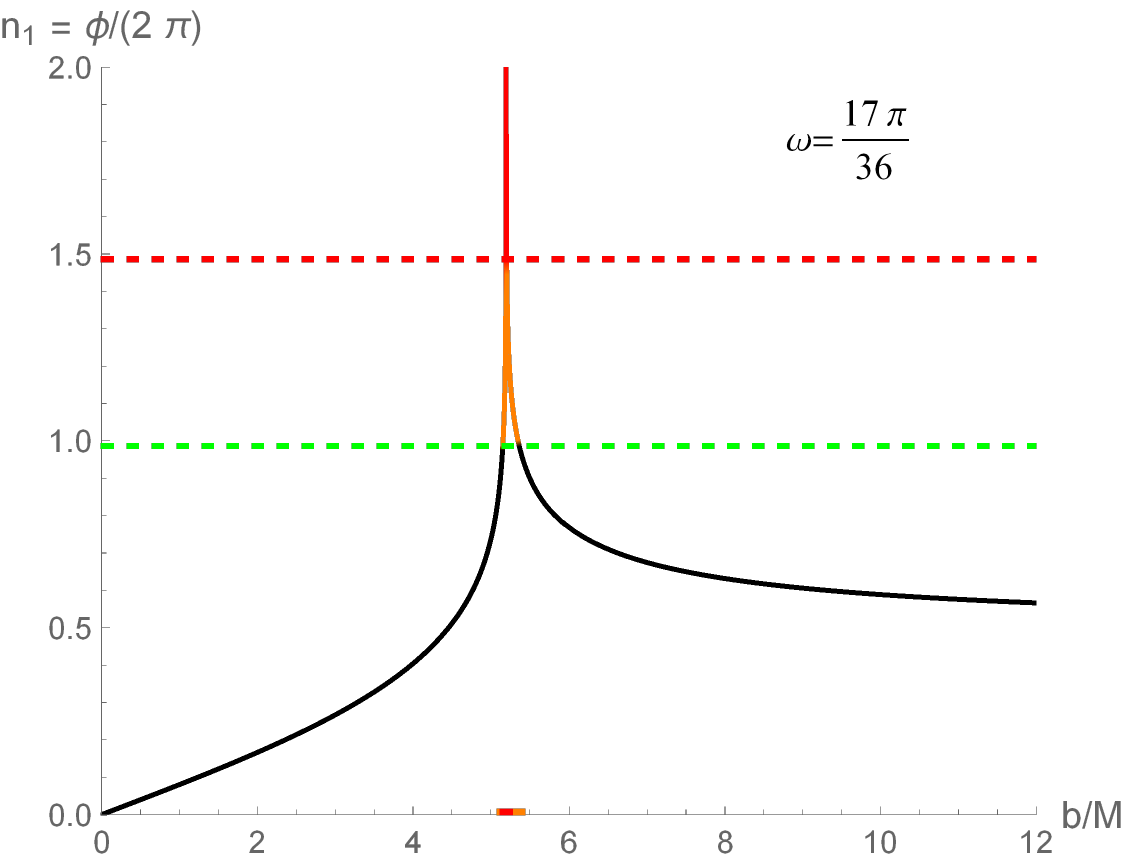}
	\end{minipage}
	\hfill
	\begin{minipage}{0.45\textwidth}
		\includegraphics[width=3in,height=5in,keepaspectratio]{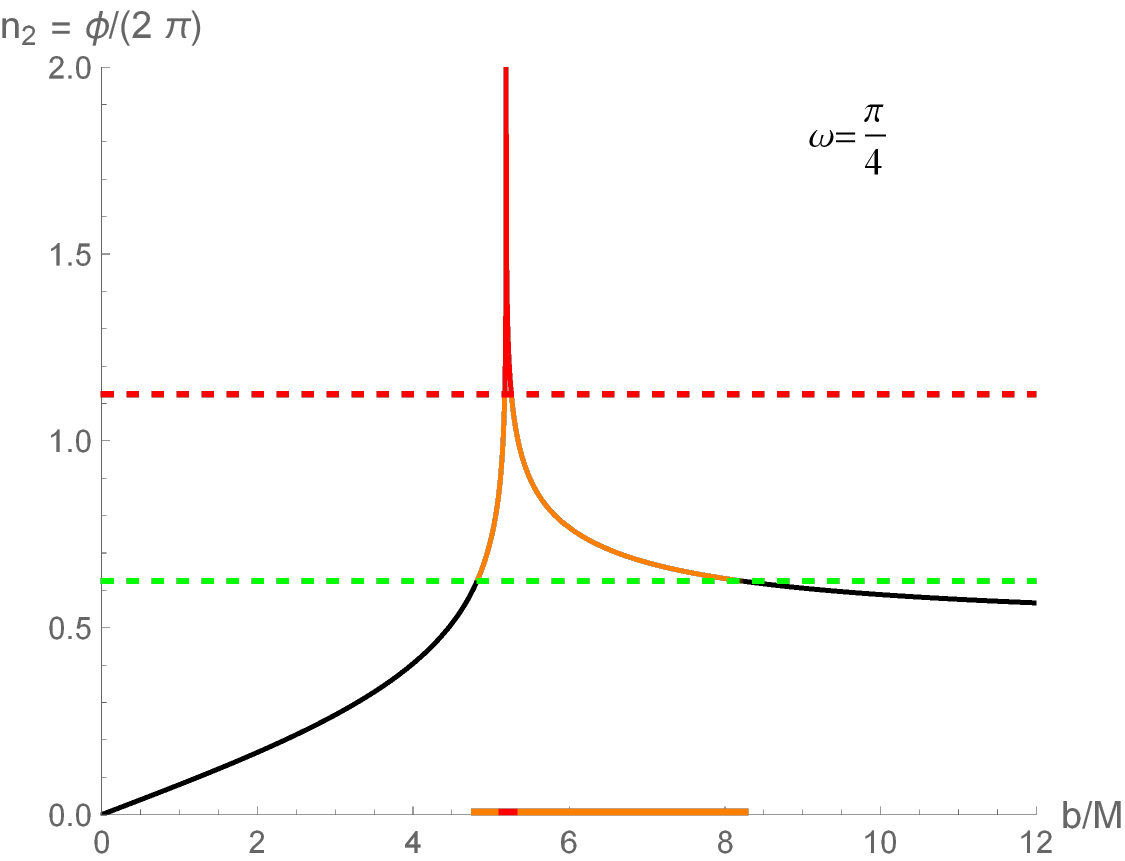}
	\end{minipage}
	\hfill
	\begin{minipage}{0.45\textwidth}
		\includegraphics[width=3in,height=5in,keepaspectratio]{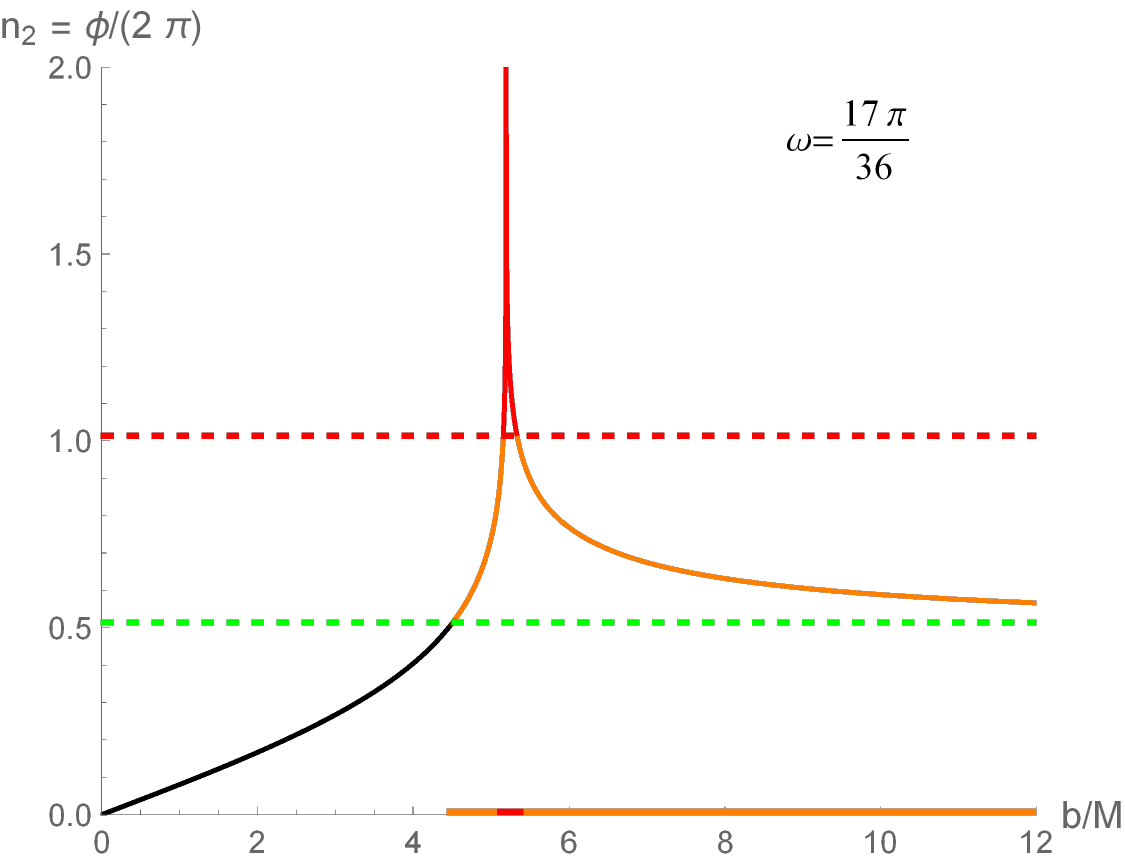}
	\end{minipage}

	\caption{The orbit numbers at fixed $\eta=\pi/2$ for varying inclination angles $\omega$. The first row shows the counter-side orbit numbers ($n_1$), while the second row displays the co-side orbit numbers ($n_2$). The black, orange, and red curves correspond to direct, lensed, and photon ring emissions respectively.}
	\label{fig_n1b}
\end{figure*}

\begin{figure*}[htbp]
	\centering
	\begin{minipage}{0.45\textwidth}
		\includegraphics[width=3in,height=5in,keepaspectratio]{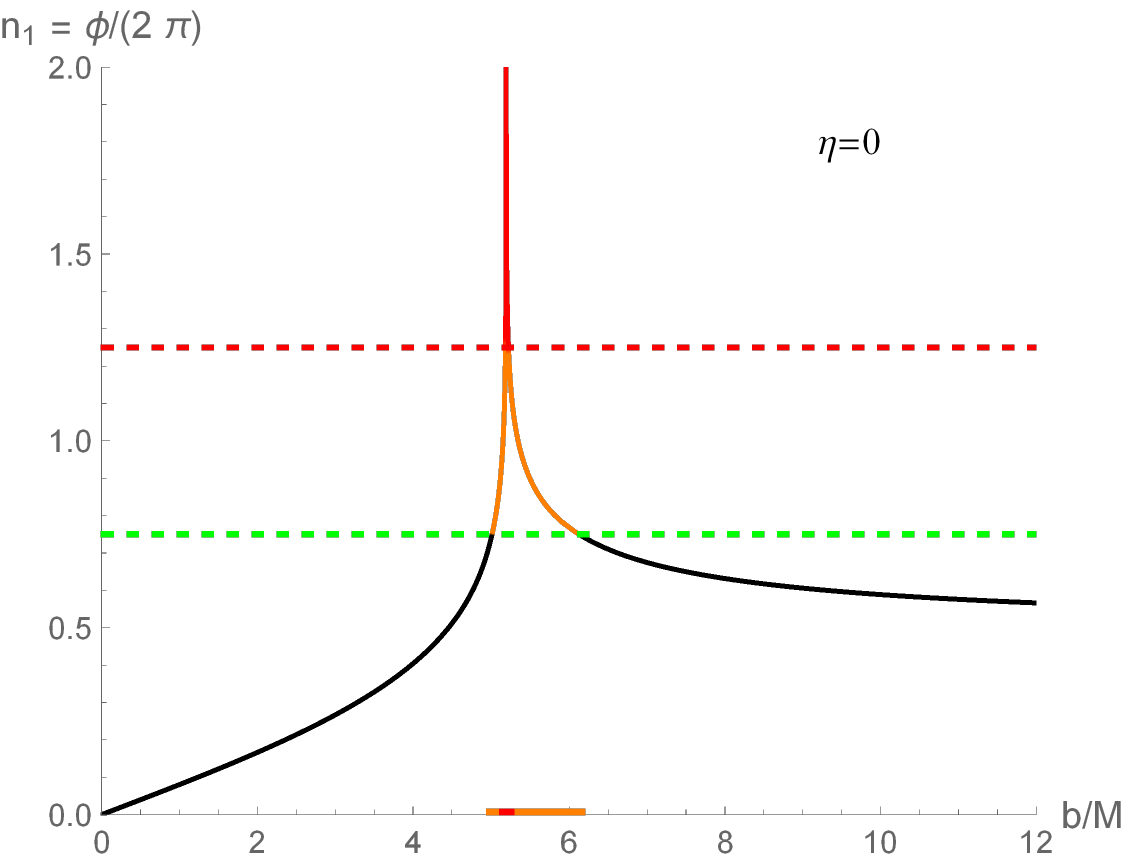}
	\end{minipage}
	\hfill
	\begin{minipage}{0.45\textwidth}
		\includegraphics[width=3in,height=5in,keepaspectratio]{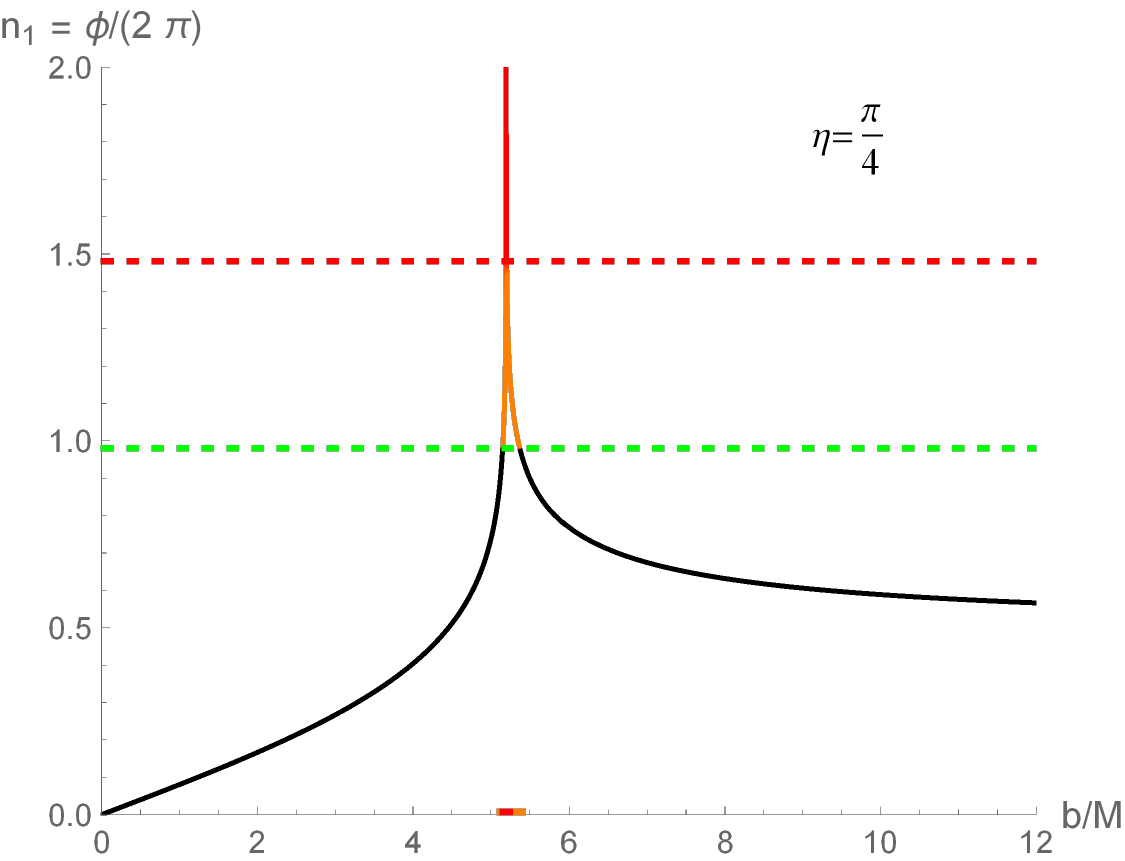}
	\end{minipage}
	\hfill
	\begin{minipage}{0.45\textwidth}
		\includegraphics[width=3in,height=5in,keepaspectratio]{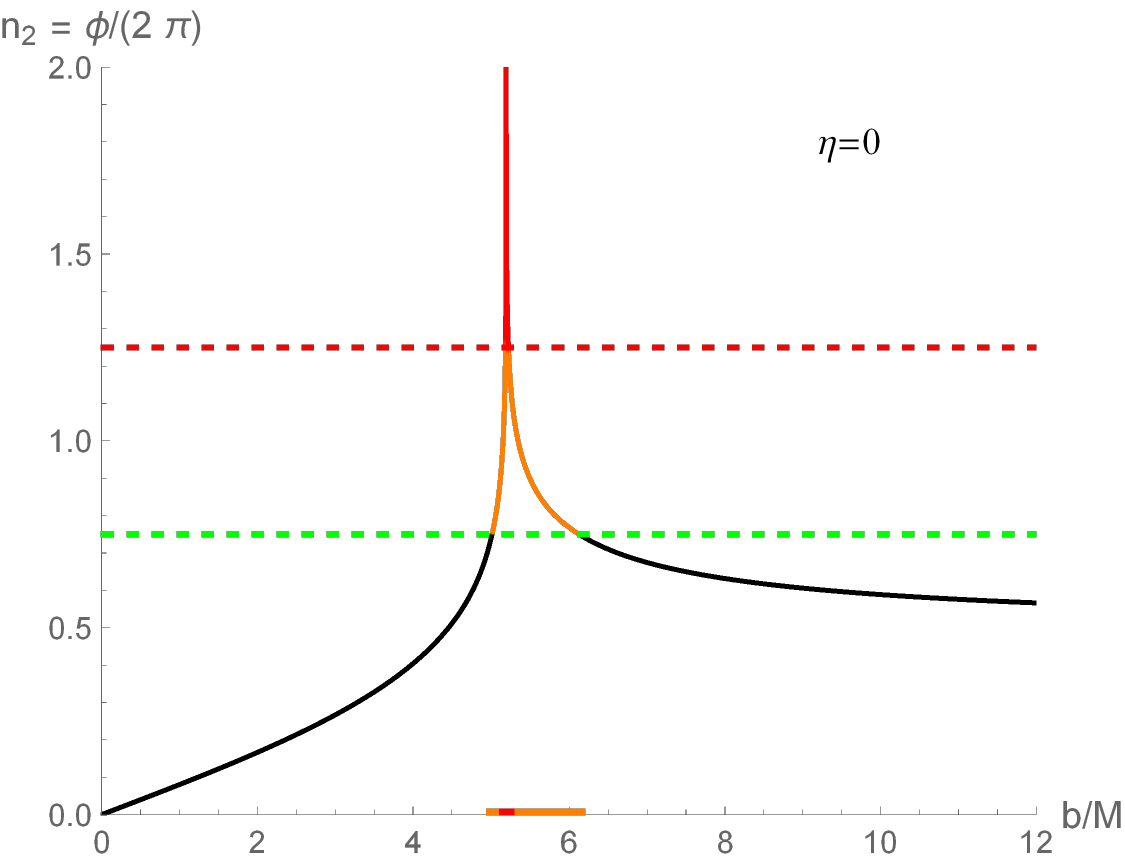}
	\end{minipage}
	\hfill
	\begin{minipage}{0.45\textwidth}
		\includegraphics[width=3in,height=5in,keepaspectratio]{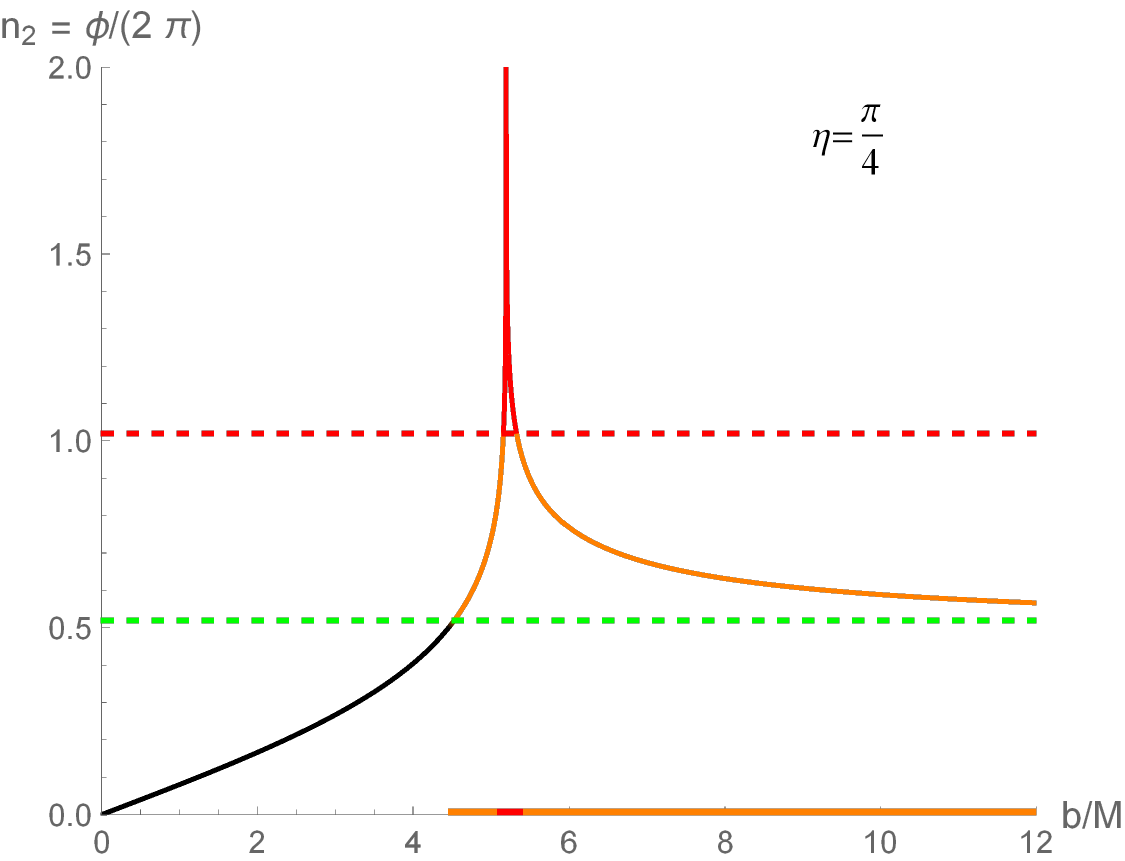}
	\end{minipage}

	\caption{The orbit numbers at fixed $\omega=17\pi/36$ for varying inclination angles $\eta$. The first row shows the counter-side orbit numbers ($n_1$), while the second row displays the co-side orbit numbers ($n_2$). The black, orange, and red curves correspond to direct, lensed, and photon ring emissions respectively.}
	\label{fig_n2b}
\end{figure*}

In Figs.~\ref{fig_n1b} and \ref{fig_n2b}, using Eqs.~\eqref{n1}, \eqref{counterside} and \eqref{coside}, we present the variations of orbit numbers $n_1$ and $n_2$ with impact parameter $b$ for different values of $\omega$ and $\eta$, respectively. In Fig.~\ref{fig_n1b}, with the polar angle fixed at $\eta=\pi/2$, the impact parameter range corresponding to the lensed emission region of the counter-side semi-disk gradually decreases with increasing viewing inclination, while that of the co-side semi-disk shows significant enlargement. Figure~\ref{fig_n2b} demonstrates similar trends at fixed inclination $\omega=17\pi/36$ across various polar angles $\eta$. Additionally, we find the orbit numbers remain invariant under inclination changes when $\eta=0$.

It is crucial to emphasize that for a given polar angle $\eta$ in the observer's plane, the counter-side and co-side semi-disks are imaged in the upper and lower half-planes of the $O'X'$ axis in the observer's coordinate system, respectively, as shown in Fig.~\ref{fig_photon_new}. Moreover, the observed image exhibits exact symmetry about the $O'Y'$ axis. This symmetry reduces our analysis to the range $0\leq \eta \leq \pi/2$ while fully characterizing the entire image. All subsequent $\eta$ values are implicitly constrained to this angular domain without repeated specification.

\subsection{Transfer functions}

To determine the optical appearance of a BH surrounded by a disk at various inclination angles, we must establish a map between each point on the accretion disk and its corresponding position on the observer's plane. Indeed, the transfer function can be introduced to characterize the map~\cite{Gralla:2019xty}. Notably, since light rays from the counter-side and co-side semi-disks undergo different azimuthal angles and cannot be treated uniformly, one needs to consider the corresponding transfer functions separately. For the counter-side semi-disk, the transfer function $r_{\rm m}(b)\equiv r(\phi,b)=1/u(\phi,b)$, as a function of the inclination angle $\omega$ and the observer's polar angle $\eta$, reads
\begin{equation}
	r_{\rm m}(b)=\frac{1}{u\left[\frac{2m-1}{2}\pi+\arctan{({\tan{\omega} \sin{\eta}})}, b\right]}, \; m=1,2,3,\cdots.
	\label{r1function}
\end{equation}
Here, $m$ denotes the $m$-th intersection between the light ray and the accretion disk. Similarly, for the co-side semi-disk, the transfer function $R_{\rm m}(b)\equiv r(\phi,b)=1/u(\phi,b)$, which is also a function of $\omega$ and $\eta$, takes the form
\begin{equation}
	R_{\rm m}(b)=\frac{1}{u\left[\frac{2m-1}{2}\pi-\arctan{({\tan{\omega} \sin{\eta}})},b \right]}, \; m=1,2,3,\cdots.
		\label{R1function}
\end{equation}
It is worth to noting that the counter-side transfer function essentially characterizes the intersection positions on the entire accretion disk for light rays originating from each point in the observer's upper half-plane, where these intersection points can also lie on the co-side disk. Conversely, the co-side transfer function describes the intersection positions on the entire accretion disk for light rays from each point in the observer's lower half-plane. These transfer functions can be obtained by numerically solving Eq.~\eqref{guiji}.

\begin{figure*}[htbp]
	\centering
	\begin{minipage}{0.45\textwidth}
		\includegraphics[width=3in,height=5in,keepaspectratio]{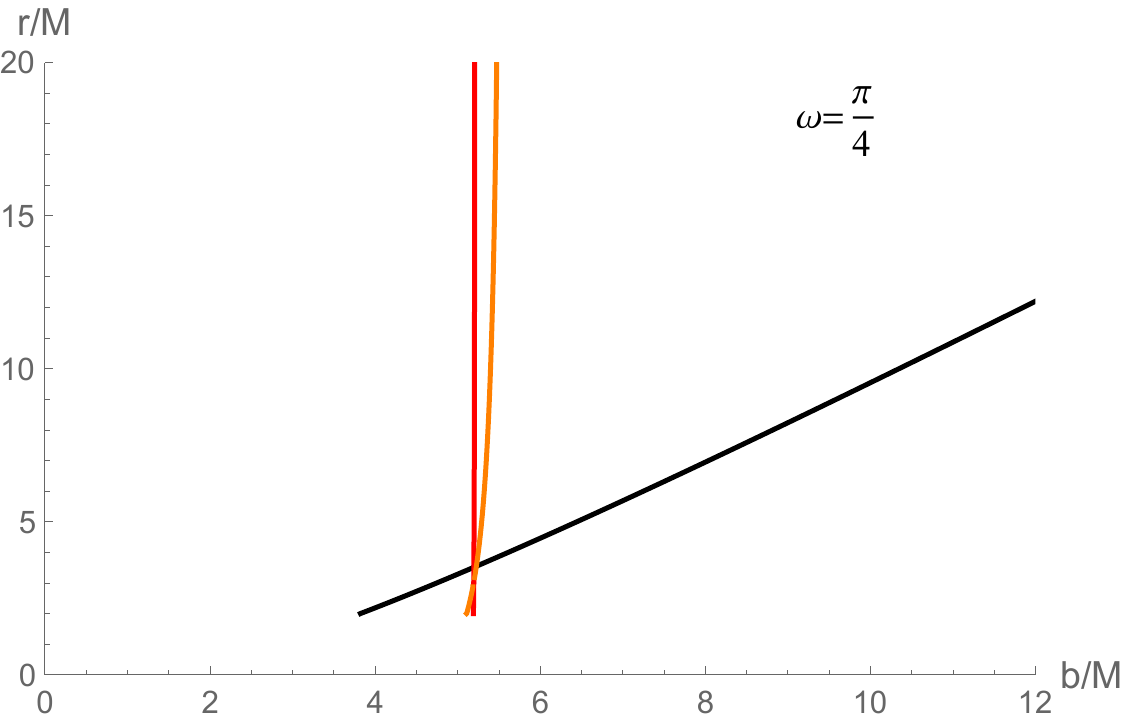}
	\end{minipage}
	\hfill
	\begin{minipage}{0.45\textwidth}
		\includegraphics[width=3in,height=5in,keepaspectratio]{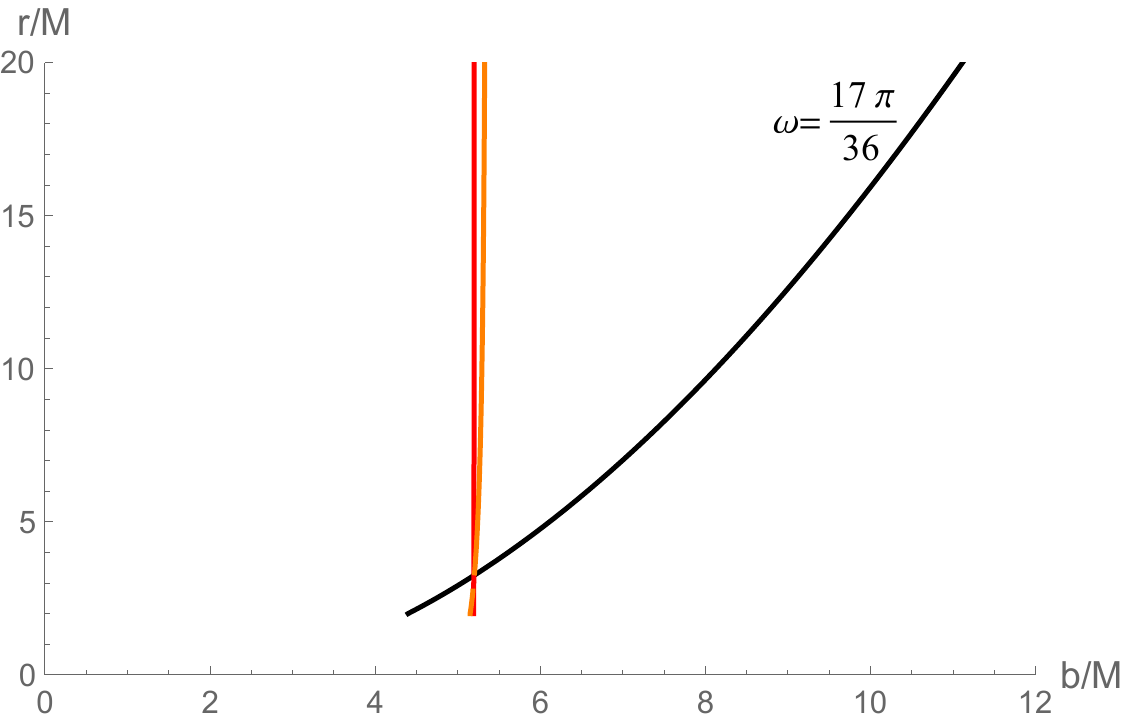}
	\end{minipage}
	\hfill
	\begin{minipage}{0.45\textwidth}
		\includegraphics[width=3in,height=5in,keepaspectratio]{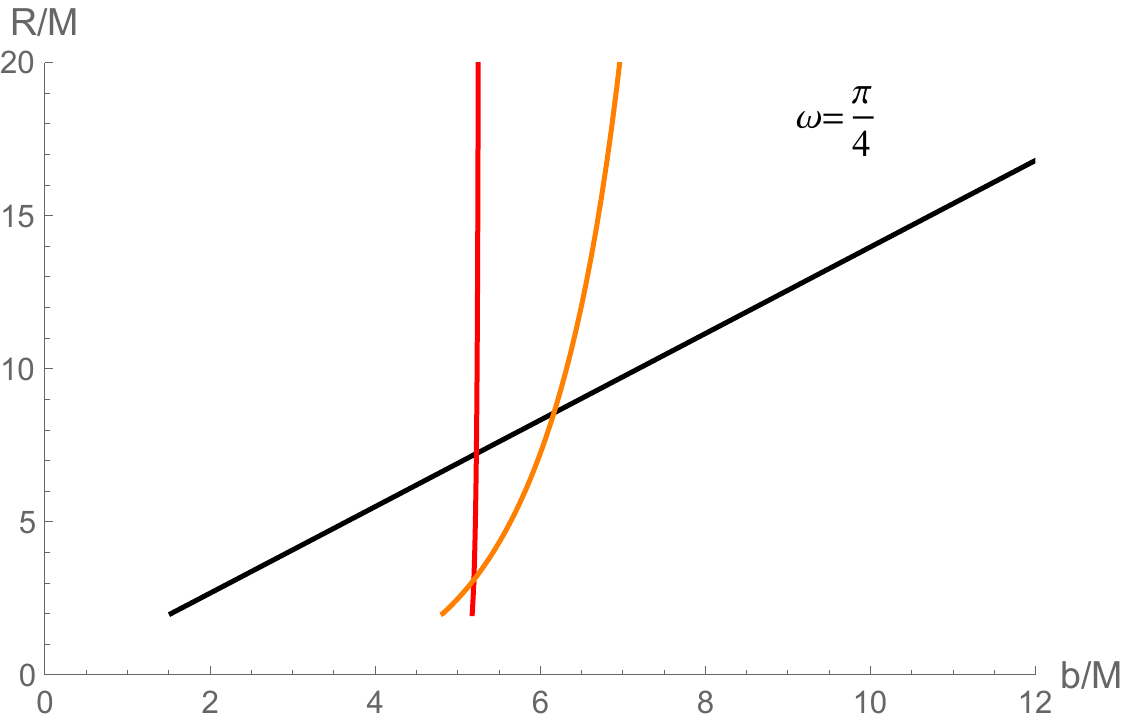}
	\end{minipage}
	\hfill
	\begin{minipage}{0.45\textwidth}
		\includegraphics[width=3in,height=5in,keepaspectratio]{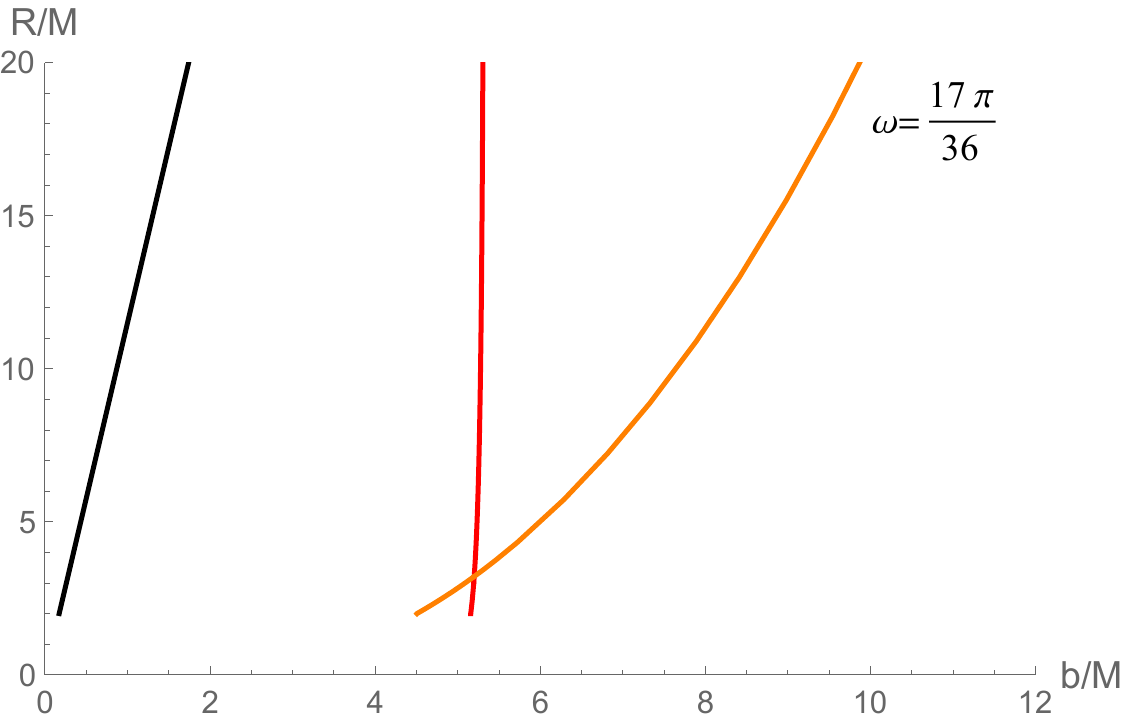}
	\end{minipage}

	\caption{Transfer functions at fixed $\eta=\pi/2$ for varying inclination angles $\omega$. The first row presents the counter-side transfer functions $r_{\rm m}(b)$, whereas the second row corresponds to the co-side transfer functions $R_{\rm m}(b)$. The black, orange, and red curves represent the transfer functions for $m=1$, $m=2$, and $m=3$ respectively.}
	\label{fig_chuan_r1b}
\end{figure*}

\begin{figure*}[htbp]
	\centering
	\begin{minipage}{0.45\textwidth}
		\includegraphics[width=3in,height=6in,keepaspectratio]{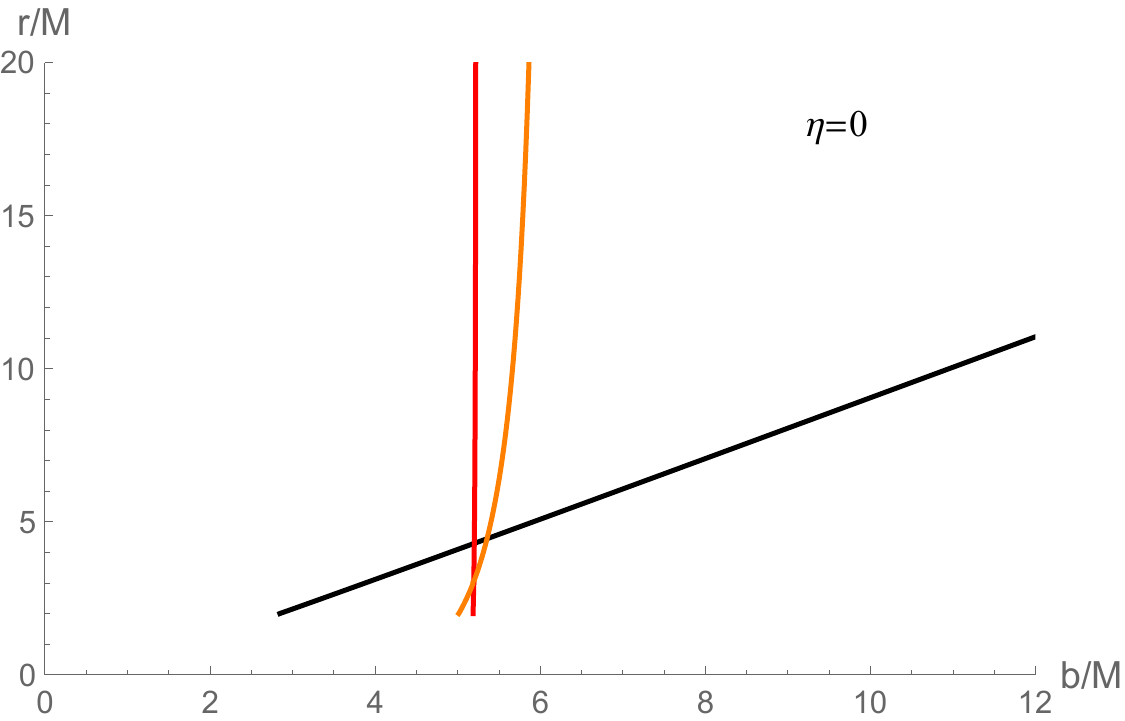}
	\end{minipage}
	\hfill
	\begin{minipage}{0.45\textwidth}
		\includegraphics[width=3in,height=6in,keepaspectratio]{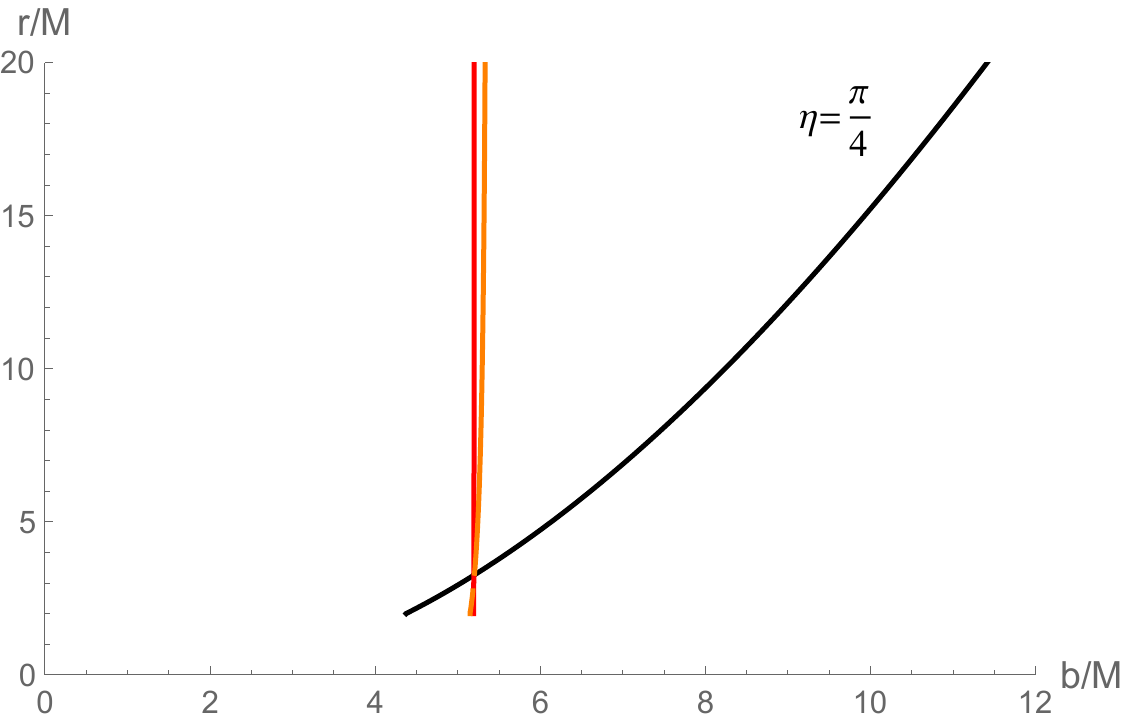}
	\end{minipage}
	\hfill
	\begin{minipage}{0.45\textwidth}
		\includegraphics[width=3in,height=6in,keepaspectratio]{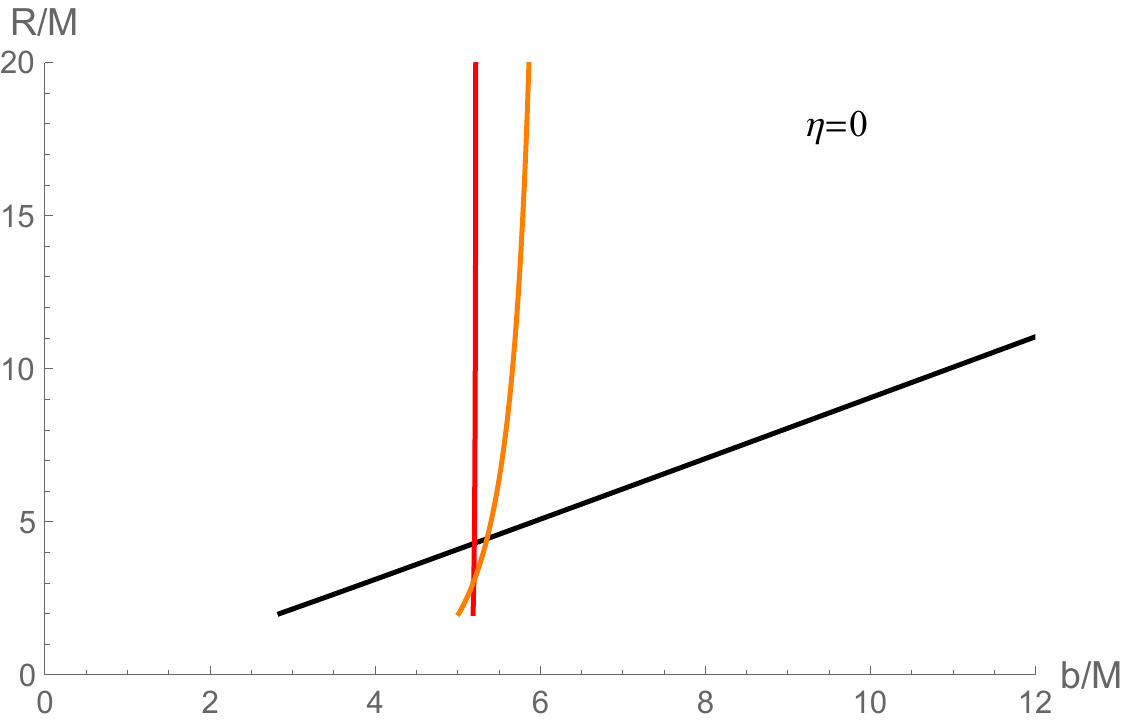}
	\end{minipage}
	\hfill
	\begin{minipage}{0.45\textwidth}
		\includegraphics[width=3in,height=6in,keepaspectratio]{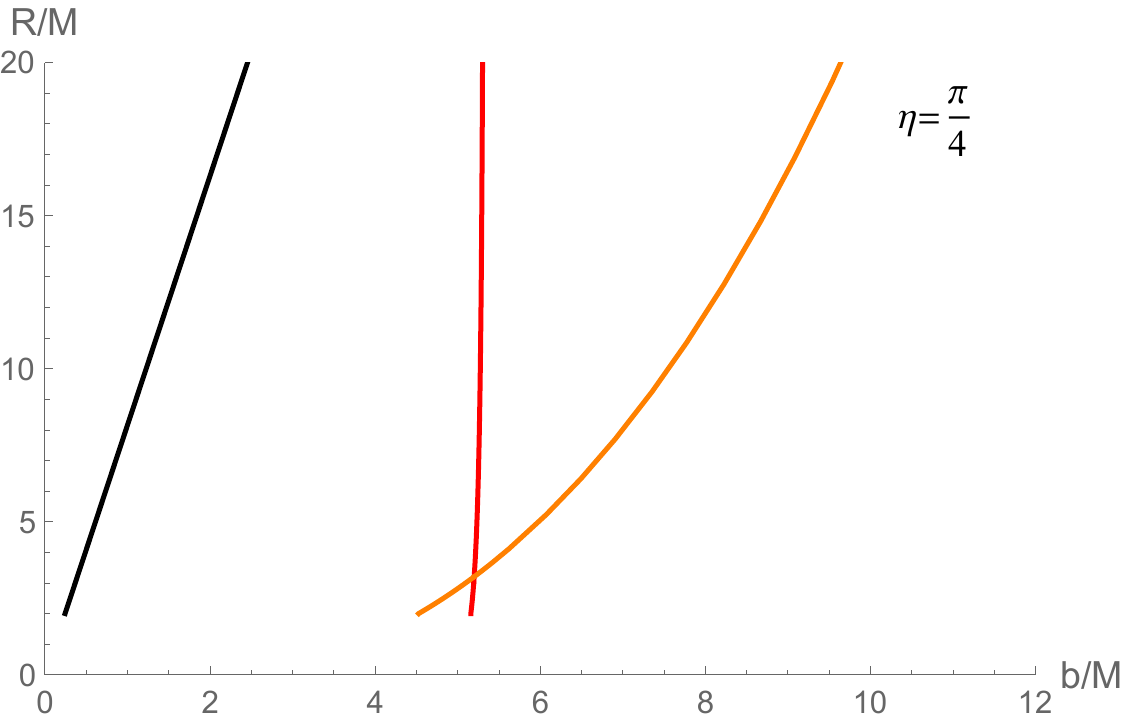}
	\end{minipage}

	\caption{Transfer functions at fixed $\omega=17\pi/36$ for varying inclination angles $\eta$. The first row presents the counter-side transfer functions $r_{\rm m}(b)$, whereas the second row corresponds to the co-side transfer functions $R_{\rm m}(b)$. The black, orange, and red curves represent the transfer functions for $m=1$, $m=2$, and $m=3$ respectively.}
	\label{fig_chuan2}
\end{figure*}

Figures~\ref{fig_chuan_r1b} and \ref{fig_chuan2} respectively present the transfer functions for the counter-side semi-disk and the first three transfer functions for the co-side semi-disk, with fixed $\eta=\pi/2$ and $\omega=17\pi/36$. Here, black, orange, and red curves correspond to the cases of $m=1$, $m=2$, and $m=3$, respectively. The results align with prior findings on orbit numbers: as inclination $\omega$ (or polar angle $\eta$) increases, the impact parameter ranges for the second ($m=2$) and third ($m=3$) transfer functions of the counter-side semi-disk shrink, while those of the co-side semi-disk expand. Moreover, the slope of the first ($m=1$) transfer function grows progressively with increasing $\omega$ and $\eta$, accompanied by a reduction in its impact parameter range, which indicates a contraction of the direct emission region on the observer's plane with a bigger $\omega$.

\section{Optical appearance of Schwarzschild black hole with a optically thin disk at different inclinations}\label{section4}

In the previous section, we established the correspondence between emission points on the accretion disk and their projected positions on the observer's plane by the transfer functions. In this section, we will consider both the local emission intensity of the optically thin disk and gravitational redshift effect, and obtain the received intensity at frequency $\nu_{o}$ originating from radius $r$ with emitted frequency $\nu_{e}$, which is given by~\cite{Gralla:2019xty}
\begin{equation}
	I_{\nu_o}^{\rm obs}=g^3 I_{\nu_e}^{\rm em}(r),
\end{equation}
where $g$ denotes the redshift factor. The redshift factor can be calculated through $g=\frac{P_{\mu} u_{o}^{\mu}}{P_{\nu} u_{e}^{\nu}}$, where $P_{\mu}$ denotes the photon's 4-momentum, while $u_{o}^{\mu}$ and $u_{e}^{\nu}$ correspond to the 4-velocities of the observer and the emitter, respectively~\cite{Yang:2024ulu}. For a static observer at infinity, its 4-velocity is given by $u_{o}^{\mu}=(1,0,0,0)$, while for a static emitter, its 4-velocity is $u_{e}^{\nu}=(1/\sqrt{f(r)},0,0,0)$. The redshift factor is thus calculated as $g=\sqrt{f(r)}$. After integrating over the frequency $\nu_{e}$, we obtain
\begin{equation}
	I^{\rm obs}=\int g^4 I_{\nu_e}^{\rm em}(r) d\nu_e=f(r)^2I_{\rm em}(r).\label{obs}
\end{equation}
The observed intensity in the observer's plane is determined by the total contribution of the emission intensities at the intersections between light rays and the accretion disk (neglecting absorption). Therefore, the final received intensities $I_{\rm obs1}(b)$ and $I_{\rm obs2}(b)$, as functions of $\omega$ and $\eta$, for the counter-side and co-side semi-disks read, respectively
\begin{equation}
	\begin{split}
	I_{\rm obs1}(b)&=\sum_{m}f(r)^2I_{\rm em}(r)|_{r=r_{\rm m} (b)},\\
	I_{\rm obs2}(b)&=\sum_{m}f(r)^2I_{\rm em}(r)|_{r=R_{\rm m} (b)}.
	\label{obs1}
	\end{split}
\end{equation}
As pointed out in~\cite{Gralla:2019xty}, the transfer function for $m=3$ has an extremely steep slope, contributing only to a drastically narrowed region of the image with negligible flux. We therefore exclude the cases of $m>3$ in our analysis.

\begin{figure}[htbp]
	\centering
	\includegraphics[width=3.2in,height=5in,keepaspectratio]{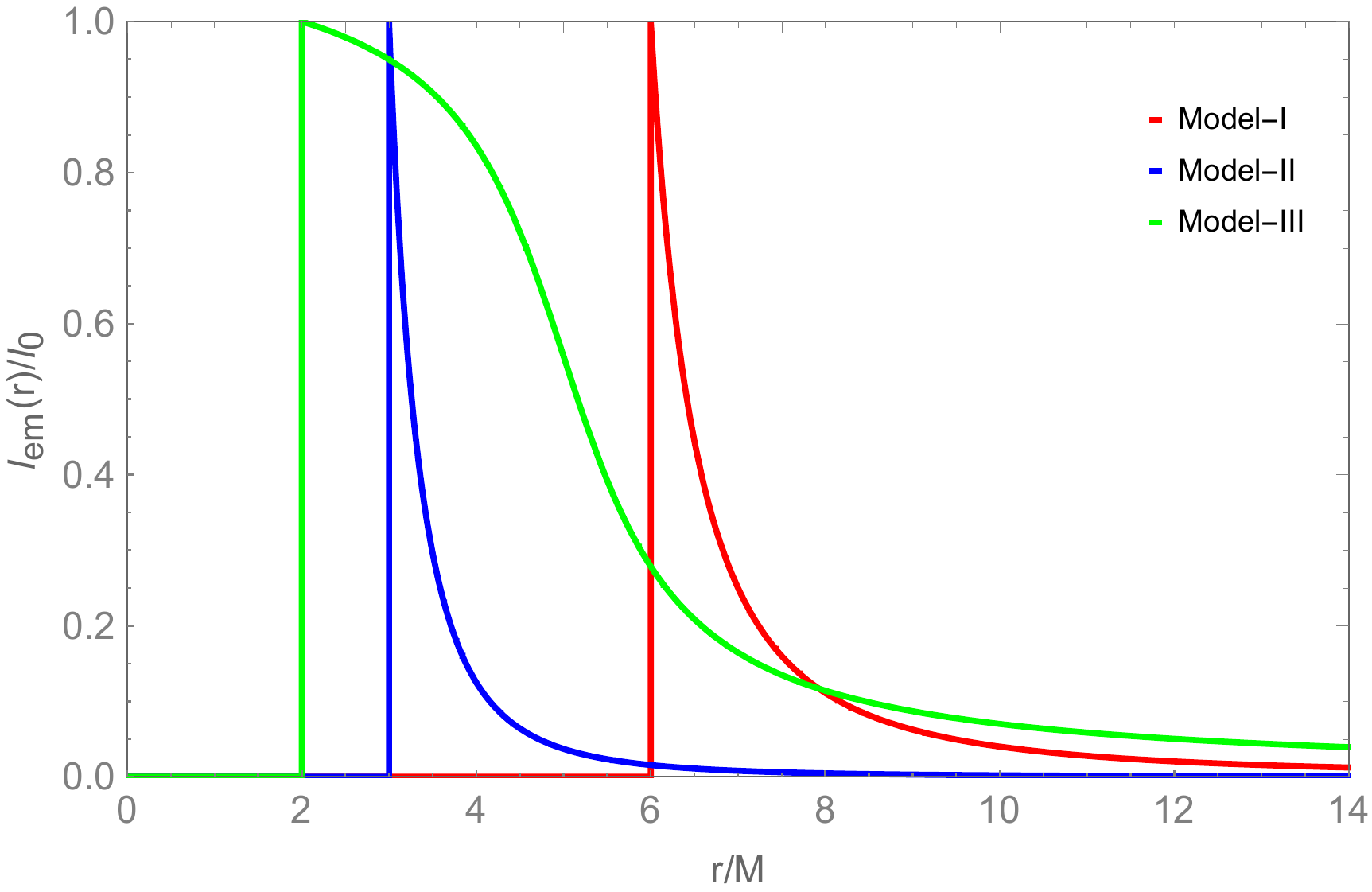}
	\caption{The emission intensity profiles of the three models, where the red, blue, and green curves represent Model-I, Model-II, and Model-III, respectively.}
	\label{fig_emit}
\end{figure}

Once the emission profile of the thin disk is specified, the corresponding received intensity can be obtained by Eq.~\eqref{obs1}. In this paper, we consider three commonly used emission models~\cite{Gralla:2019xty,Wang:2022yvi}
\begin{align}
	&\text{Model-I:}\quad I_{\rm em}(r):=
	\begin{cases}
		I_0\left[\frac{1}{r-(r_{\rm isco}-1)}\right]^2, &\hspace{0.6cm} r>r_{\rm isco}\\
		0,&\hspace{0.6cm} r \leqslant r_{\rm isco}
	\end{cases},\label{inten1}\\
	&\text{Model-II:}\quad I_{\rm em}(r):=
	\begin{cases}
		I_0\left[\frac{1}{r-(r_{\rm ph}-1)}\right]^3, &\hspace{0.9cm} r>r_{\rm ph}\\
		0,&\hspace{0.9cm} r\leqslant r_{\rm ph}
	\end{cases},\label{inten2}\\
	&\text{Model-III:}\quad I_{\rm em}(r):=
	\begin{cases}
		I_0\frac{\frac{\pi}{2}-\arctan[r-(r_{\rm isco}-1)]}{\frac{\pi}{2}-\arctan[r_{\rm h}-(r_{\rm isco}-1)]}, &\hspace{0.1cm} r>r_{\rm h}\\
		0,&\hspace{0.1cm} r\leqslant r_{\rm h}
	\end{cases},\label{inten3}
\end{align}
where $I_0$ is the maximum emission intensity, and the quantities $r_{\rm h}$, $r_{\rm ph}$, and $r_{\rm isco}$ denote the event horizon radius, photon sphere radius, and innermost stable circular orbit (ISCO) radius of the Schwarzschild BH, respectively. The values of these quantities are omitted here (see~\cite{Wang:2023vcv,Yang:2022btw,Chen:2025ifv} for details). Figure~\ref{fig_emit} presents the emission intensity profiles for these three distinct models, which correspond to emission profiles initiating their decay at $r_{\rm isco}$, $r_{\rm ph}$, and $r_{\rm h}$, respectively.

\begin{figure*}[htb]
	\centering
	\begin{minipage}{0.33\textwidth}
		\includegraphics[height=3.9cm,keepaspectratio]{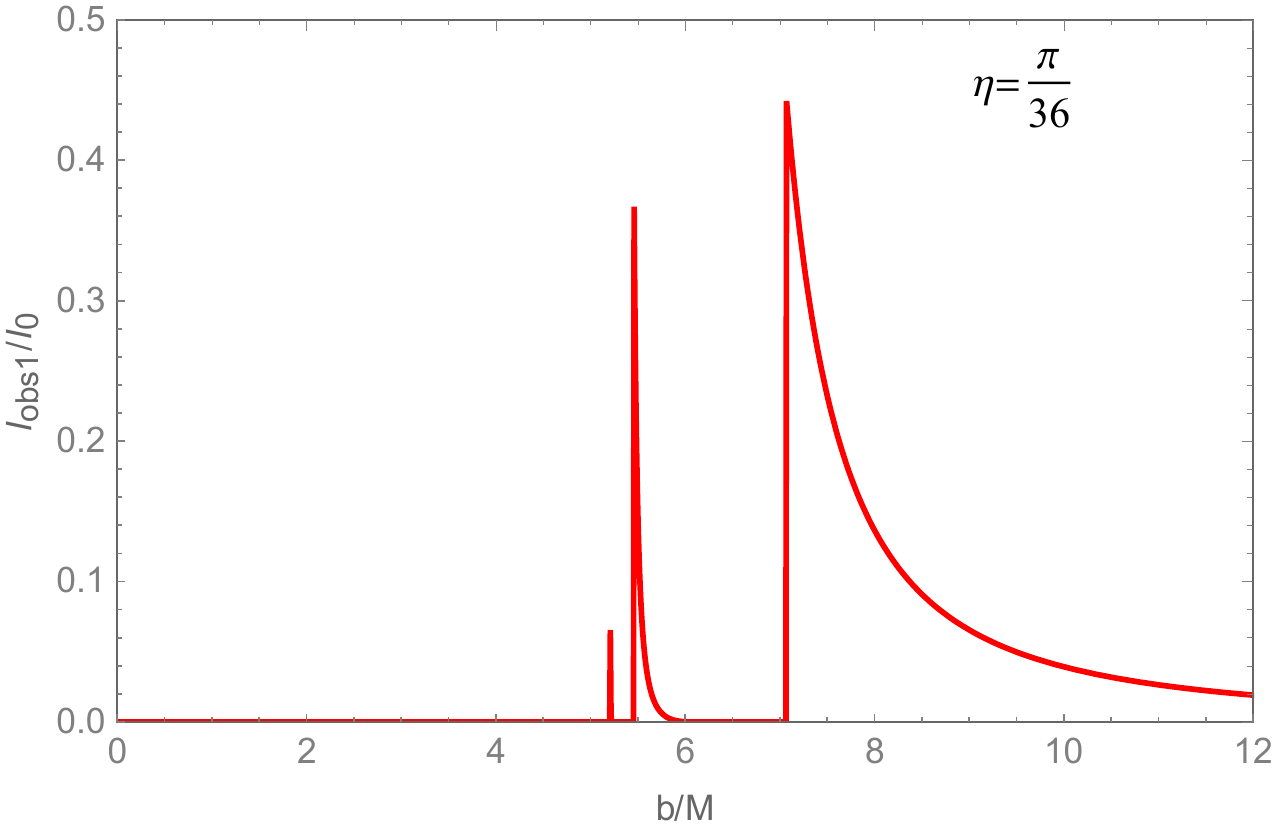}
	\end{minipage}
	\hfill
	\begin{minipage}{0.33\textwidth}
		\includegraphics[height=3.9cm]{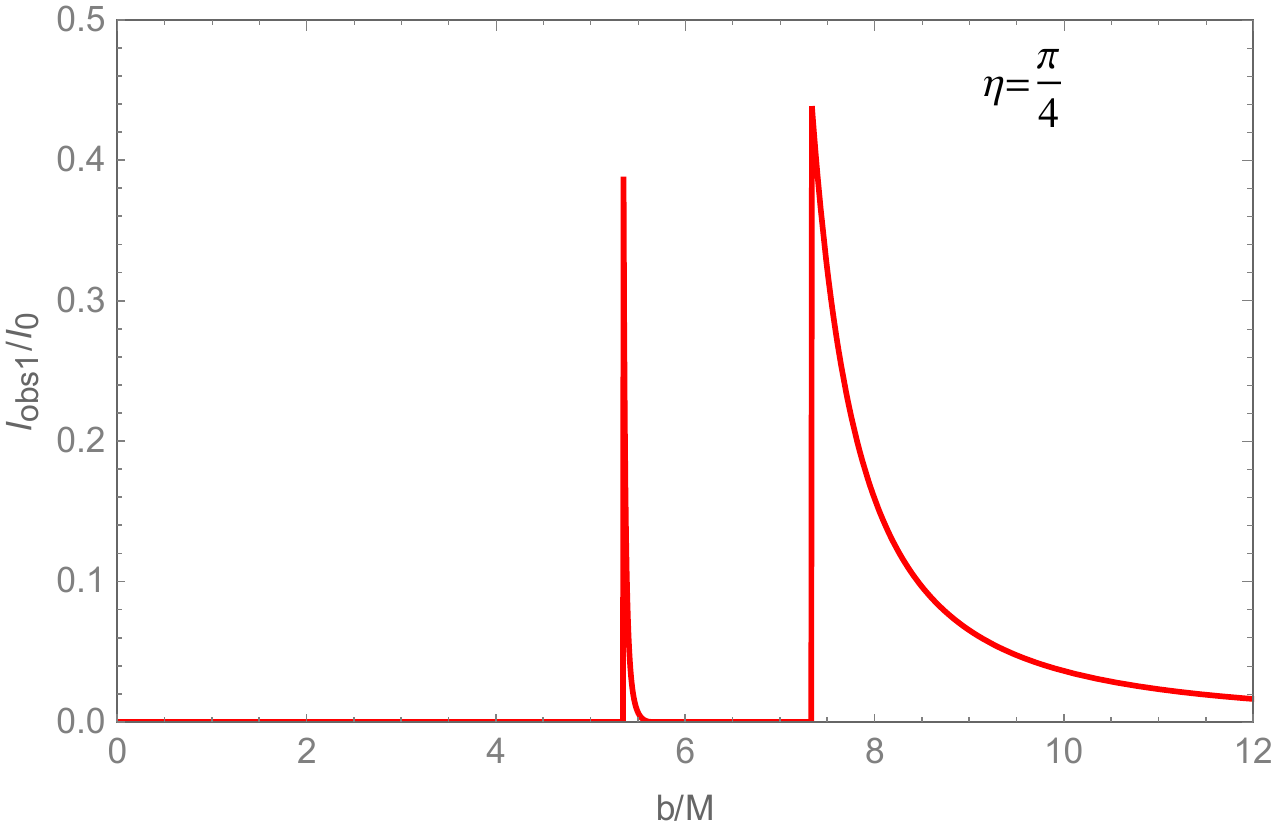}
	\end{minipage}
	\hfill
	\begin{minipage}{0.33\textwidth}
		\includegraphics[height=3.9cm,keepaspectratio]{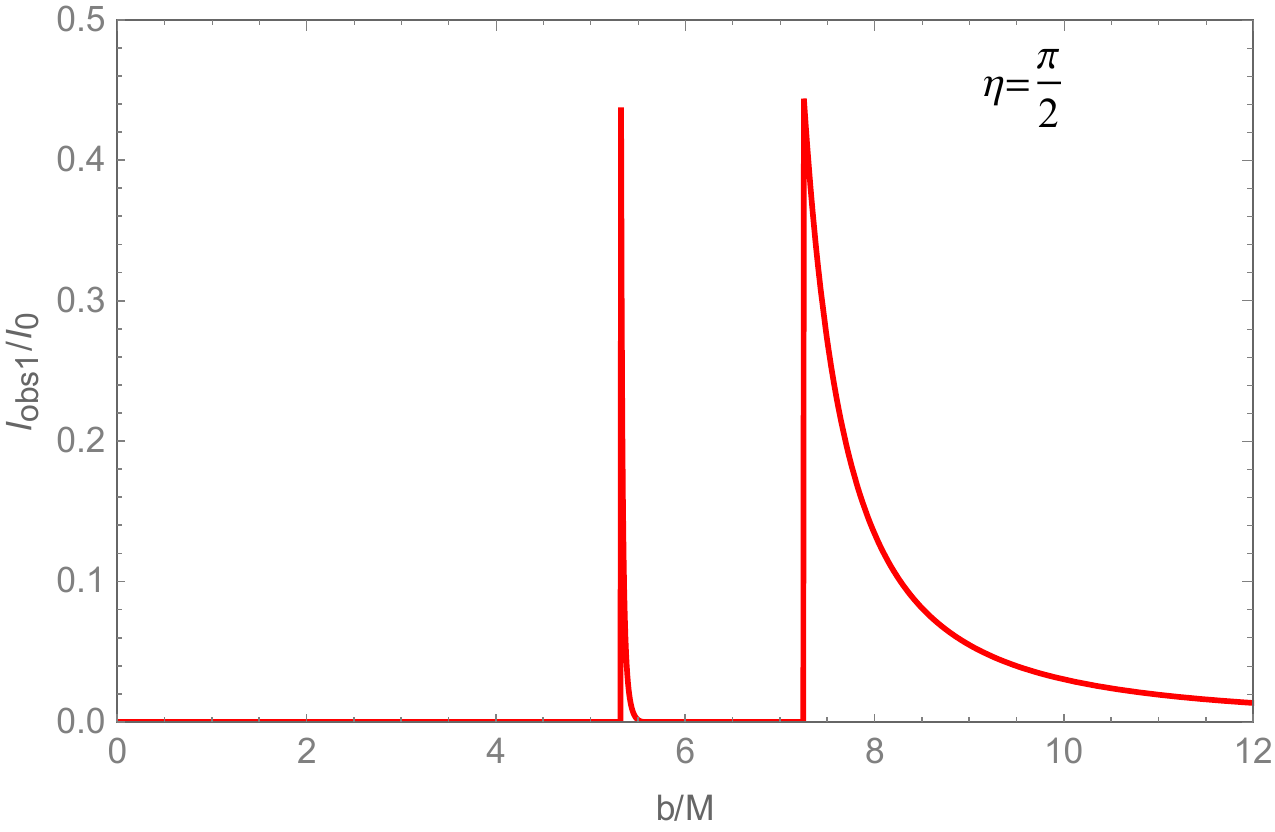}
	\end{minipage}
	\hfill
	\begin{minipage}{0.33\textwidth}
		\includegraphics[height=3.9cm,keepaspectratio]{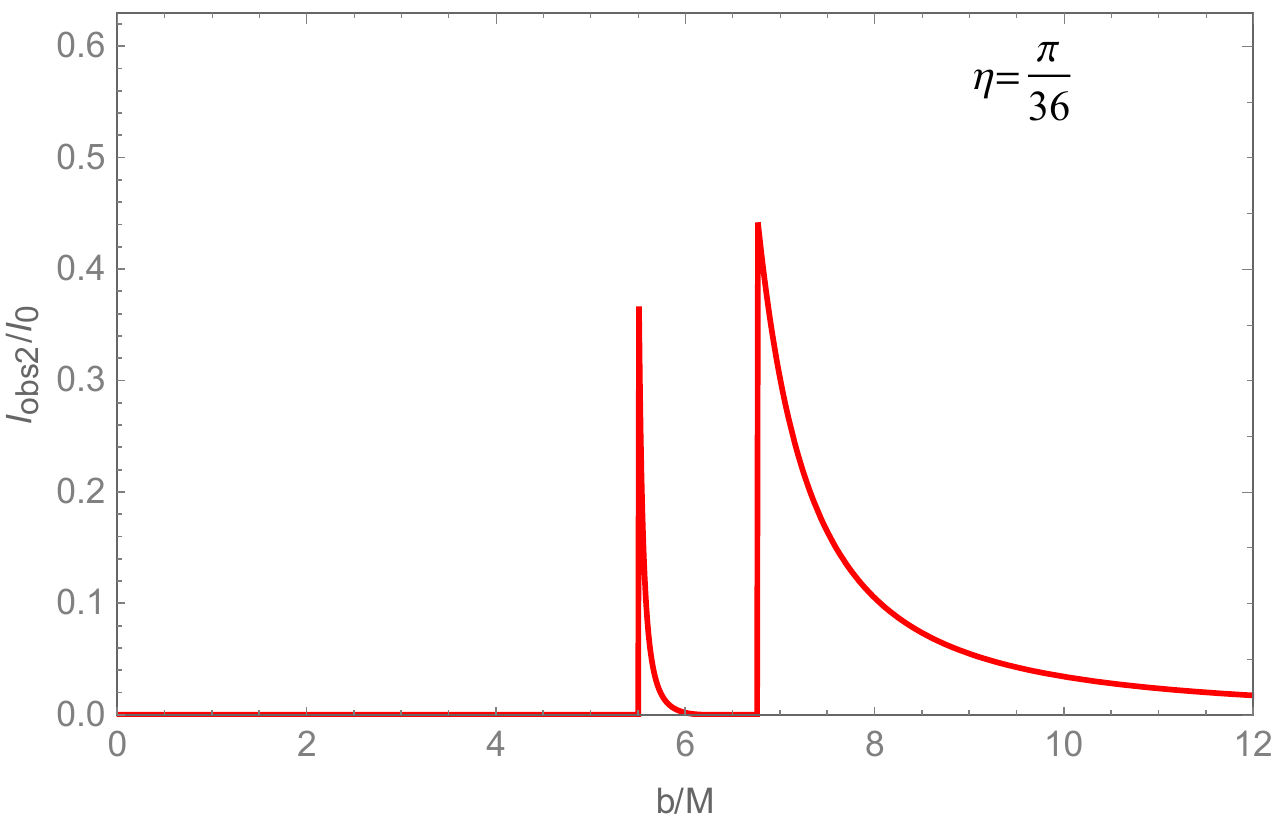}
	\end{minipage}
	\hfill
	\begin{minipage}{0.33\textwidth}
		\includegraphics[height=3.9cm]{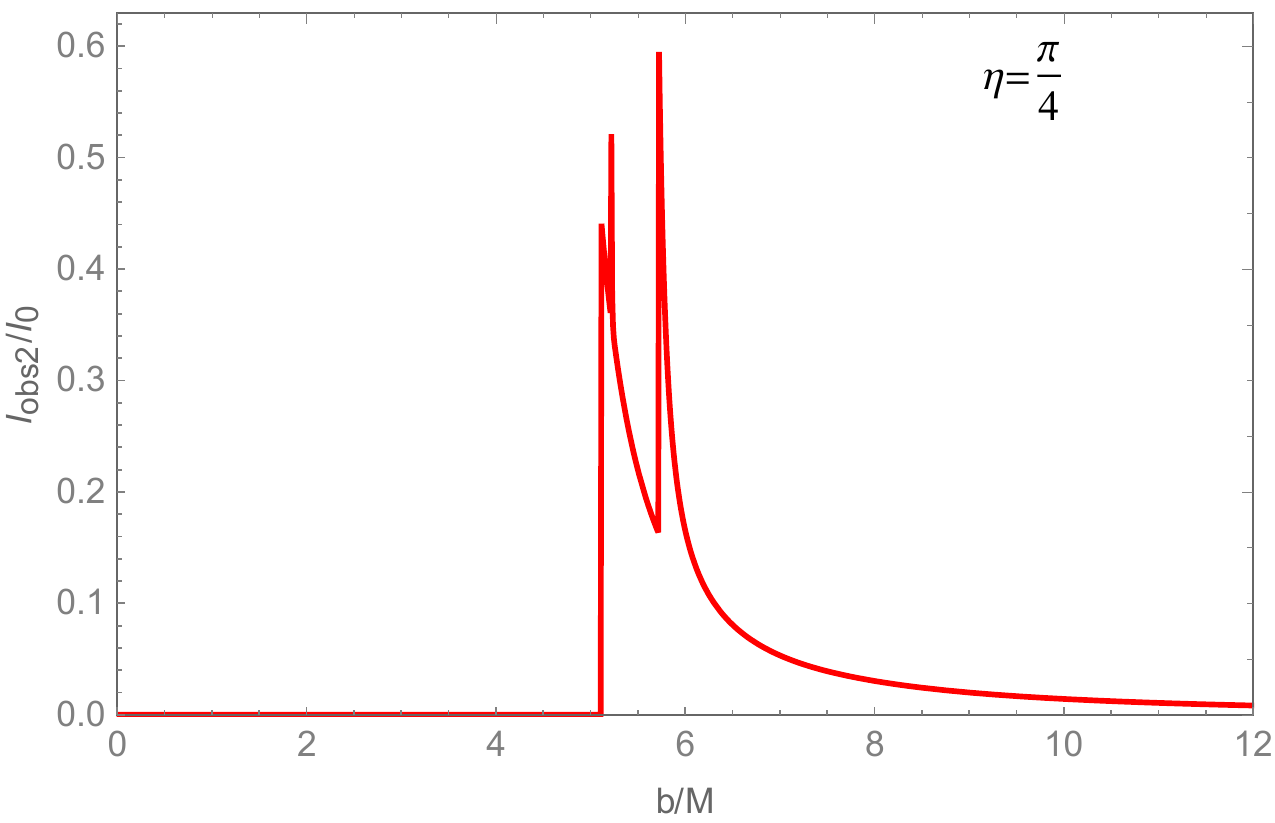}
	\end{minipage}
	\hfill
	\begin{minipage}{0.33\textwidth}
		\includegraphics[height=3.9cm,keepaspectratio]{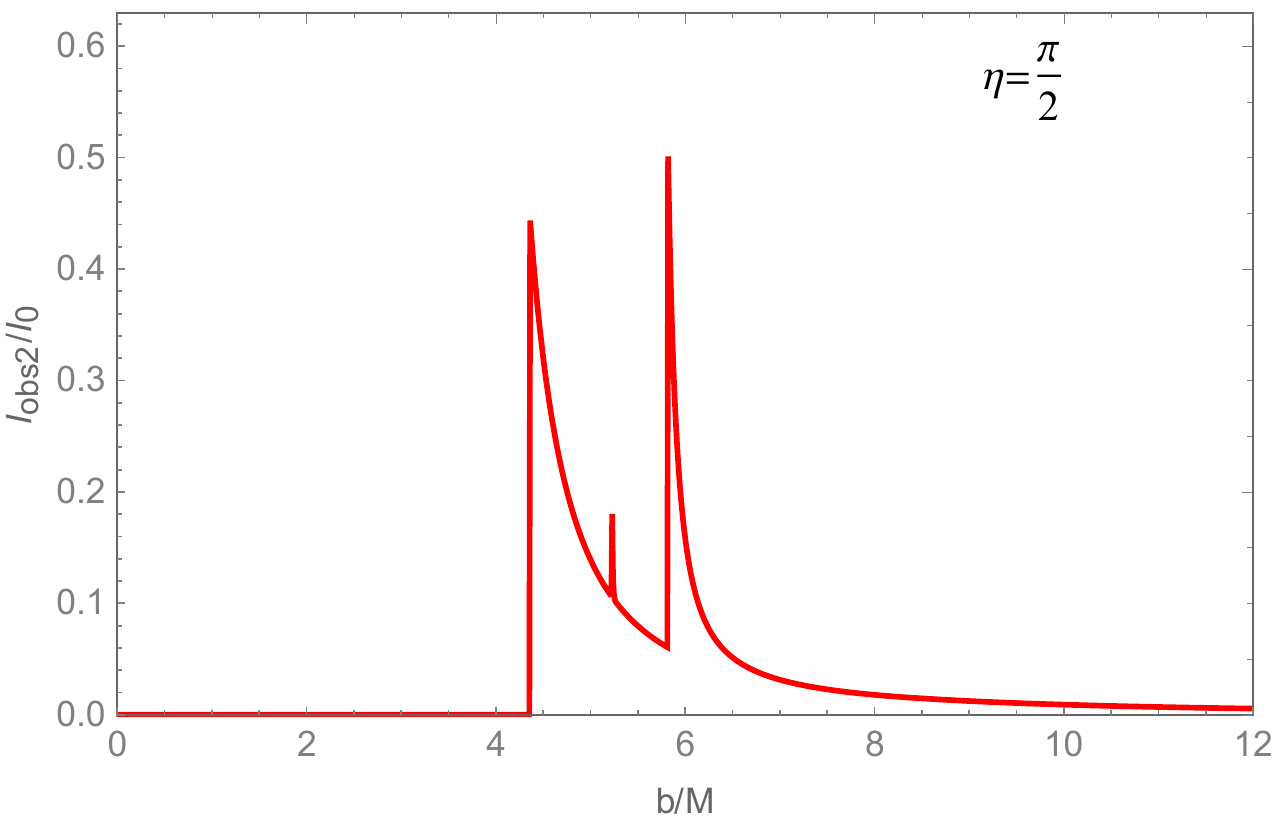}
	\end{minipage}

	\caption{For Model-I at $\pi/4$ inclination, the counter-side and co-side intensities are shown for varying polar angles $\eta$. The upper and lower rows show $I_{\text{obs1}}$ and $I_{\text{obs2}}$ respectively.}
	\label{fig:model_jieshou45}
\end{figure*}

\begin{figure*}[htb]
	\centering
	\begin{minipage}{0.33\textwidth}
		\includegraphics[height=3.9cm,keepaspectratio]{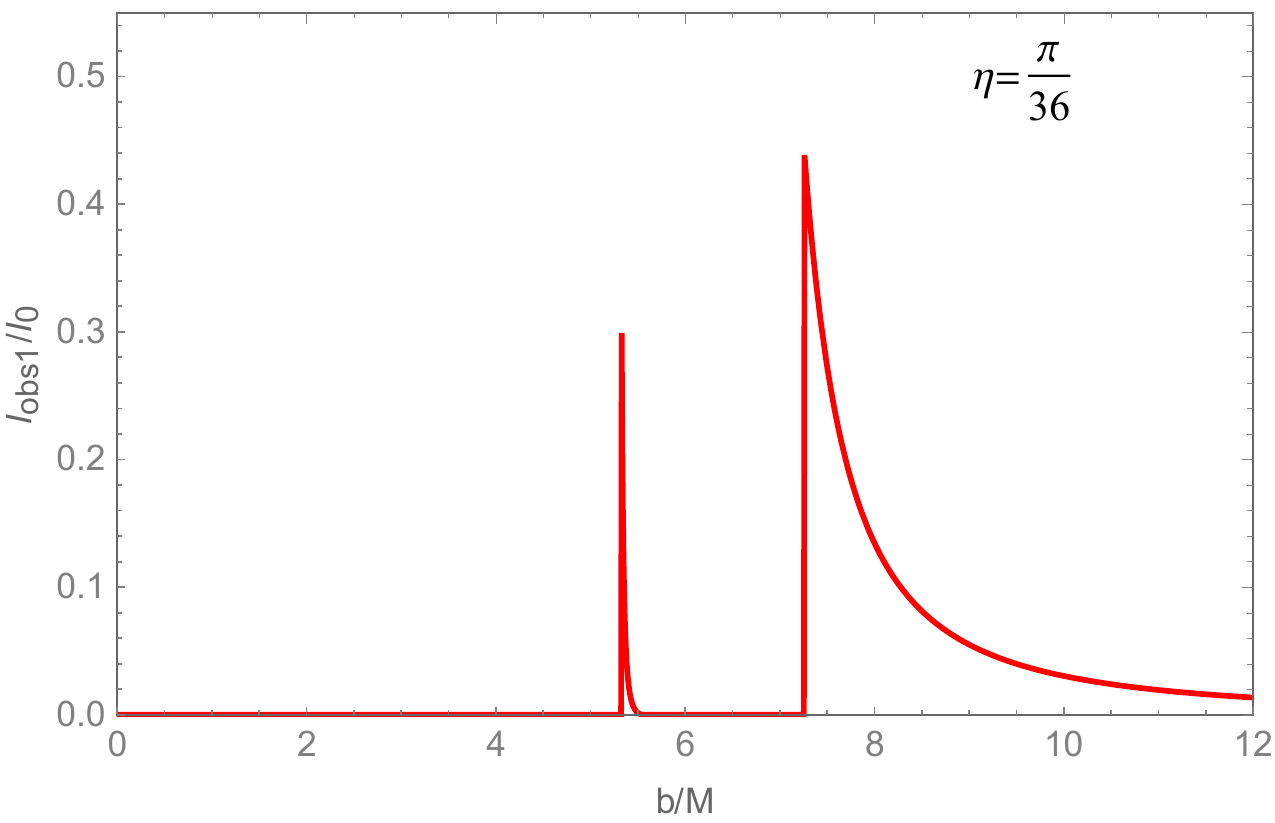}
	\end{minipage}
	\hfill
	\begin{minipage}{0.33\textwidth}
		\includegraphics[height=3.9cm]{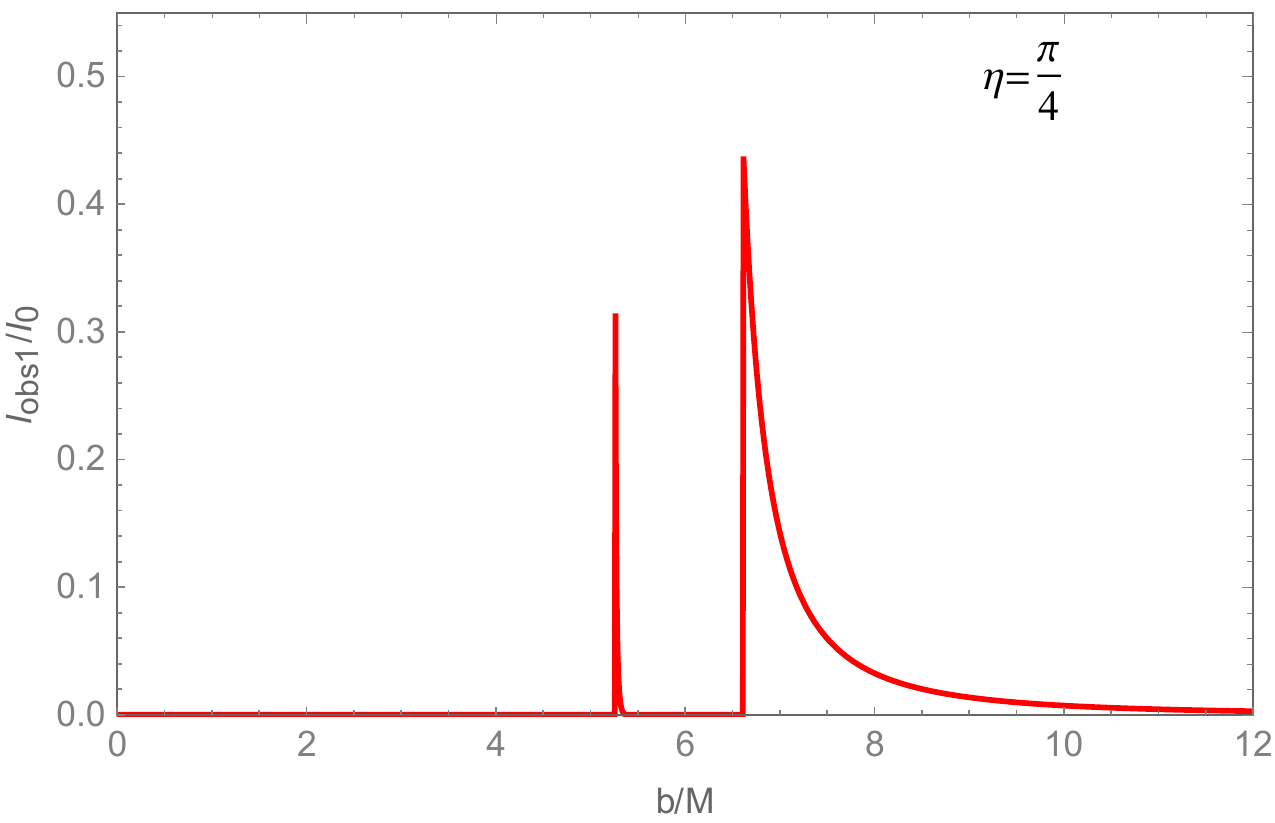}
	\end{minipage}
	\hfill
	\begin{minipage}{0.33\textwidth}
		\includegraphics[height=3.9cm,keepaspectratio]{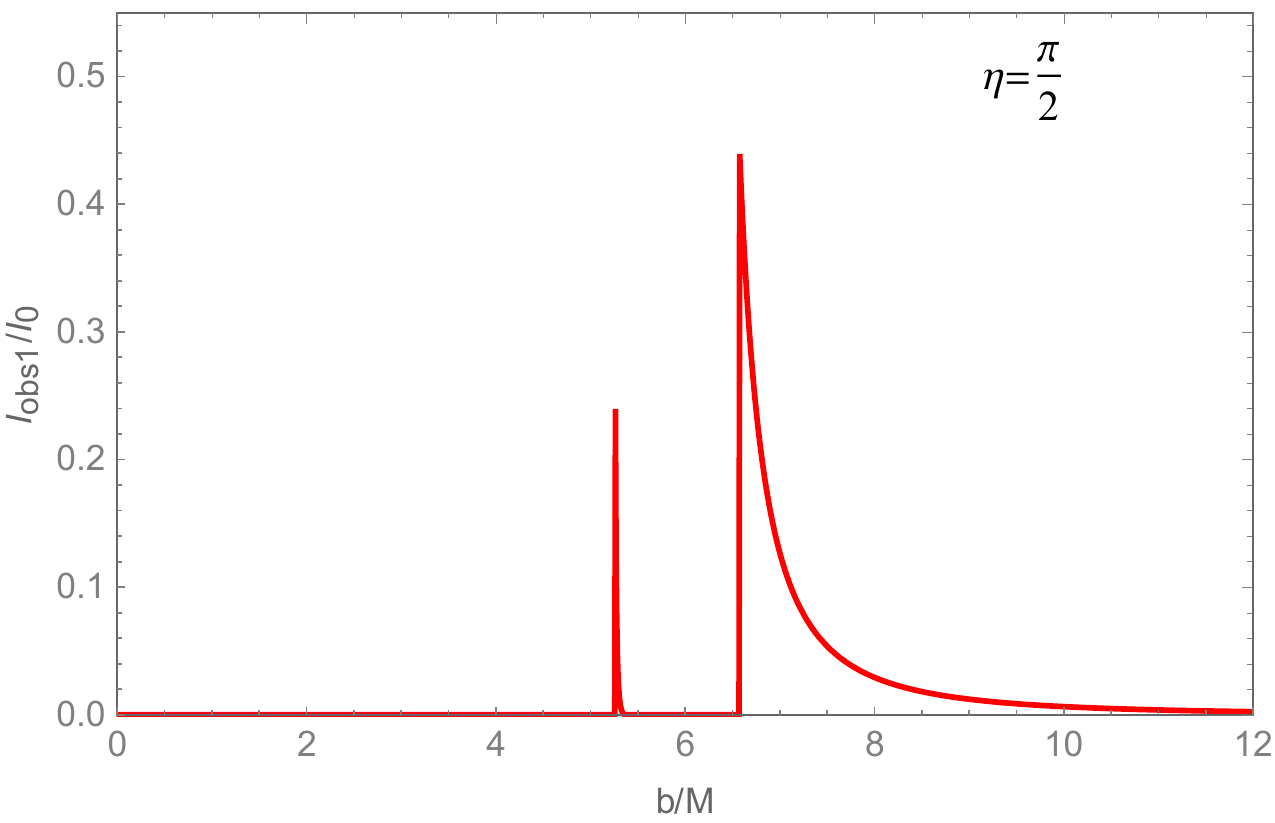}
	\end{minipage}
	\hfill
	\begin{minipage}{0.33\textwidth}
		\includegraphics[height=3.9cm,keepaspectratio]{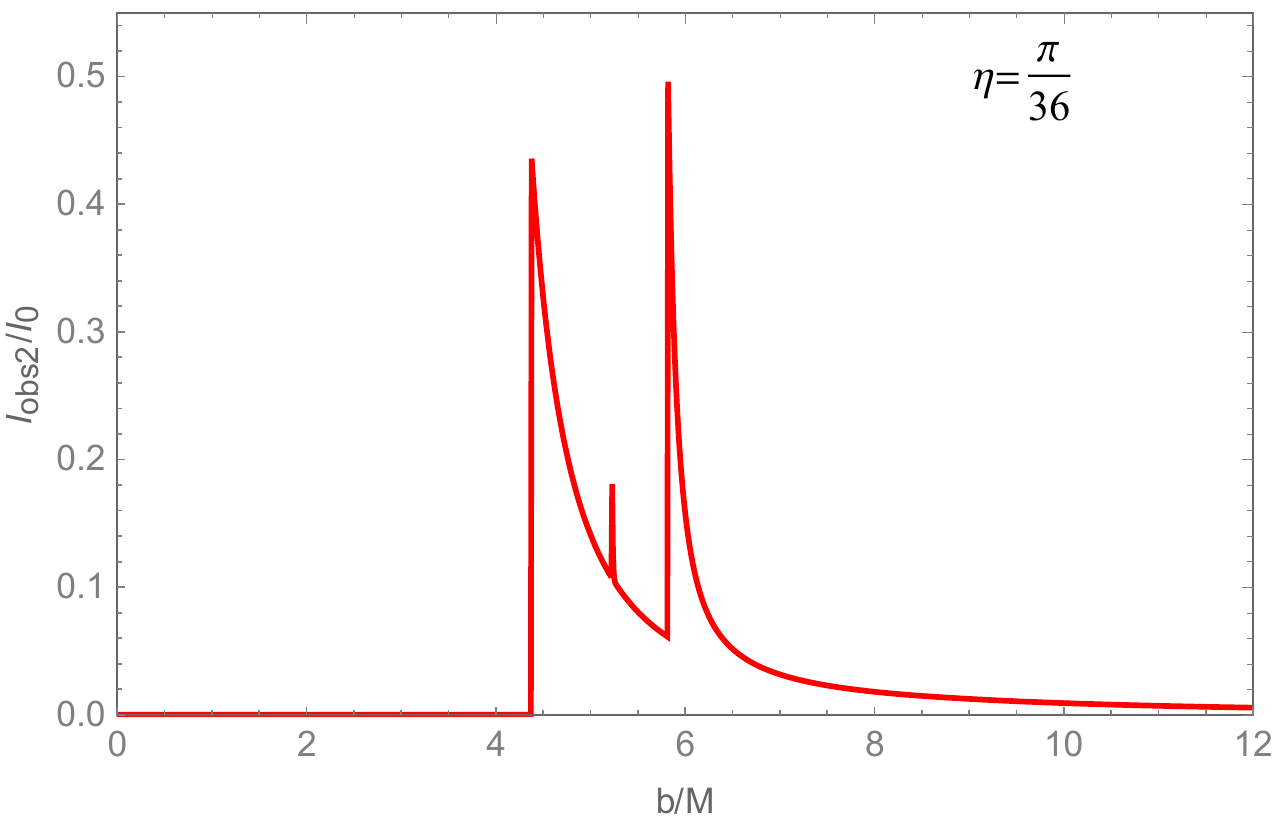}
	\end{minipage}
	\hfill
	\begin{minipage}{0.33\textwidth}
		\includegraphics[height=3.9cm]{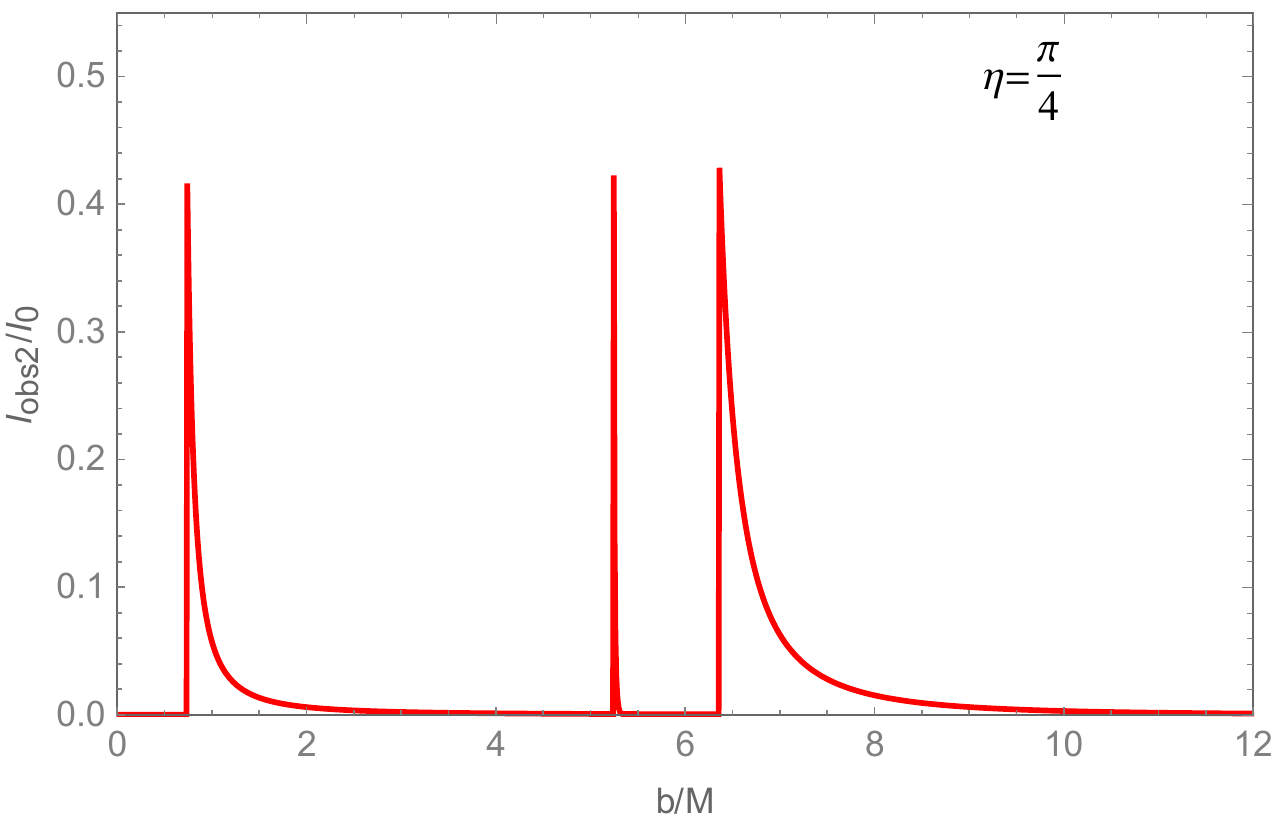}
	\end{minipage}
	\hfill
	\begin{minipage}{0.33\textwidth}
		\includegraphics[height=3.9cm,keepaspectratio]{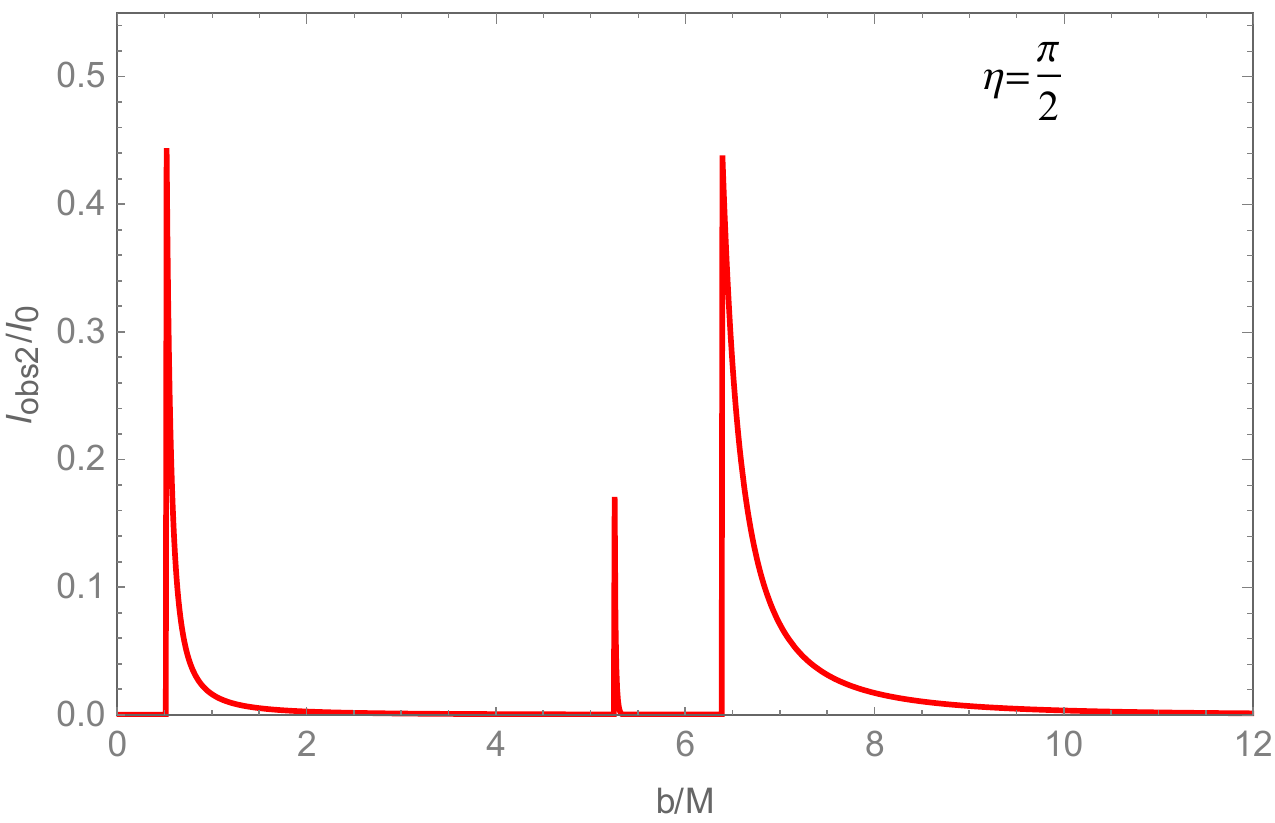}
	\end{minipage}

	\caption{For Model-I at $17\pi/36$ inclination, the counter-side and co-side intensities are shown for varying polar angles $\eta$. The upper and lower rows show $I_{\text{obs1}}$ and $I_{\text{obs2}}$ respectively.}
	\label{fig:model_jieshou85}
\end{figure*}

Combining Eqs.~\eqref{r1function}, \eqref{R1function}, and \eqref{obs1}, we calculate the received intensity on the observer's plane for different $\eta$ at inclination angle $\omega$ with values $\pi/4$ and $17\pi/36$ for the emission intensity characterized by the Model-I, and plotted the results in Fig.~\ref{fig:model_jieshou45} and \ref{fig:model_jieshou85}. From the figures, we observe that for the counter-side semi-disk, the number of intensity peaks in $I_{\rm obs1}$ shows no significant variation with polar angle $\eta$, while the distance between adjacent peaks displays only minor adjustments. Meanwhile, increasing the inclination angle reduces both the intensity at fixed $\eta$ and the peak separation distance. In contrast, the co-side intensity $I_{\rm obs2}$ exhibits more complex intensity profiles, developing additional emission peaks. For Model-II and Model-III, we briefly present the received intensities for select polar angles at inclination angles $\omega=\pi/4$ and $\omega=17\pi/36$ in Appendix \ref{appendix}.

\begin{figure*}[htbp]
	\centering
	\subfigure[Model-I:$\pi/36$]{\includegraphics[width=0.31\textwidth]{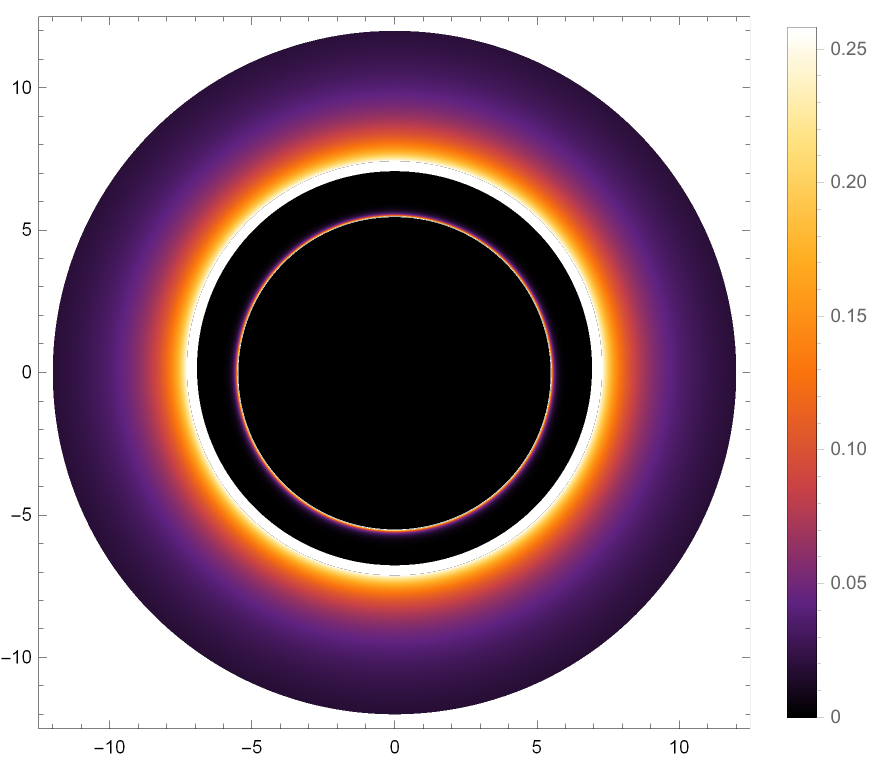}}
	\hfill
	\subfigure[Model-II:$\pi/36$]{\includegraphics[width=0.31\textwidth]{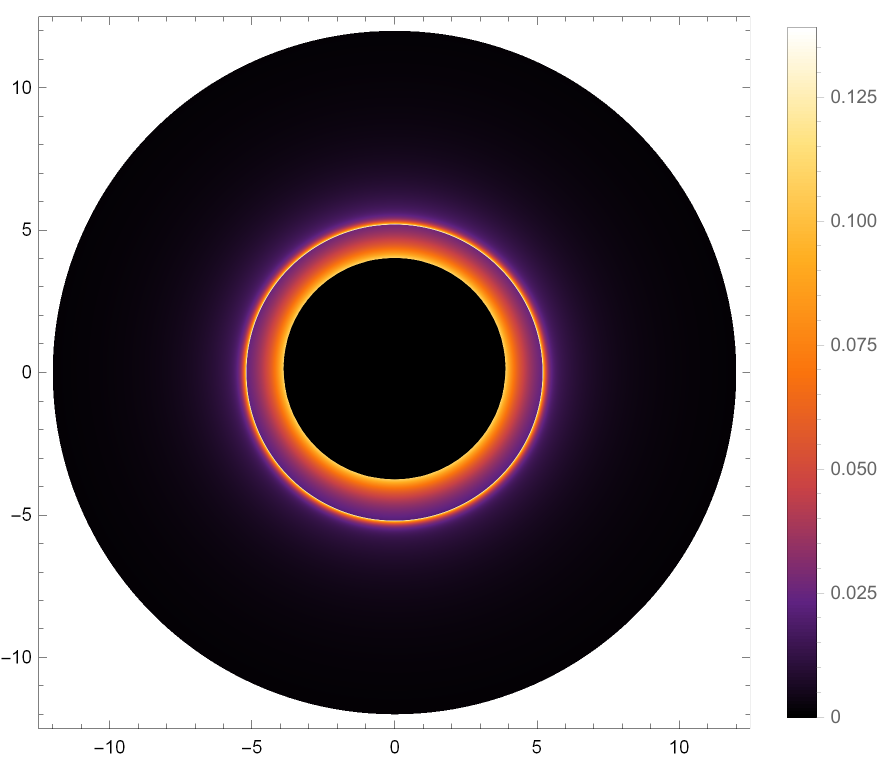}}
	\hfill
	\subfigure[Model-III:$\pi/36$]{\includegraphics[width=0.31\textwidth]{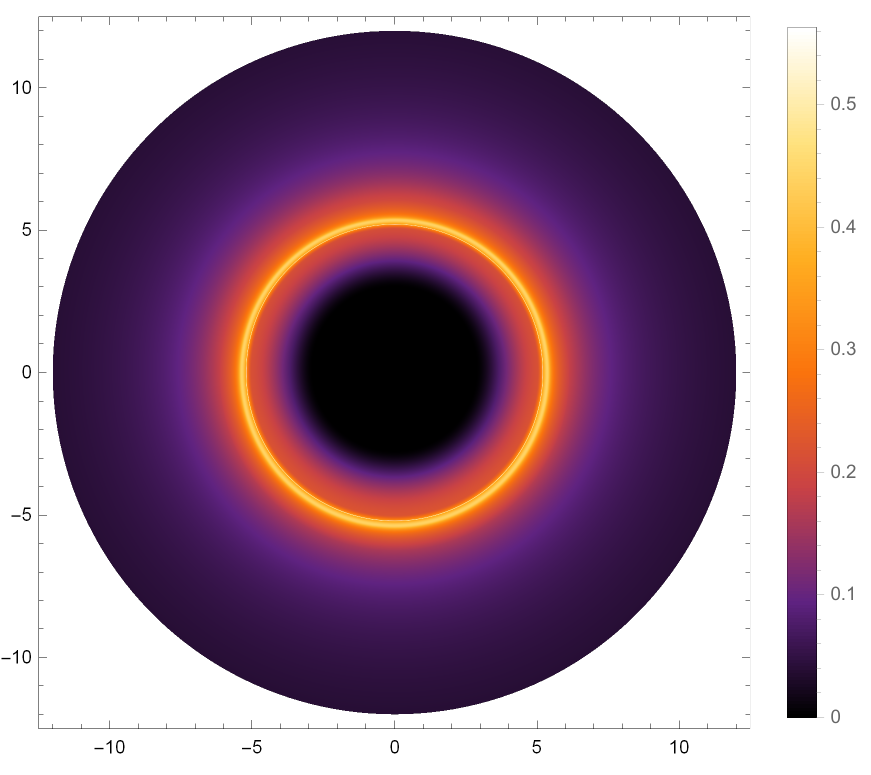}}
	\hfill
	\subfigure[Model-I:$17\pi/180$]{\includegraphics[width=0.31\textwidth]{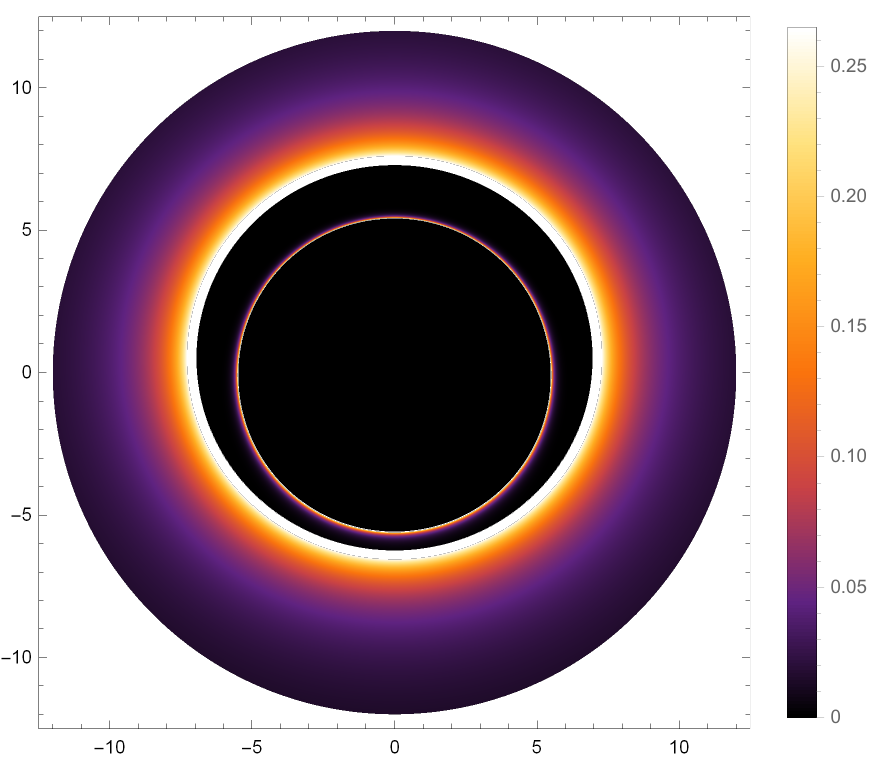}}
	\hfill
	\subfigure[Model-II:$17\pi/180$]{\includegraphics[width=0.31\textwidth]{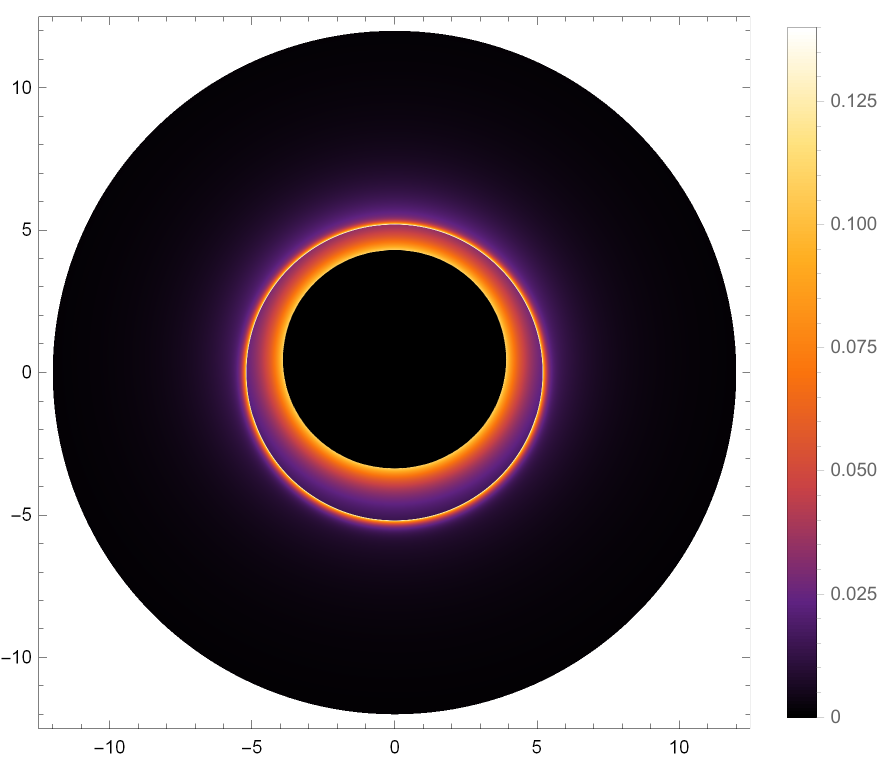}}
	\hfill
	\subfigure[Model-III:$17\pi/180$]{\includegraphics[width=0.31\textwidth]{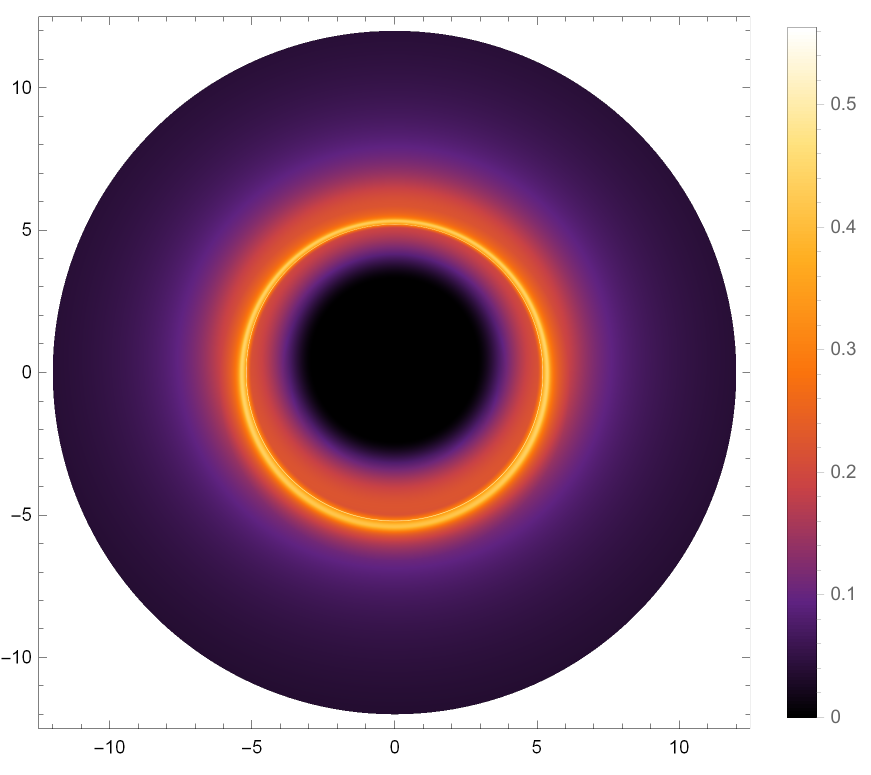}}
	\hfill
	\subfigure[Model-I:$\pi/4$]{\includegraphics[width=0.31\textwidth]{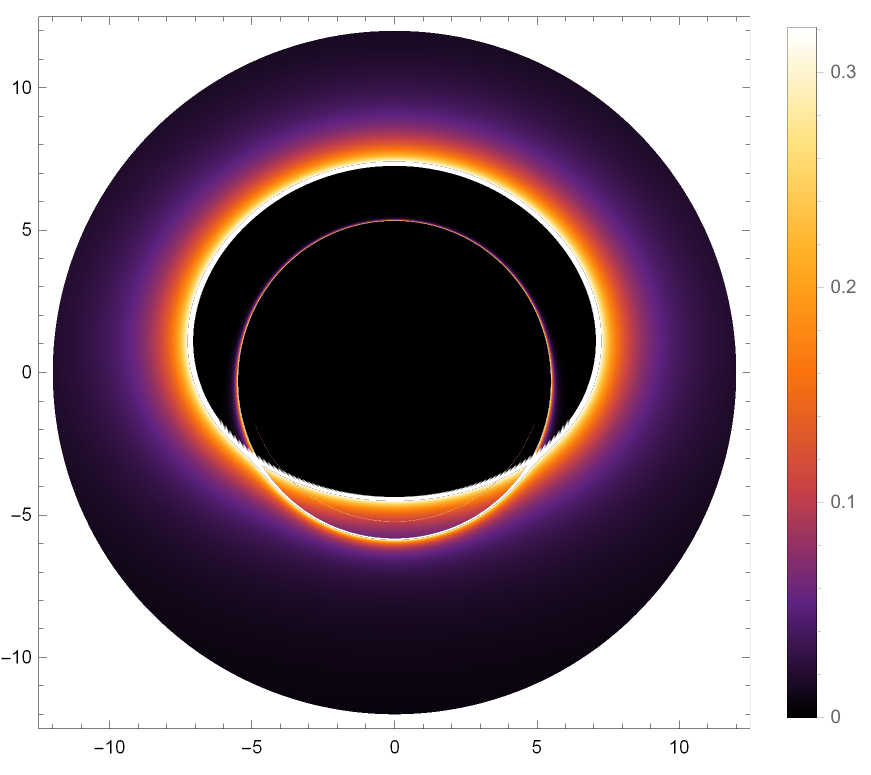}}
	\hfill
	\subfigure[Model-II:$\pi/4$]{\includegraphics[width=0.31\textwidth]{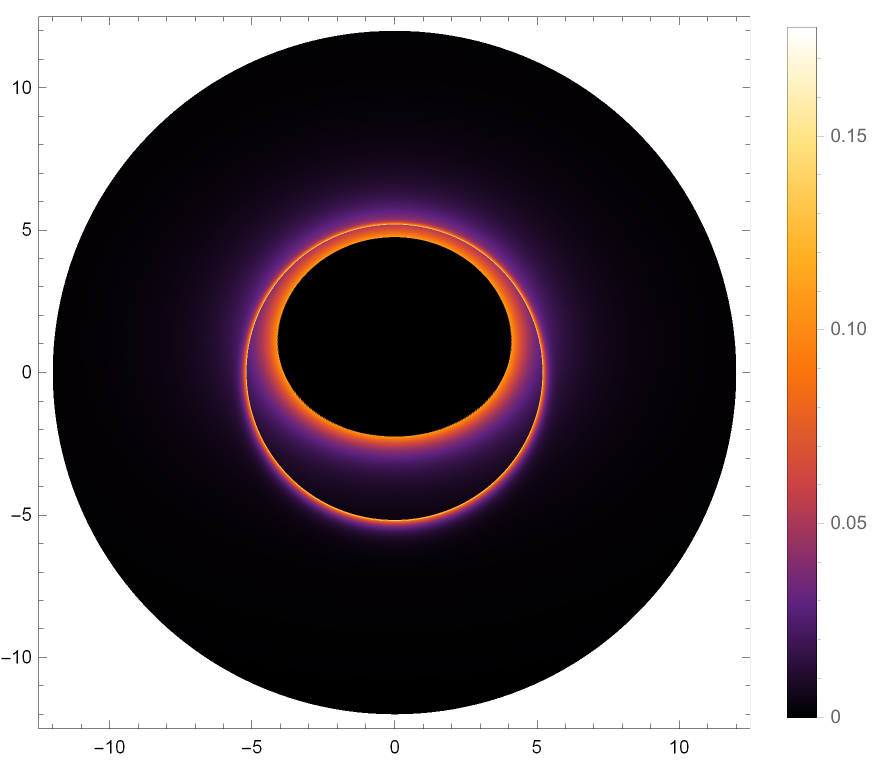}}
	\hfill
	\subfigure[Model-III:$\pi/4$]{\includegraphics[width=0.31\textwidth]{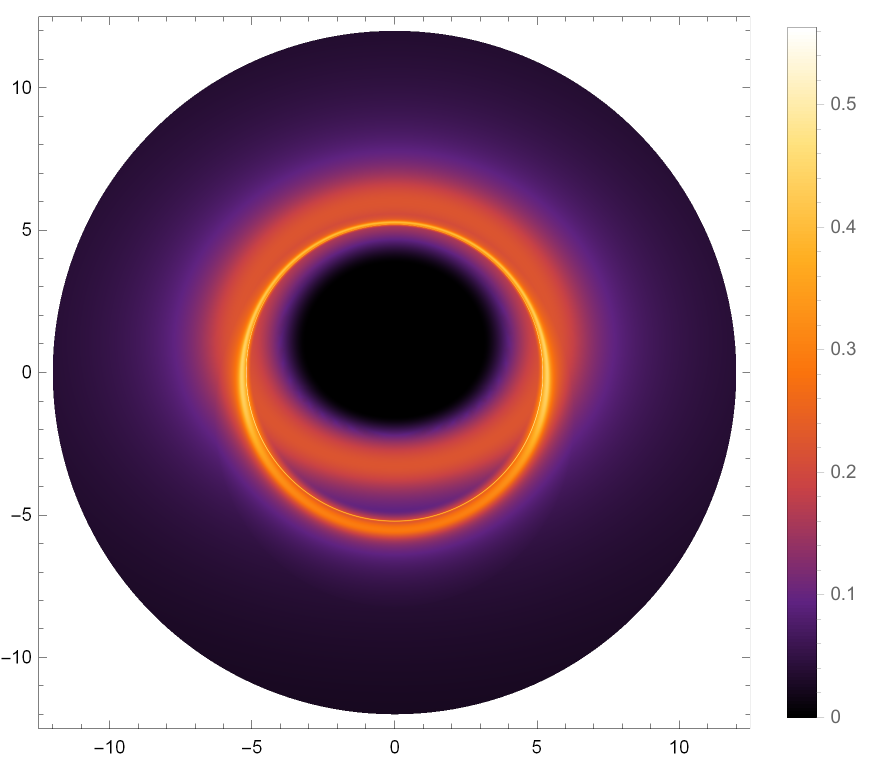}}
	\hfill
	\subfigure[Model-I:$17\pi/36$]{\includegraphics[width=0.31\textwidth]{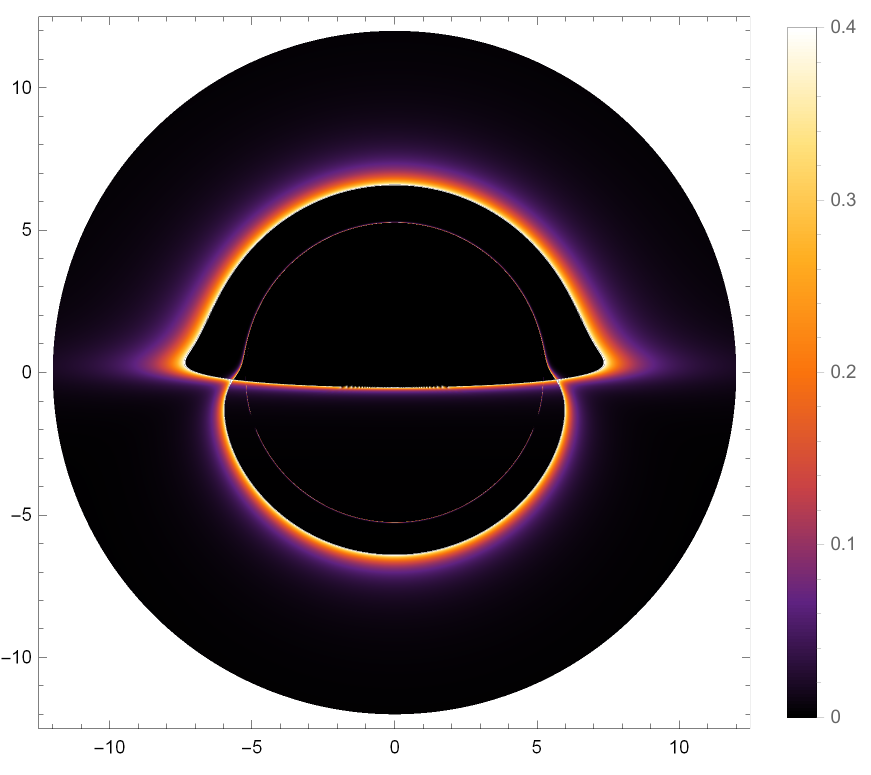}}
	\hfill
	\subfigure[Model-II:$17\pi/36$]{\includegraphics[width=0.31\textwidth]{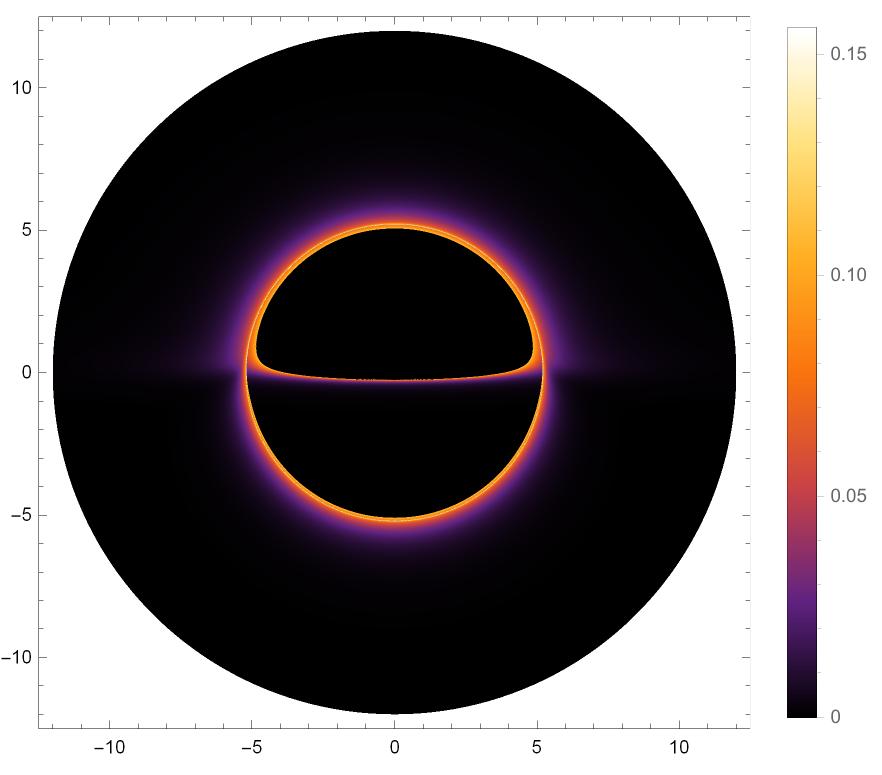}}
	\hfill
	\subfigure[Model-III:$17\pi/36$]{\includegraphics[width=0.31\textwidth]{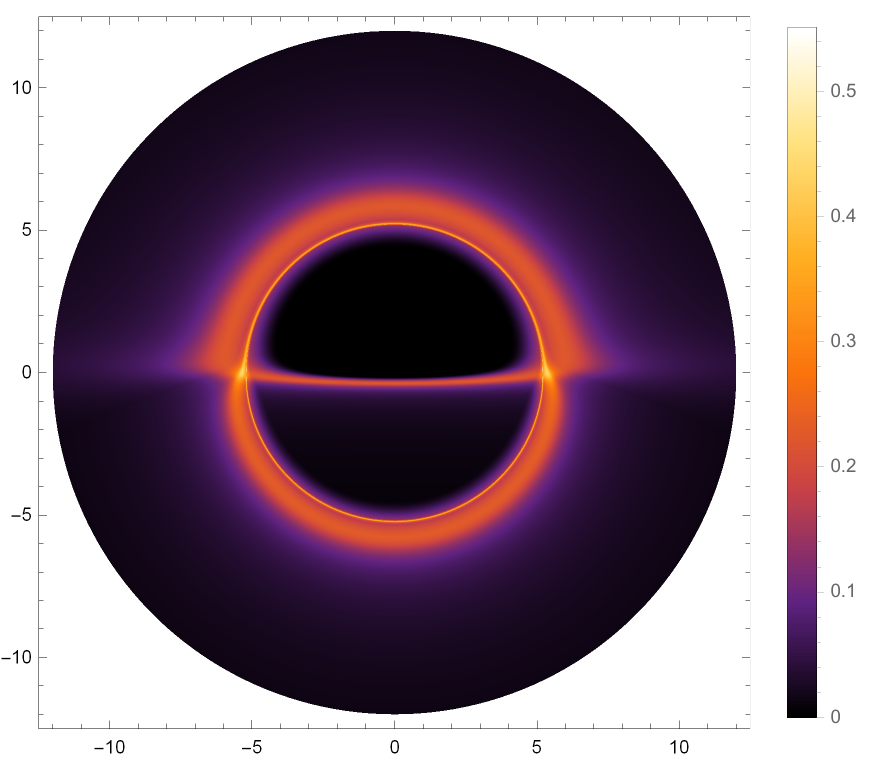}}

	\caption{Optical appearances of a Schwarzschild BH illuminated by three types of geometrically and optically thin accretion disks at varying inclination angles. Columns 1-3 display images for Model-I, Model-II, and Model-III at inclinations $\omega=\pi/36$, $17\pi/180$, $\pi/4$, and $17\pi/36$, respectively.}
	\label{fig:chuan_guang1}
\end{figure*}

We then compute the received intensity for multiple polar angles at each inclination, and combine these results with the symmetry properties to project the intensity distribution onto a two-dimensional plane. This yields the final optical appearance of a thin-disk-illuminated Schwarzschild BH at various inclinations. As shown in Fig.~\ref{fig:chuan_guang1}, we present the optical appearances for three thin disk models at inclination angles of $\pi/36$, $17\pi/180$, $\pi/4$, and $17\pi/36$. And we find that as the inclination angle $\omega$ increases, the partial bright rings in the optical appearance of thin accretion disks for all three models become progressively compressed and deviate from circularity.

\section{Optical appearance of Schwarzschild black hole with a optically thick disk at different inclinations}\label{section5}

For accretion disks, when scattering of photons by the disk is neglected, the absorption degree of radiation is typically characterized by the ``true" absorption optical depth parameter $\bar{\tau}$, defined as~\cite{Abramowicz:2011xu,Shakura:1972te}
\begin{equation}
	\bar{\tau}=\sigma_{\rm ff} U_0.\label{depth_parameter}
\end{equation}
Here $\sigma_{\rm ff}$ and $U_0$ denote the free-free absorption coefficient and the surface density of the disk, respectively. In general, an optical depth $\bar{\tau} > 1$ signifies strong absorption by the disk, indicating that the disk is opaque (i.e., optically thick). Guided by this definition, we next investigate the optical appearance of a Schwarzschild BH surrounded by optically thick versions of the three accretion disk models. Unlike the optically thin case, the optically thick disk obstructs photon propagation, preventing some rays from reaching the observer and resulting in darker images~\cite{Abramowicz:2011xu}. Therefore, to determine the final appearance, the spatial region of the thick accretion disk has to be specified. Based on the emission profiles of the three models, we define the disk's inner boundaries at locations where the emission intensity drops to zero for the three emission models. Furthermore, due to the rapid emission decay in all models, we consider finite outer boundaries for these accretion disks. In what follows, we consider the regions of the optically thick disks with radius $r$ as
\begin{itemize}
	\item Model-I:\quad $ r_{\rm isco} \leq r\leq 12$.

	\item Model-II:\quad $ r_{\rm ph} \leq r\leq 12$.

	\item Model-III:\quad $ r_{\rm h} \leq r\leq 12$.
\end{itemize}
Additionally, we neglect self-irradiation effects (i.e., disk heating via internal photon absorption~\cite{Luminet:1979nyg}), assuming time-independent emission profiles throughout.

The analysis from previous sections remains applicable here. For Model-I, the impact parameter range in the observer's plane, corresponding to the first transfer function $r_{1}(b)$ and $R_{1}(b)$ in $[r_{\rm isco},12]$, has no contributions from lensed or photon ring emission due to obstruction by the thick disk. However, outside this impact parameter range, the lensed and photon ring emission will contribute to the image intensity. Similar results hold for Model-II and Model-III. We therefore follow the earlier methodology, using Eq.~\eqref{obs1} to calculate the received intensity for each transfer function.

\begin{figure*}[htb]
	\centering
	\subfigure[Model-I:$17\pi/36$ (thin)]{\includegraphics[width=0.31\textwidth]{Sch_model1_85new.pdf}}
	\hfill
	\subfigure[Model-II:$17\pi/36$ (thin)]{\includegraphics[width=0.31\textwidth]{Sch_model2_85.pdf}}
	\hfill
	\subfigure[Model-III:$17\pi/36$ (thin)]{\includegraphics[width=0.31\textwidth]{Sch_model3_85.pdf}}
	\hfill
	\subfigure[Model-I:$17\pi/36$ (thick)]{\includegraphics[width=0.31\textwidth]{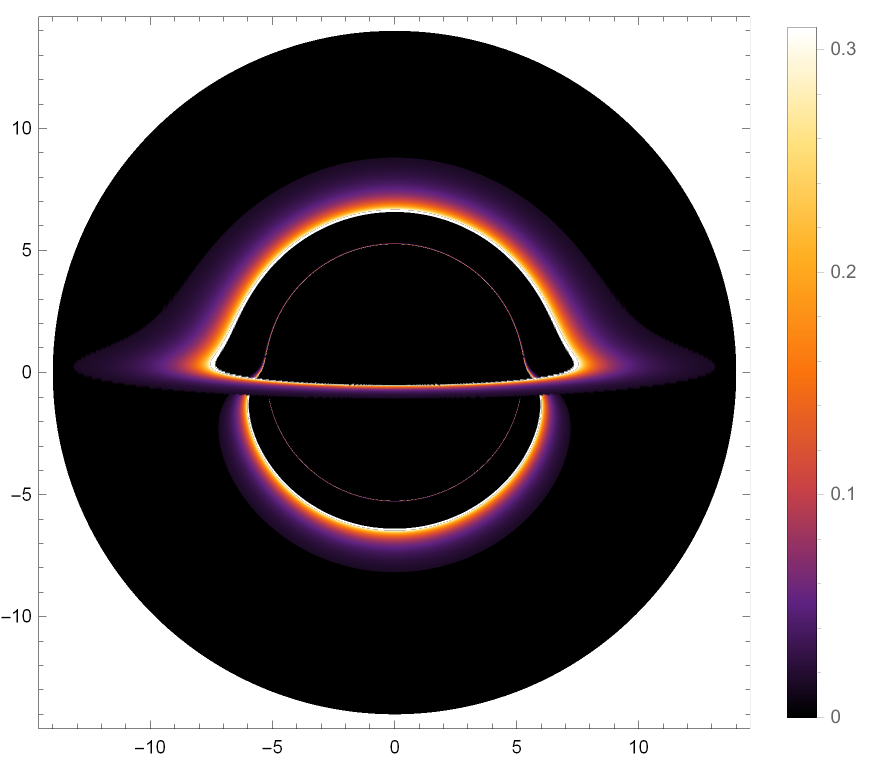}}
	\hfill
	\subfigure[Model-II:$17\pi/36$ (thick)]{\includegraphics[width=0.31\textwidth]{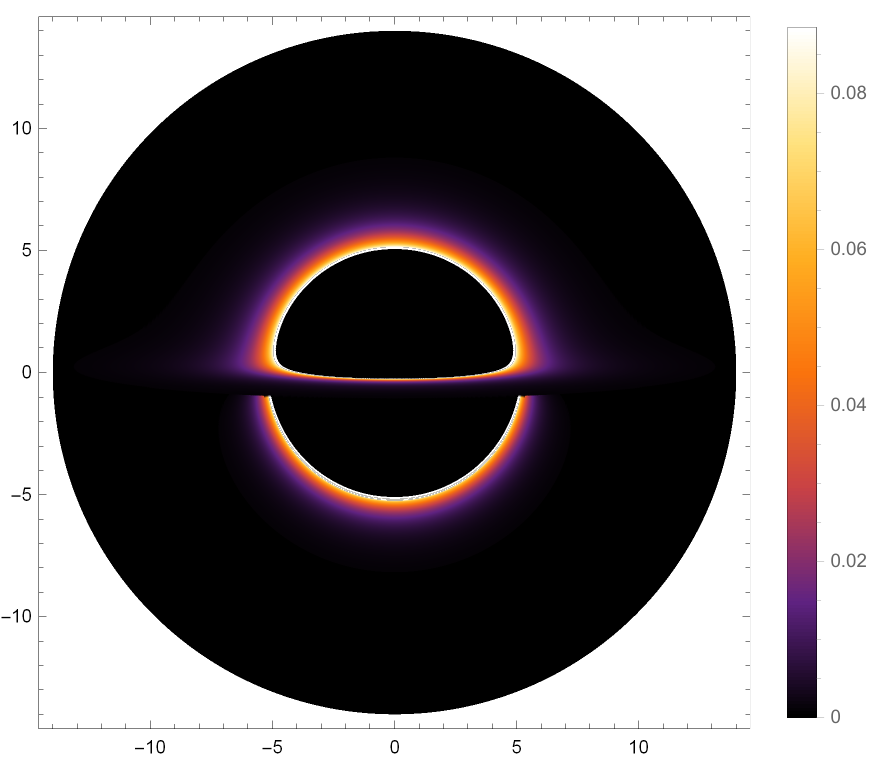}}
	\hfill
	\subfigure[Model-III:$17\pi/36$ (thick)]{\includegraphics[width=0.31\textwidth]{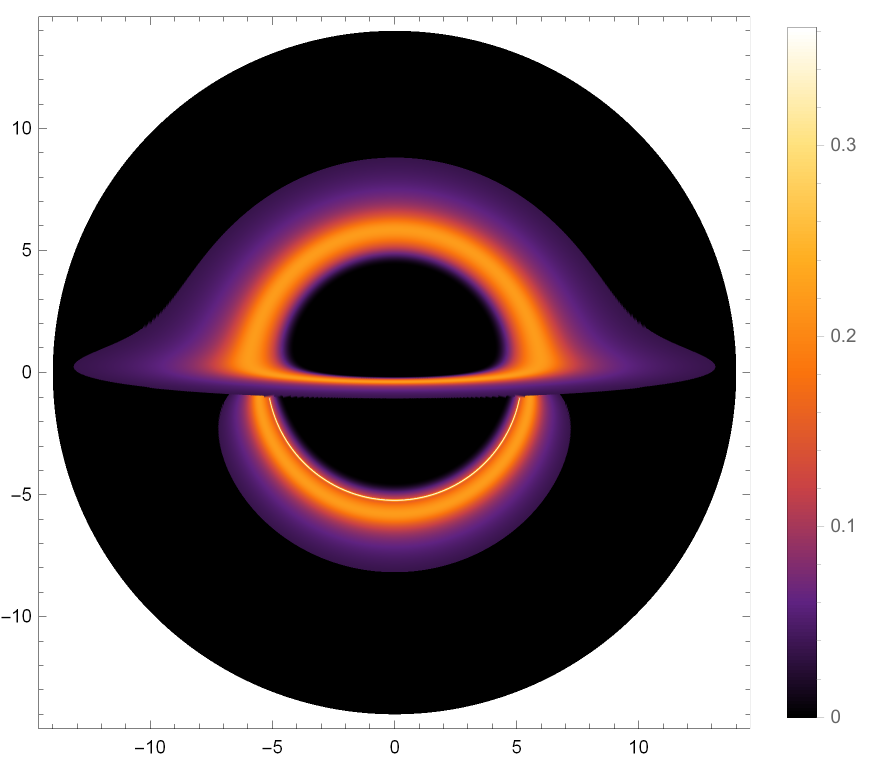}}

	\caption{Optical appearances of a Schwarzschild BH illuminated by the three accretion disk models at an inclination angle of $\omega = 17\pi/36$. The top row displays images for the three optically thin disk models, while the bottom row shows their optically thick counterparts.}
	\label{fig:thick_disk1}
\end{figure*}

In Fig.~\ref{fig:thick_disk1}, we take $\omega=17\pi/36$ as an example to demonstrate the optical appearance of a Schwarzschild BH illuminated by thick accretion disks under the three models, with direct comparison to the thin-disk cases. The top row of Fig.~\ref{fig:thick_disk1} displays the images for the three thin-disk models, while the bottom row shows their thick-disk counterparts. Two key effects are evident: (i) the bright rings disappear in the thick-disk images due to disk obscuration, and (ii) the overall intensities of the thick disks are lower than those of the thin disks.

\section{Summary}\label{section6}

In this paper, we investigate the optical appearance of a Schwarzschild BH surrounded by three geometrically thin accretion disk models as observed at different inclinations. Based on geometric configuration, we divided the accretion disk into two semi-disks: the co-side semi-disk (facing the observer) and the counter-side semi-disk (opposite the observer), and this classification is consistently used throughout our analysis.

Following the methodology of~\cite{Gralla:2019xty}, we categorized light rays intersecting the disk once, twice, and three or more times as direct, lensed, and photon ring trajectories, respectively. We then presented the photon trajectories in Fig.~\ref{fig_photon_new} and orbit numbers in Figs.~\ref{fig_n1b} and \ref{fig_n2b} for both semi-disks at fixed inclination $\omega$ and polar angle $\eta$. The results reveal striking differences between the two semi-disks: the lensed emission region is larger on the co-side, and its size increases markedly with inclination, while the corresponding region on the counter-side shrinks under the same conditions.

To establish the relationship between the accretion disk and points on the observer's plane, we introduced transfer functions $r_{\rm m}(b)$ in Eq.~\eqref{r1function} and $R_{\rm m}(b)$ in Eq.~\eqref{R1function} for the counter-side and co-side semi-disks, respectively. These functions characterize the disk intersection positions of light rays received at each point on the observer's plane. We presented the variation of these transfer functions with inclination angle $\omega$ and polar angle $\eta$ in Figs.~\ref{fig_chuan_r1b} and \ref{fig_chuan2}. The first ($m=1$) transfer functions for both semi-disks exhibit progressively steeper slopes with increasing $\omega$ and $\eta$, accompanied by a reduction in the corresponding impact parameter ranges. In contrast, the second ($m=2$) and third ($m=3$) transfer functions of both semi-disks exhibit markedly distinct behaviors with increasing $\omega$ and $\eta$. Notably, when both the $\omega$ and $\eta$ approach $\pi/2$, the lensed emission region of the co-side semi-disk dominates the image contribution.

Subsequently, by specifying the emission profiles of the thin disk models and incorporating the transfer functions, we calculated the received intensity at different polar angles $\eta$ for a fixed inclination $\omega$, with partial results presented in Fig.~\ref{fig:model_jieshou45} and \ref{fig:model_jieshou85}. Our analysis demonstrates that the co-side semi-disk exhibits more complex peak structures in the received intensity across varying polar angles $\eta$, whereas the counter-side semi-disk shows comparatively minor intensity variations. Moreover, we calculated the received intensity for each model at various $\omega$ and projected these results onto a two-dimensional plane to obtain the optical appearances of the three thin disk models under different inclinations, as shown in Fig.~\ref{fig:chuan_guang1}.

Finally, we extended our discussion to optically thick versions of the three models, comparing their images with the thin disk cases using $\omega=17\pi/36$ as a representative example in Fig.~\ref{fig:thick_disk1}. The results demonstrate that, compared to thin disks, the thick disk's obscuration effect eliminates certain bright rings and reduces the overall intensity.

In conclusion, this study extends the face-on imaging analysis of three thin accretion disk models to general inclination angles, better matching real astronomical observation conditions. We have obtained optical images for both thin and thick disk scenarios across various inclinations, enriching theoretical understanding and providing valuable references for future observational studies. The framework developed in this paper can be applied to a broader range of accretion disk models, though with increased complexity. For example, extending from geometrically thin to thick disks requires replacing the two-dimensional disk surface with a three-dimensional emitting structure. This not only introduces additional radiation from the sides of the disk but also enables the disk to obstruct the propagation of light, thereby altering the visibility of characteristic structures such as photon rings. Furthermore, while the current method is well-suited for static, spherical BHs, extending it to the Kerr spacetime (which describes rotating BHs) remains a challenge due to the additional complexity introduced by rotation. The axisymmetric and complex structure of such spacetimes breaks existing simplifications and increases the computational difficulty of ray tracing and image synthesis. The primary objective of our future work is to address this challenge by extending the present framework, with the goal of producing more comprehensive and observationally faithful BH optical images.
%-------------------------------------------------------------
\begin{acknowledgments}
This work is supported in part by NSFC Grants No. 12165005 and No. 11961131013.
\end{acknowledgments}
%-------------------------------------------------------------

%-------------------------------------------------------------
\appendix

\section{Received intensities of Model-II and Model-III}\label{appendix}

\begin{figure*}[htb]
	\centering
	\begin{minipage}{0.33\textwidth}
		\includegraphics[height=3.9cm,keepaspectratio]{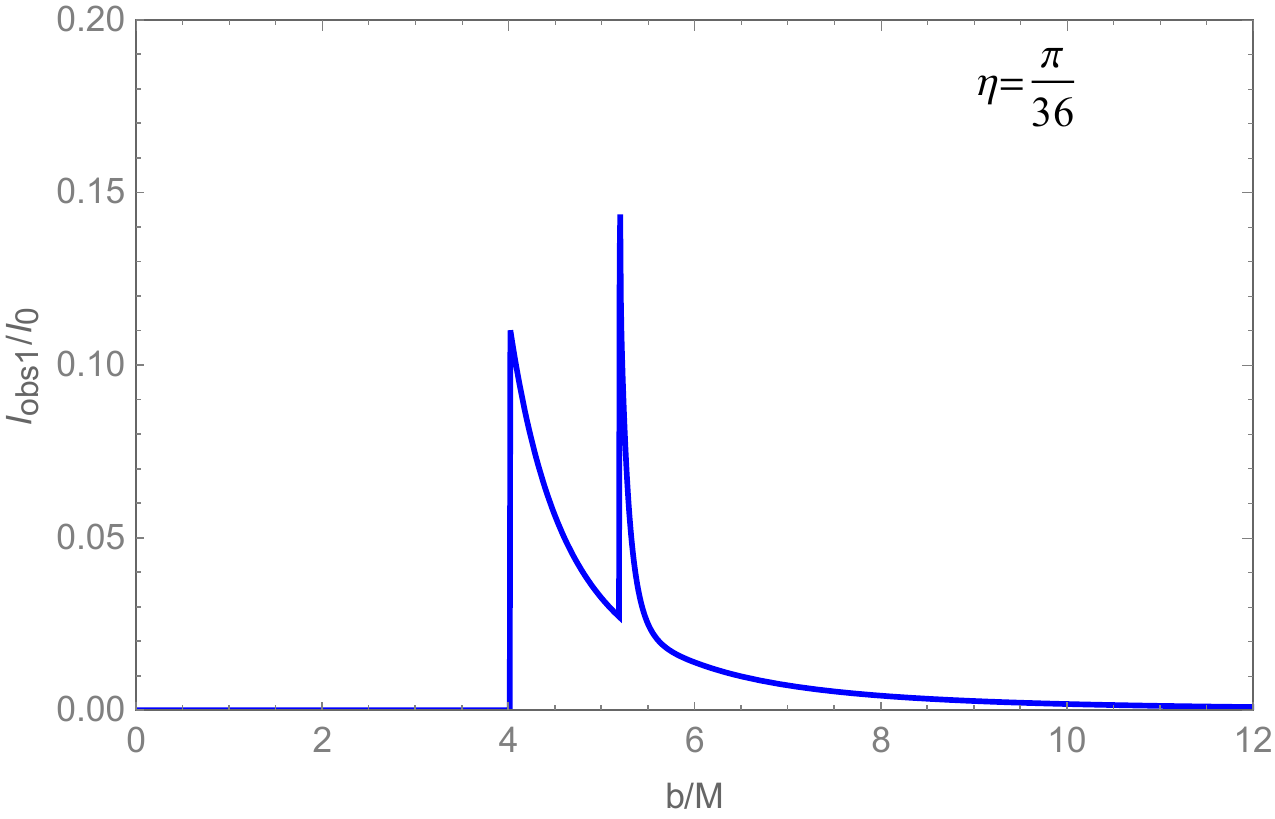}
	\end{minipage}
	\hfill
	\begin{minipage}{0.33\textwidth}
		\includegraphics[height=3.9cm]{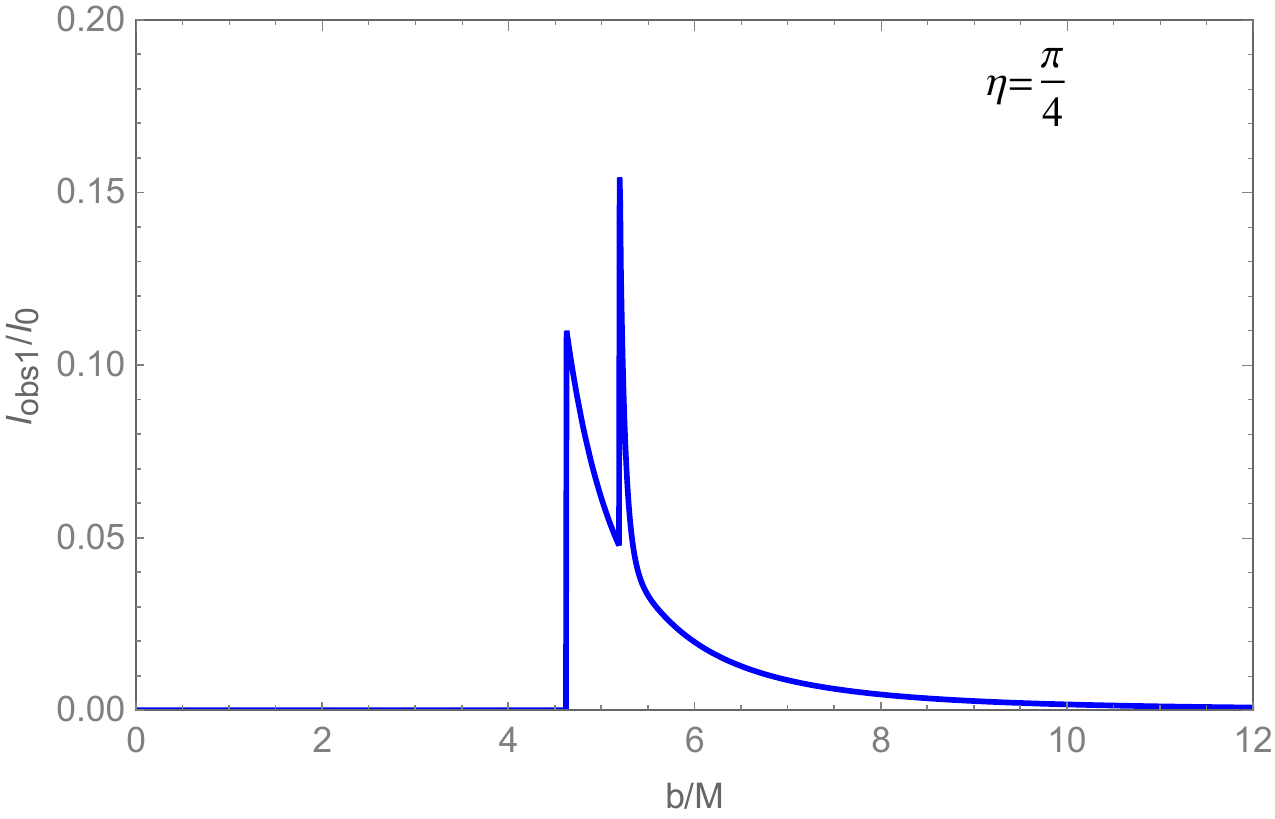}
	\end{minipage}
	\hfill
	\begin{minipage}{0.33\textwidth}
		\includegraphics[height=3.9cm,keepaspectratio]{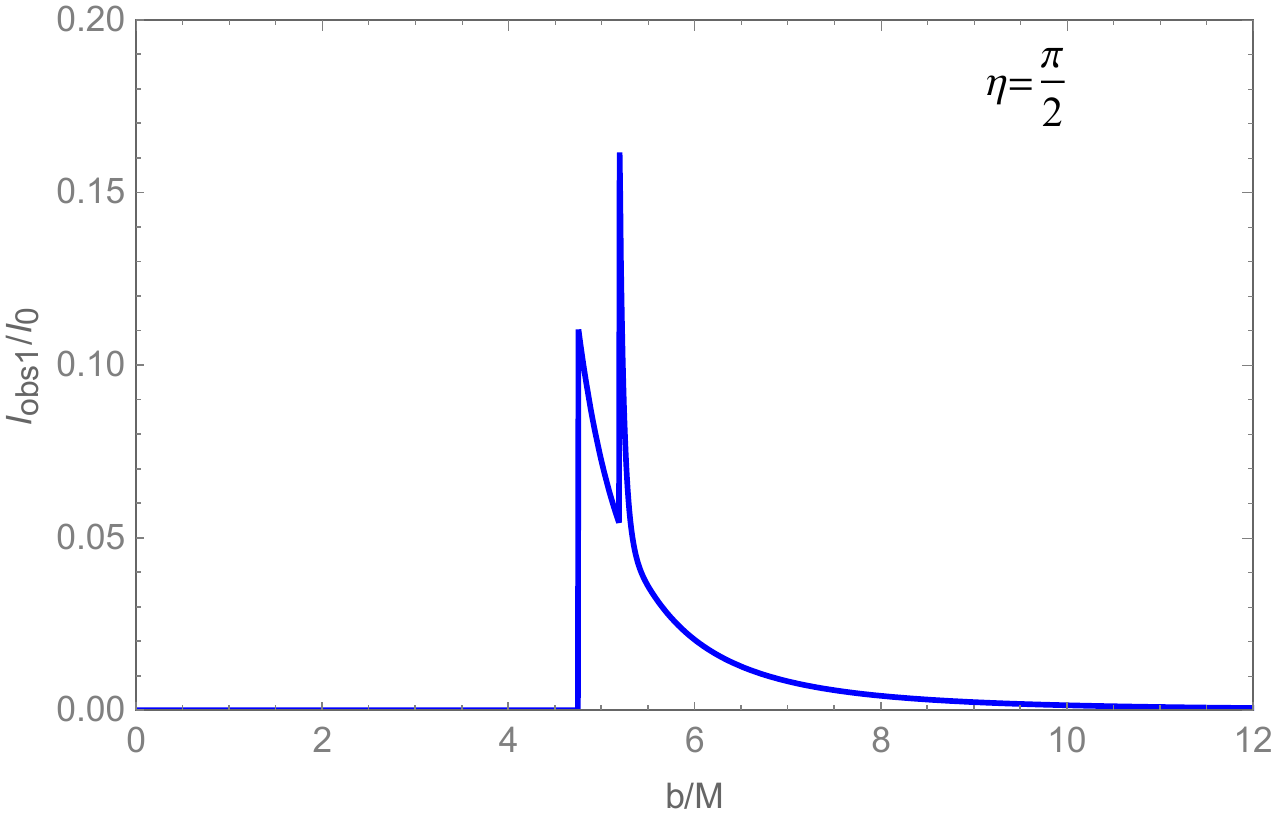}
	\end{minipage}
	\hfill
	\begin{minipage}{0.33\textwidth}
		\includegraphics[height=3.9cm,keepaspectratio]{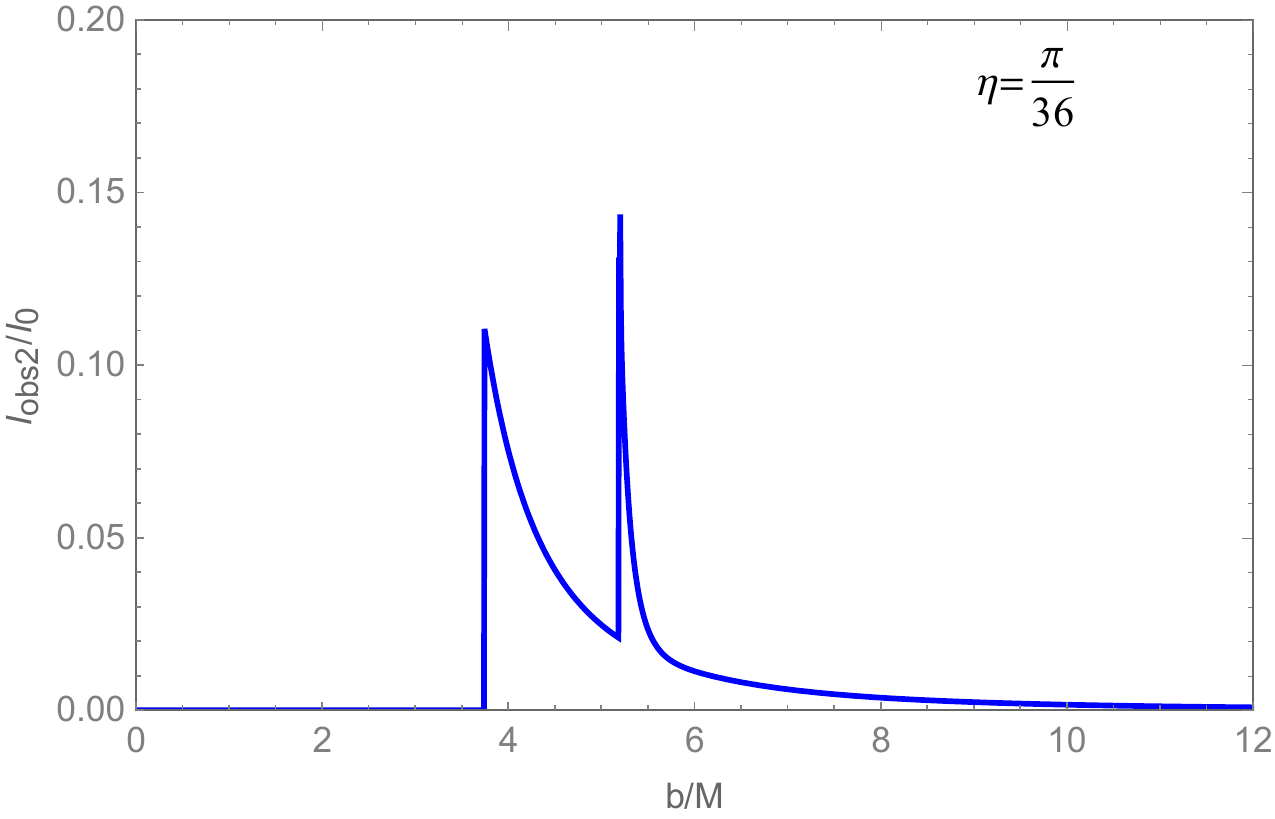}
	\end{minipage}
	\hfill
	\begin{minipage}{0.33\textwidth}
		\includegraphics[height=3.9cm]{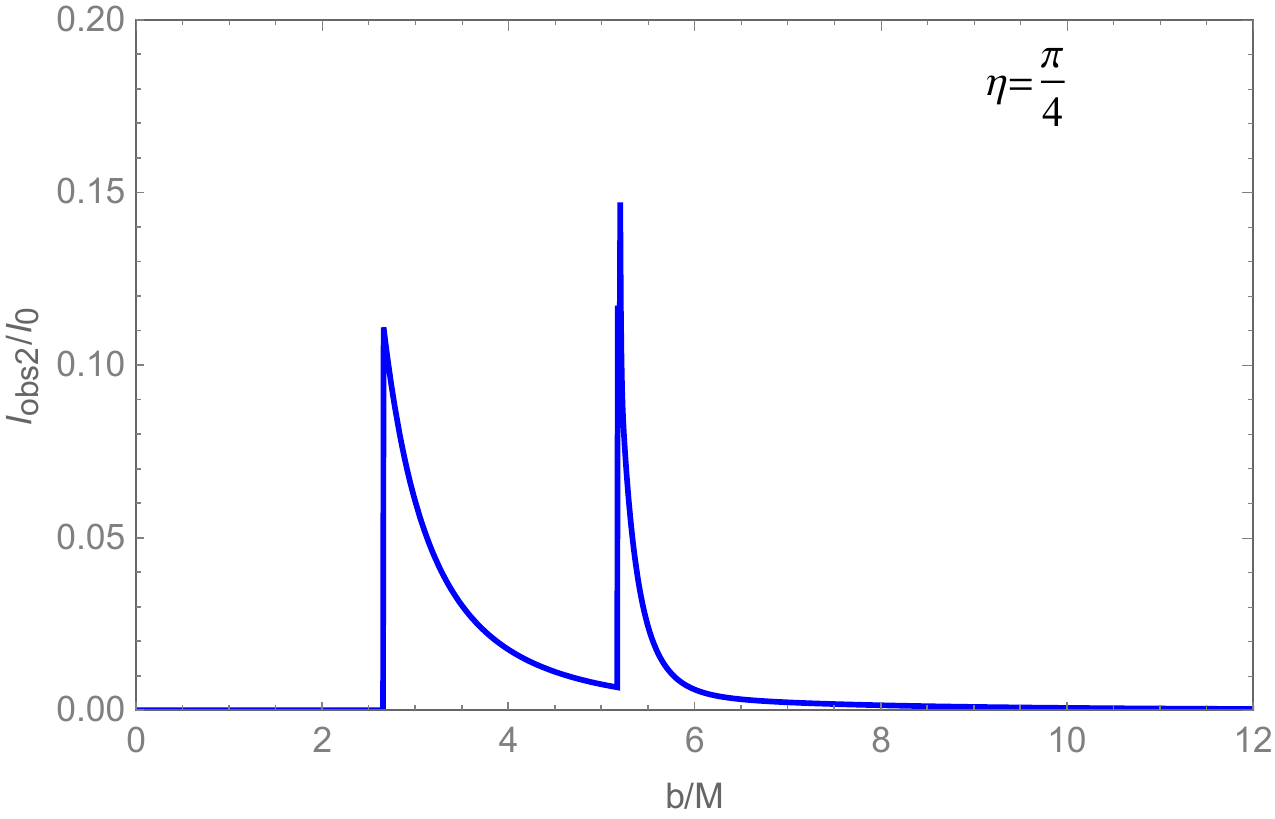}
	\end{minipage}
	\hfill
	\begin{minipage}{0.33\textwidth}
		\includegraphics[height=3.9cm,keepaspectratio]{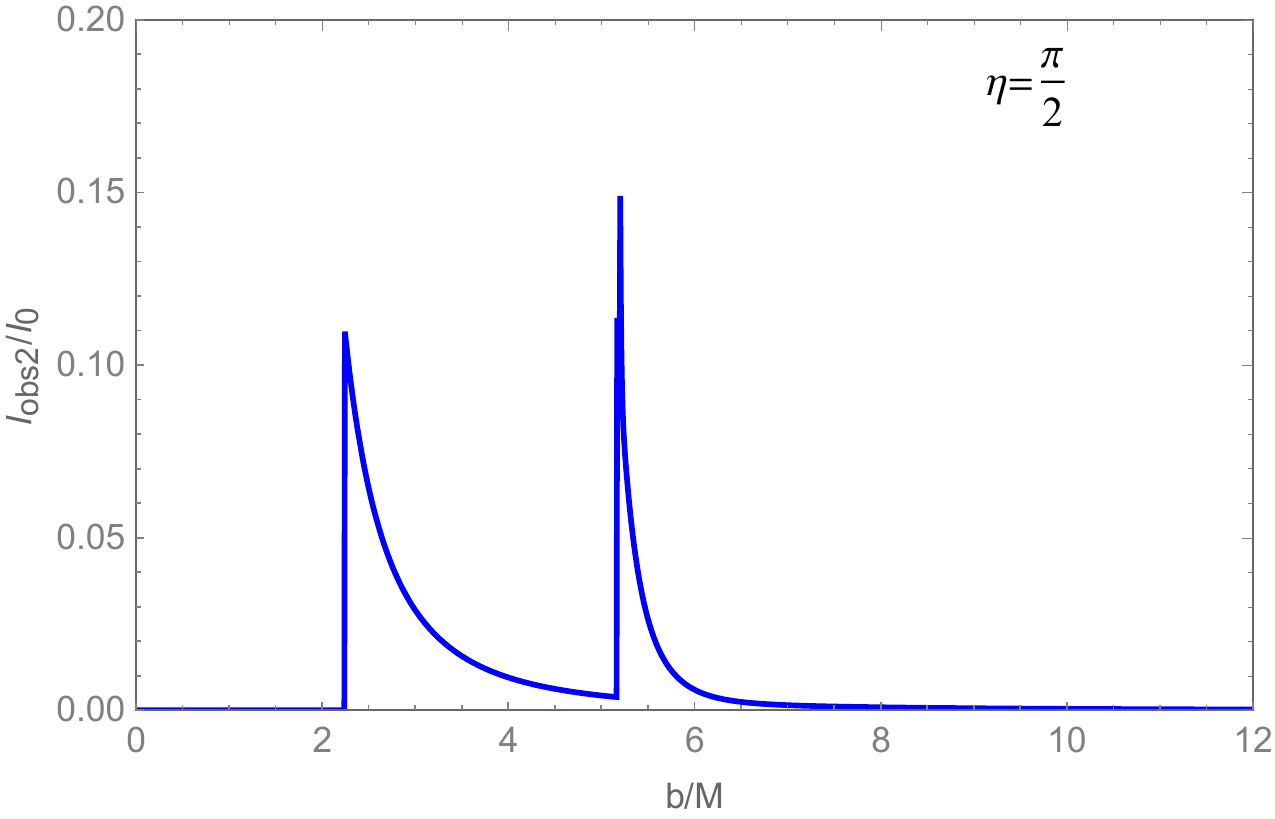}
	\end{minipage}

	\caption{For Model-II at $\pi/4$ inclination, the counter-side and co-side intensities are shown for varying polar angles $\eta$. The upper and lower rows show $I_{\text{obs1}}$ and $I_{\text{obs2}}$ respectively.}
	\label{fig:model2_jieshou45}
\end{figure*}

\begin{figure*}[htb]
	\centering
	\begin{minipage}{0.33\textwidth}
		\includegraphics[height=3.9cm,keepaspectratio]{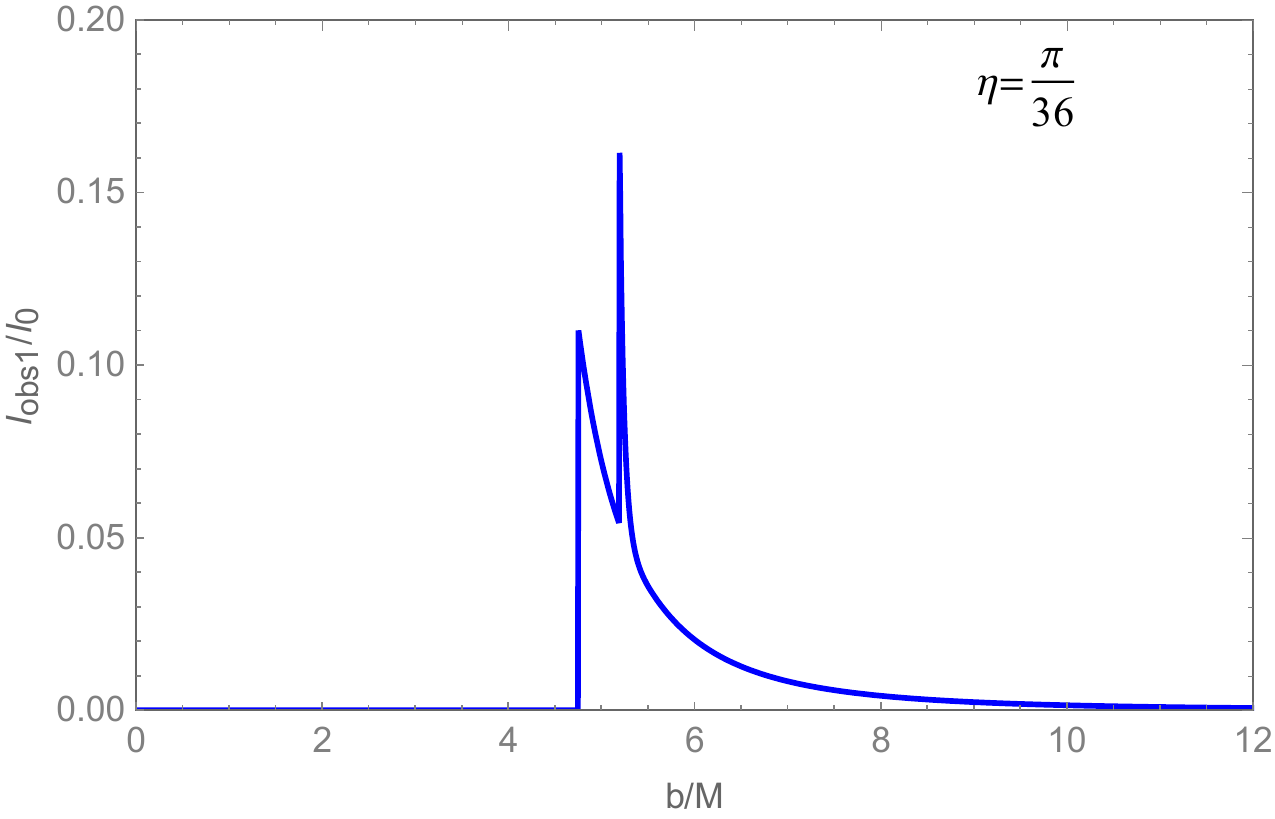}
	\end{minipage}
	\hfill
	\begin{minipage}{0.33\textwidth}
		\includegraphics[height=3.9cm]{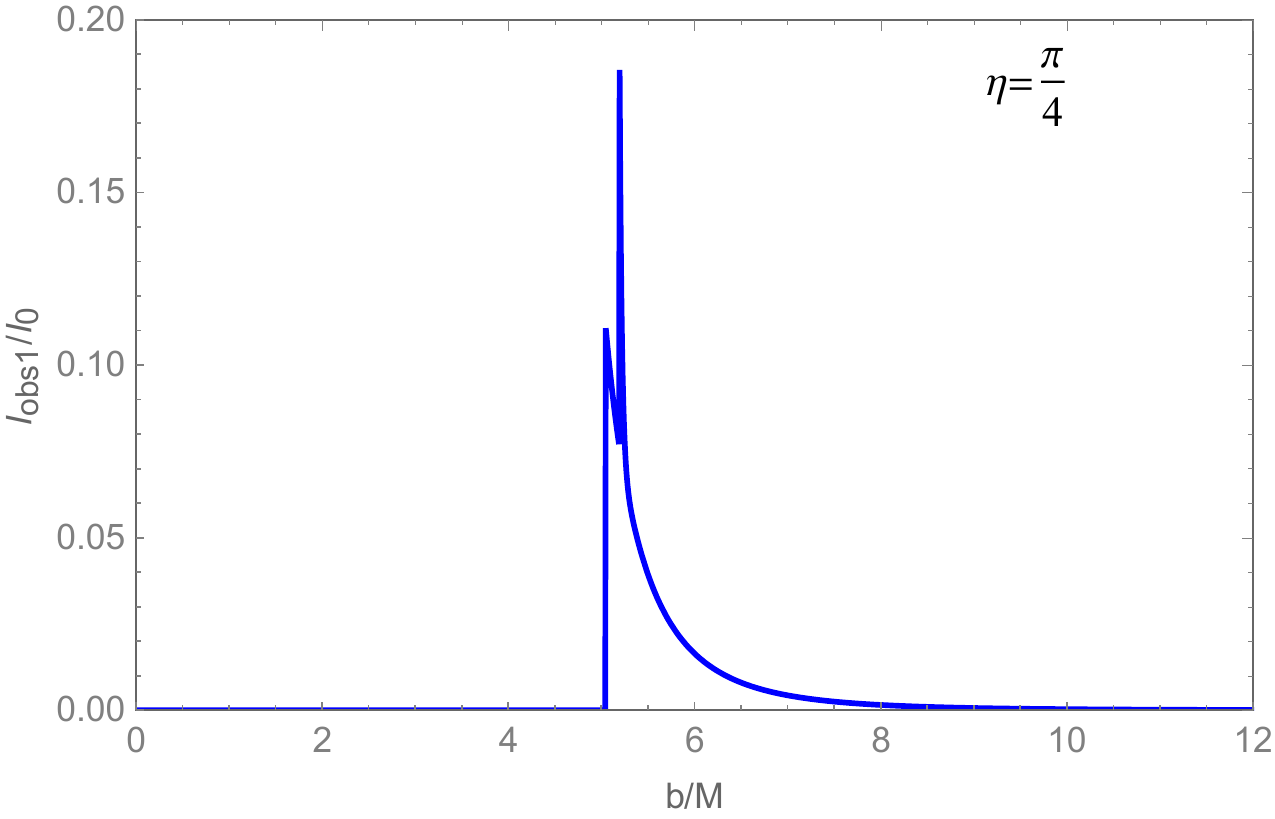}
	\end{minipage}
	\hfill
	\begin{minipage}{0.33\textwidth}
		\includegraphics[height=3.9cm,keepaspectratio]{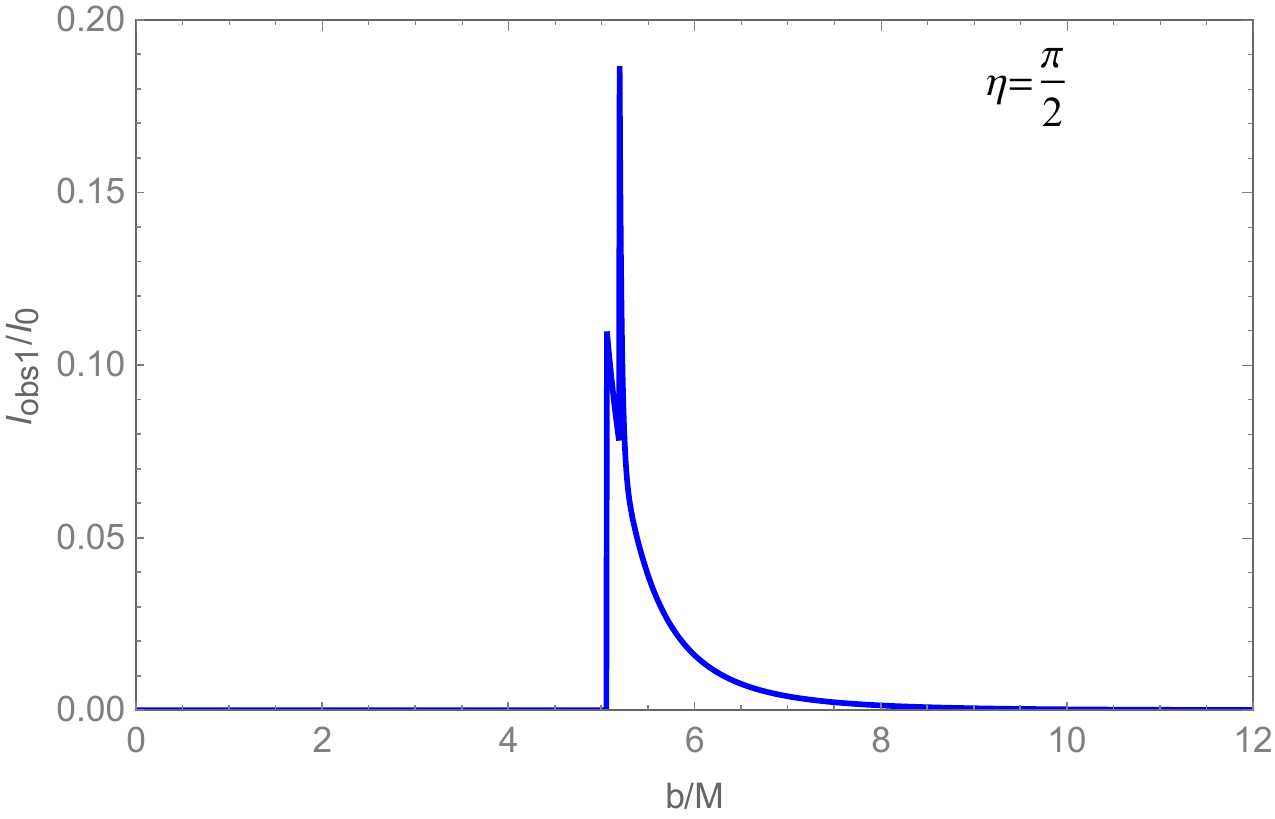}
	\end{minipage}
	\hfill
	\begin{minipage}{0.33\textwidth}
		\includegraphics[height=3.9cm,keepaspectratio]{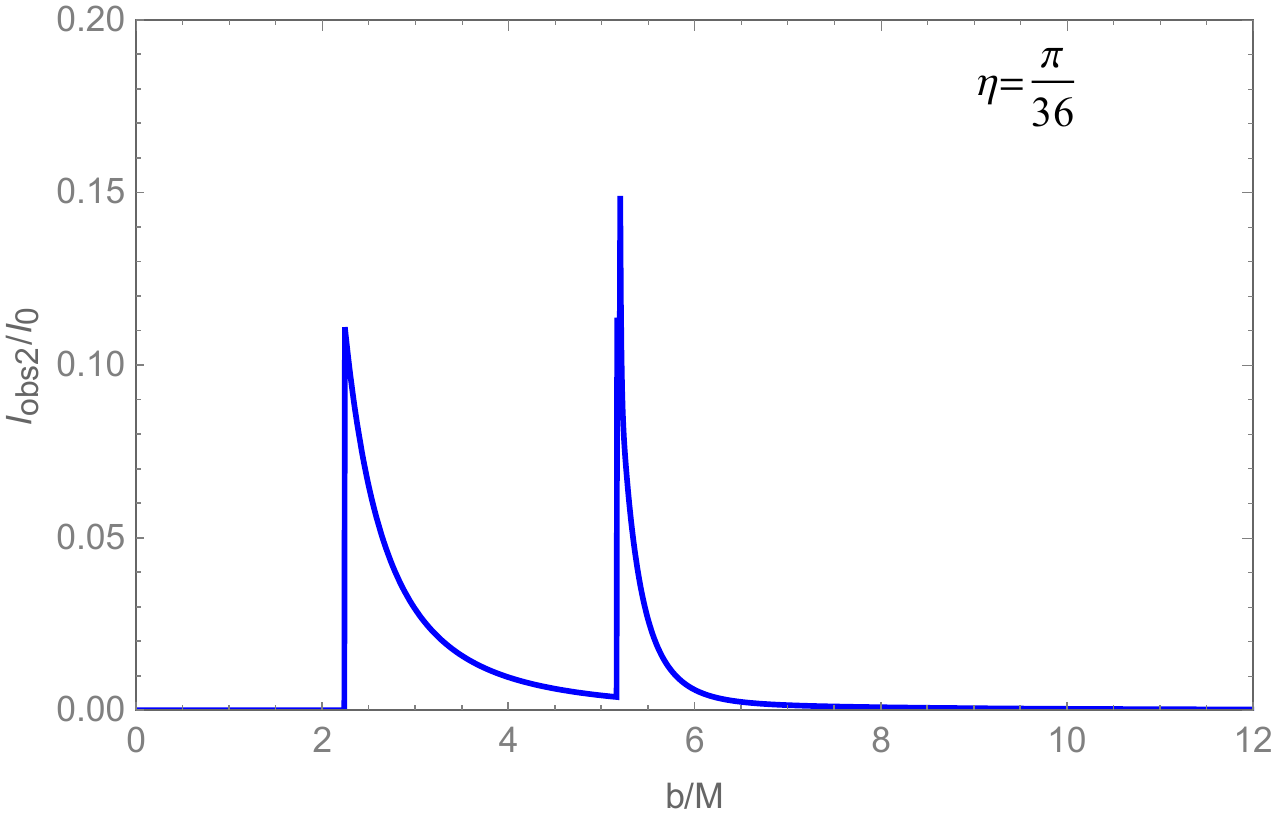}
	\end{minipage}
	\hfill
	\begin{minipage}{0.33\textwidth}
		\includegraphics[height=3.9cm]{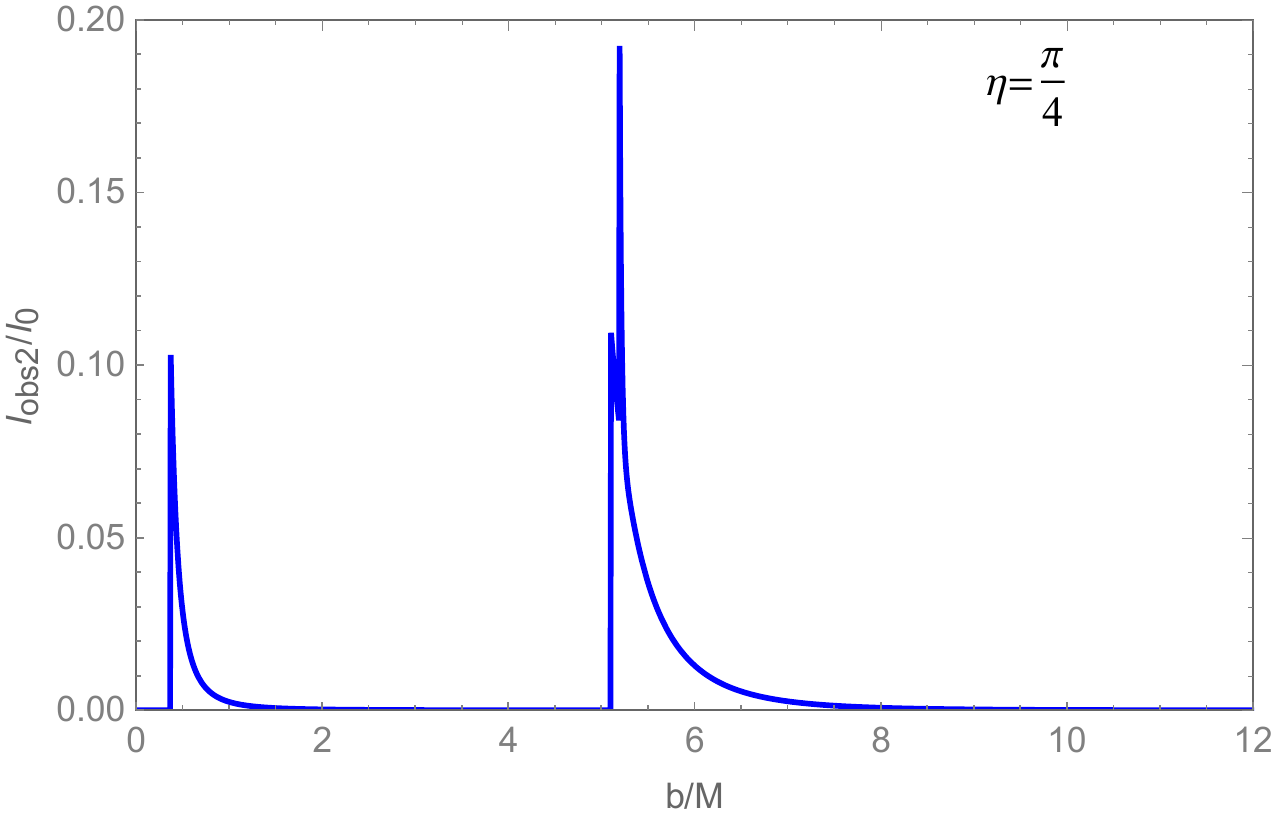}
	\end{minipage}
	\hfill
	\begin{minipage}{0.33\textwidth}
		\includegraphics[height=3.9cm,keepaspectratio]{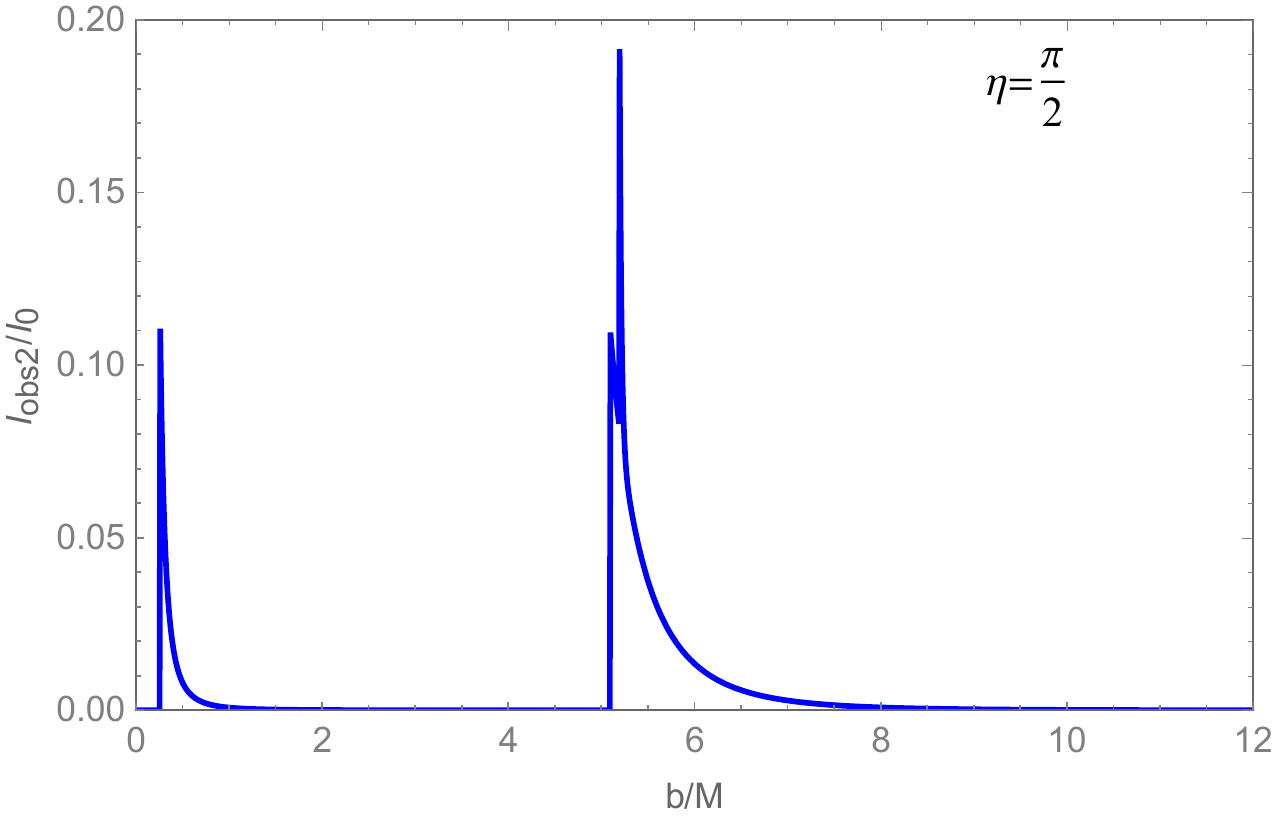}
	\end{minipage}

	\caption{For Model-II at $17\pi/36$ inclination, the counter-side and co-side intensities are shown for varying polar angles $\eta$. The upper and lower rows show $I_{\text{obs1}}$ and $I_{\text{obs2}}$ respectively.}
	\label{fig:model2_jieshou85}
\end{figure*}

\begin{figure*}[htb]
	\centering
	\begin{minipage}{0.33\textwidth}
		\includegraphics[height=3.9cm,keepaspectratio]{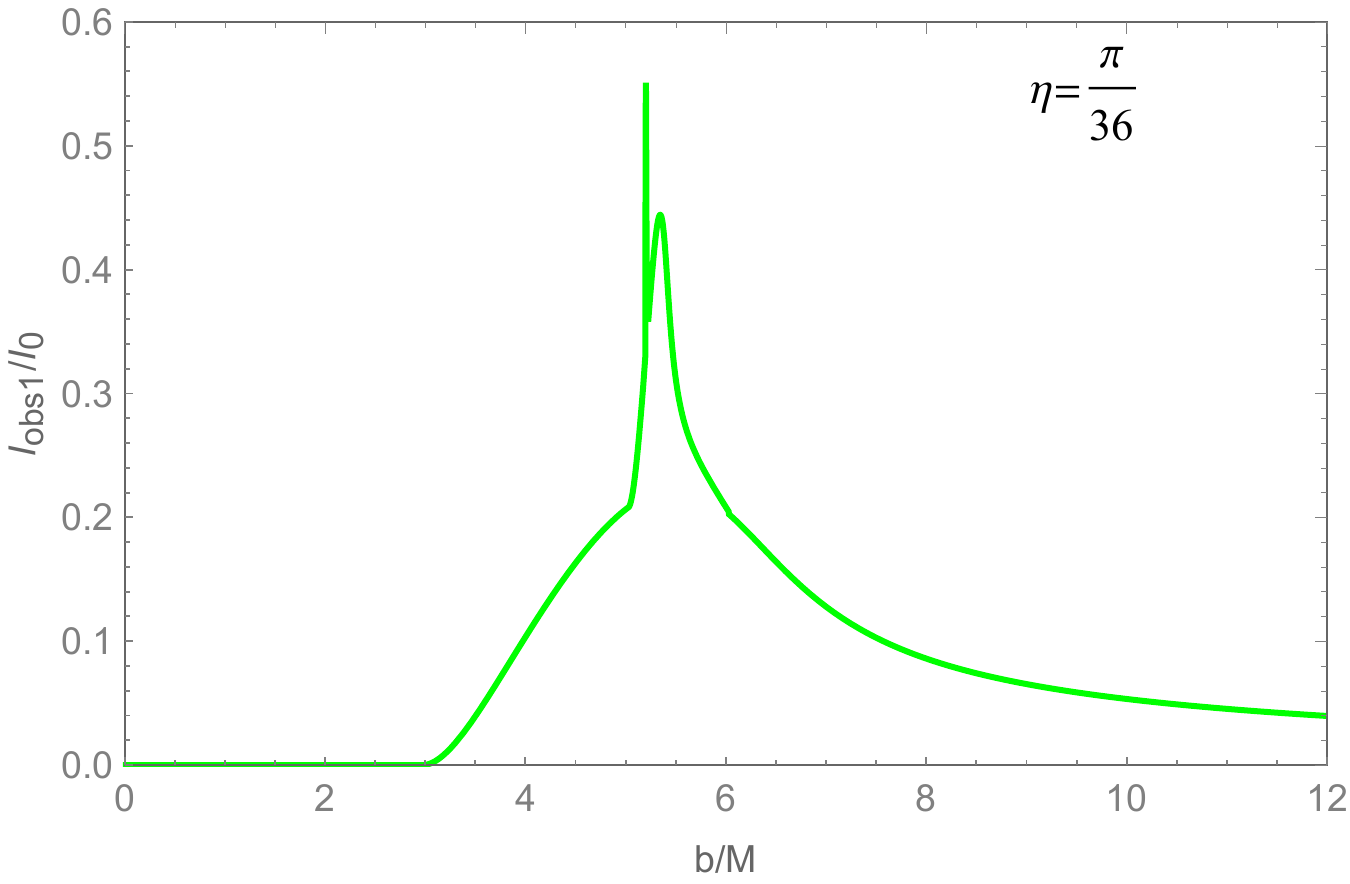}
	\end{minipage}
	\hfill
	\begin{minipage}{0.33\textwidth}
		\includegraphics[height=3.9cm]{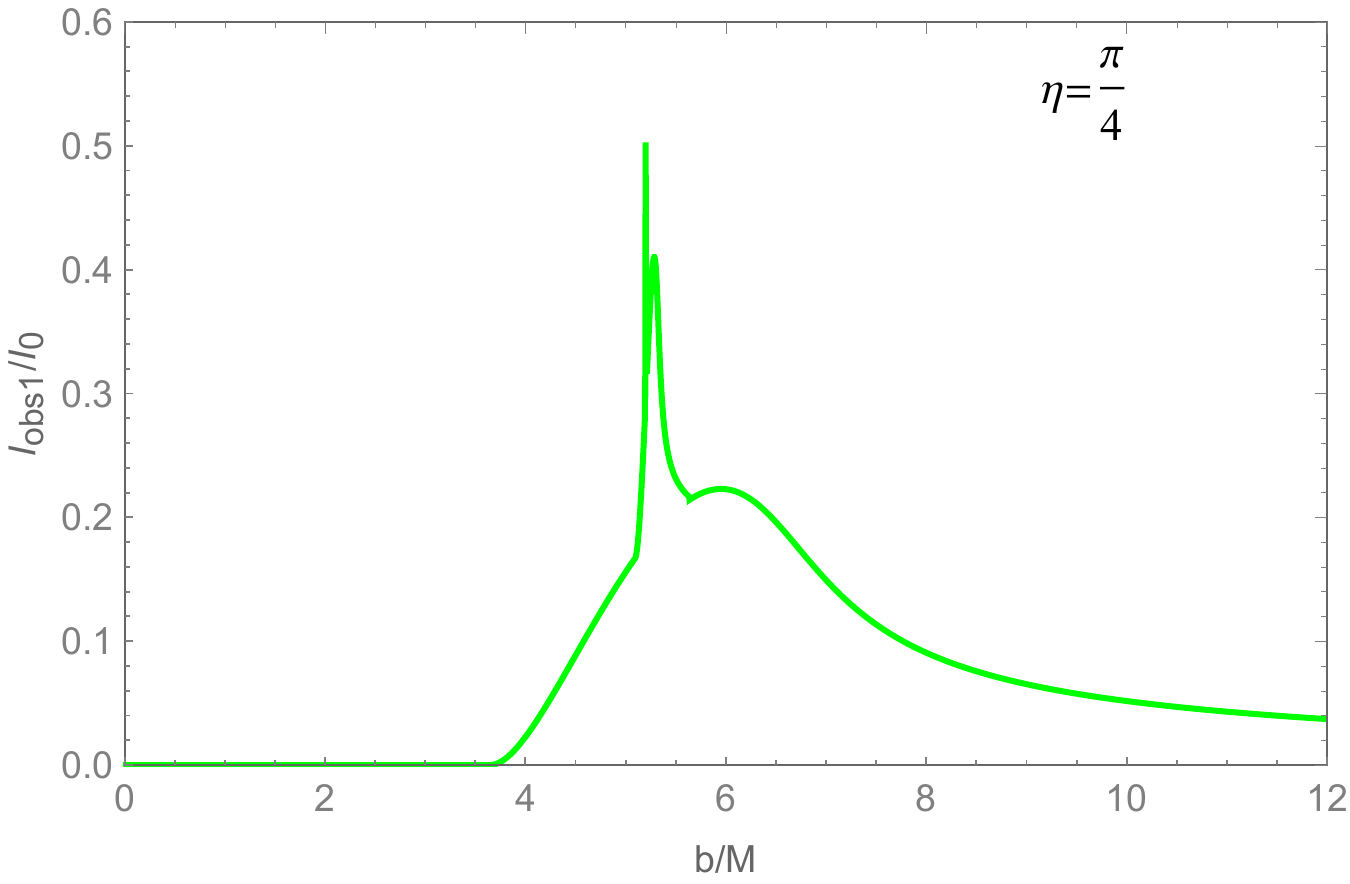}
	\end{minipage}
	\hfill
	\begin{minipage}{0.33\textwidth}
		\includegraphics[height=3.9cm,keepaspectratio]{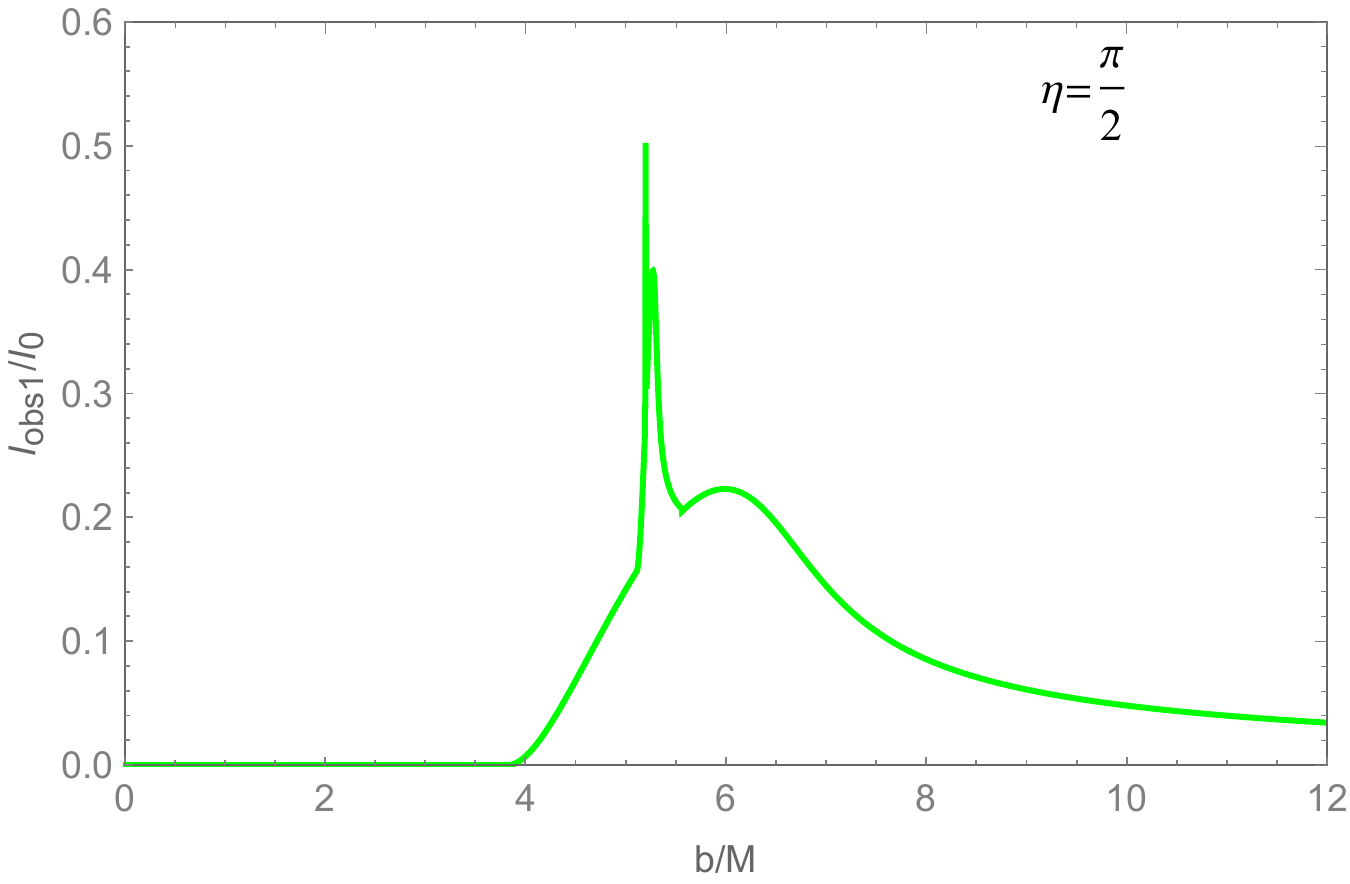}
	\end{minipage}
	\hfill
	\begin{minipage}{0.33\textwidth}
		\includegraphics[height=3.9cm,keepaspectratio]{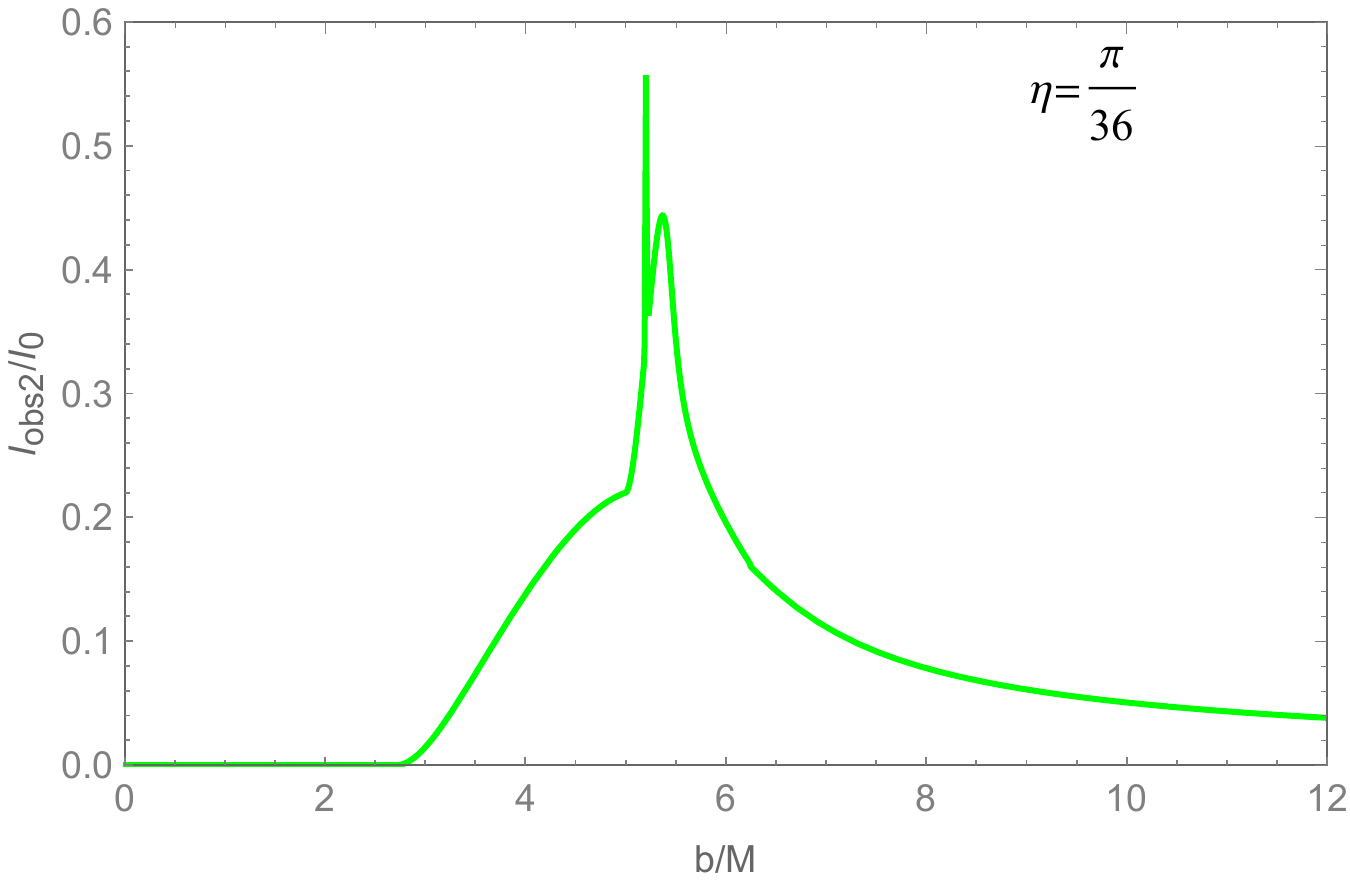}
	\end{minipage}
	\hfill
	\begin{minipage}{0.33\textwidth}
		\includegraphics[height=3.9cm]{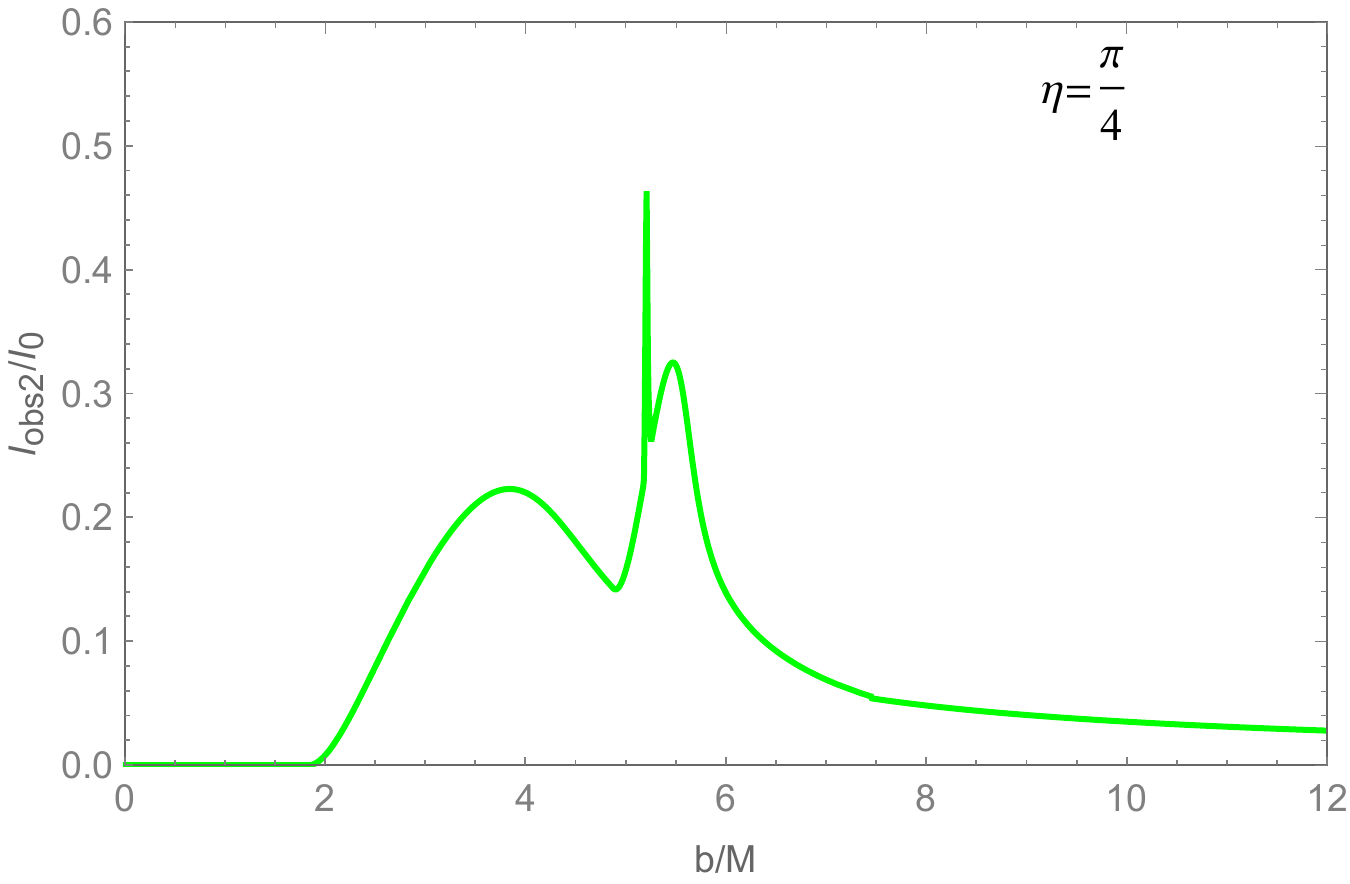}
	\end{minipage}
	\hfill
	\begin{minipage}{0.33\textwidth}
		\includegraphics[height=3.9cm,keepaspectratio]{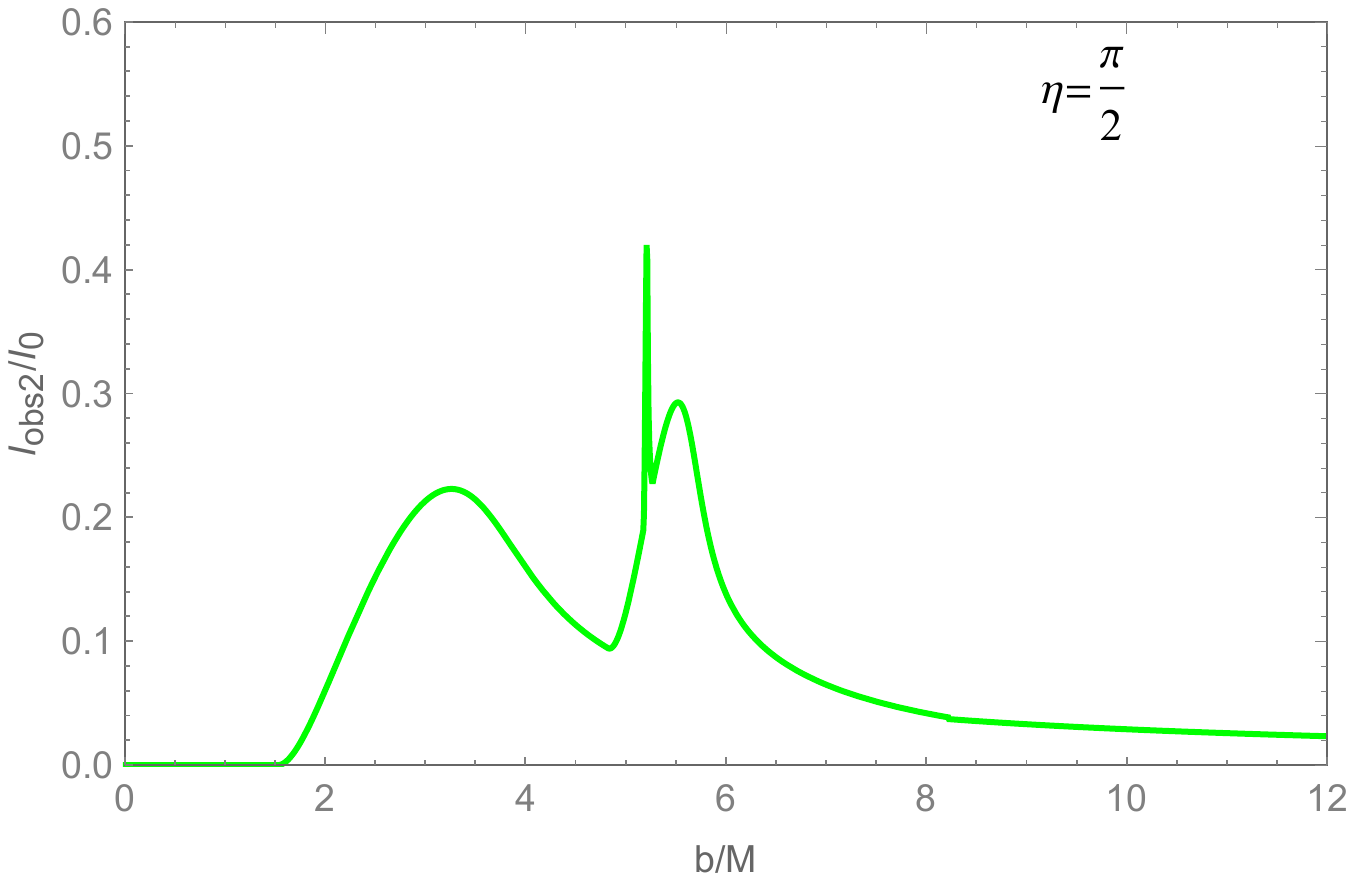}
	\end{minipage}

	\caption{For Model-III at $\pi/4$ inclination, the counter-side and co-side intensities are shown for varying polar angles $\eta$. The upper and lower rows show $I_{\text{obs1}}$ and $I_{\text{obs2}}$ respectively.}
	\label{fig:model3_jieshou45}
\end{figure*}

\begin{figure*}[htb]
	\centering
	\begin{minipage}{0.33\textwidth}
		\includegraphics[height=3.9cm,keepaspectratio]{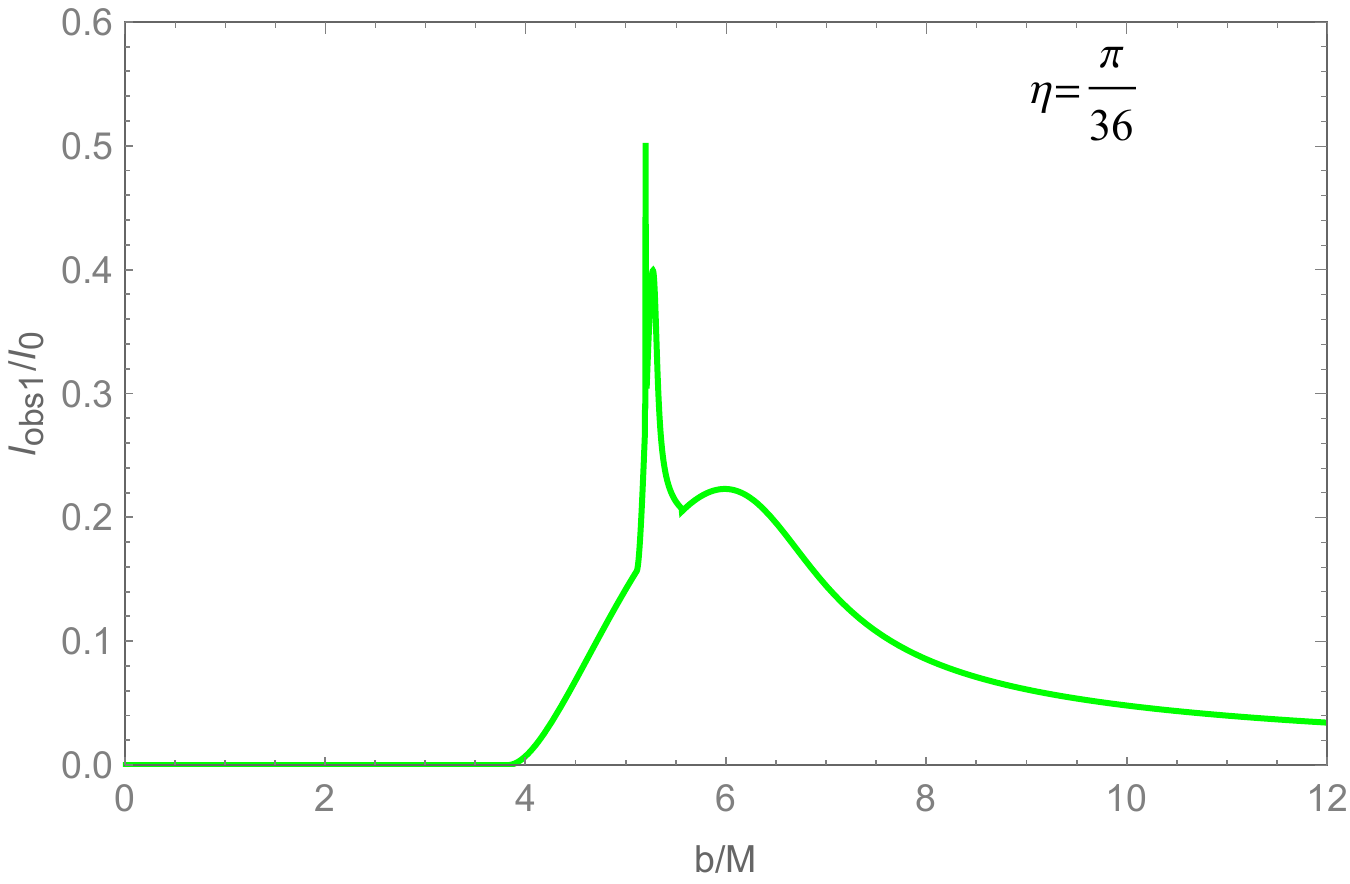}
	\end{minipage}
	\hfill
	\begin{minipage}{0.33\textwidth}
		\includegraphics[height=3.9cm]{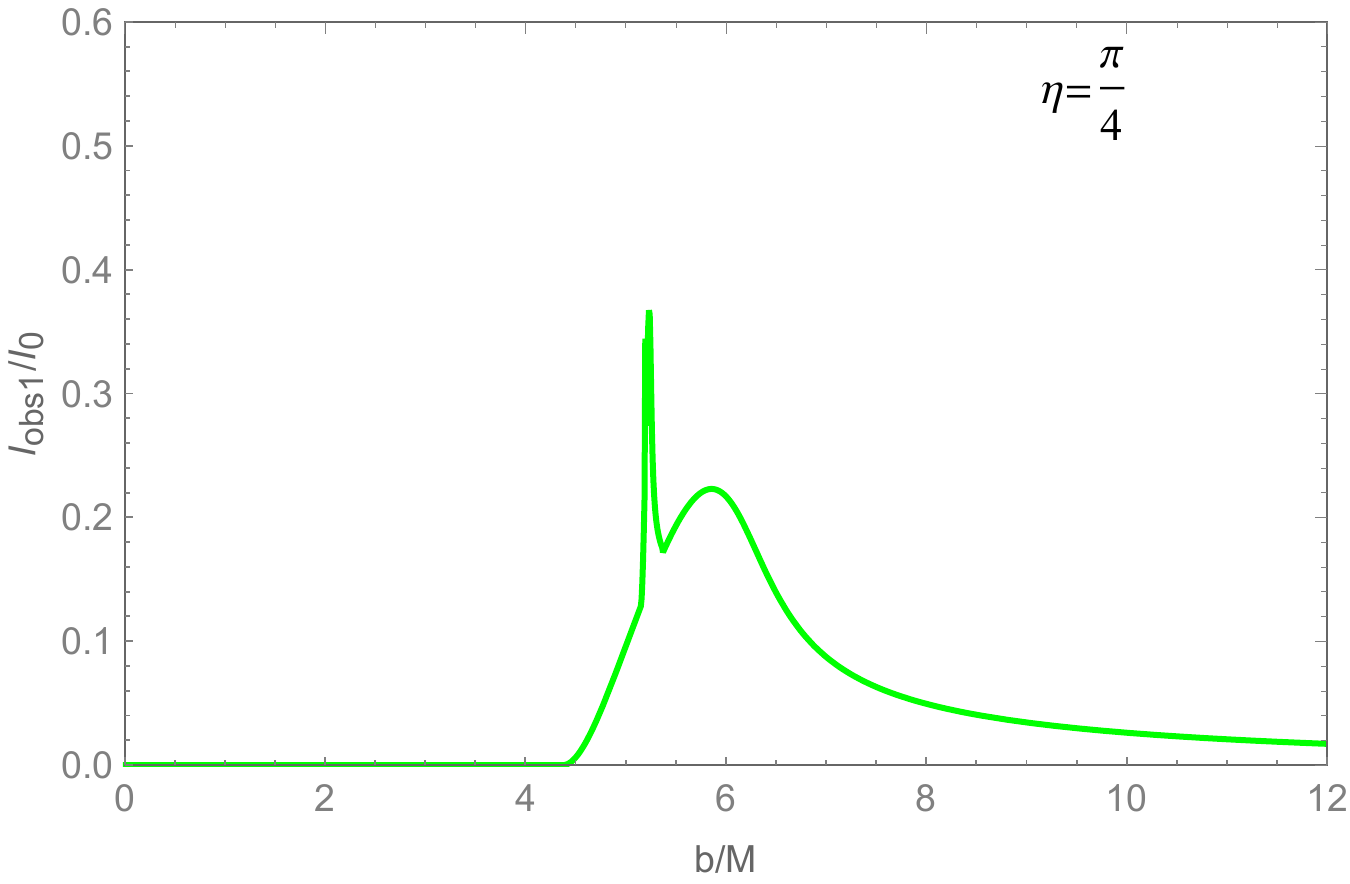}
	\end{minipage}
	\hfill
	\begin{minipage}{0.33\textwidth}
		\includegraphics[height=3.9cm,keepaspectratio]{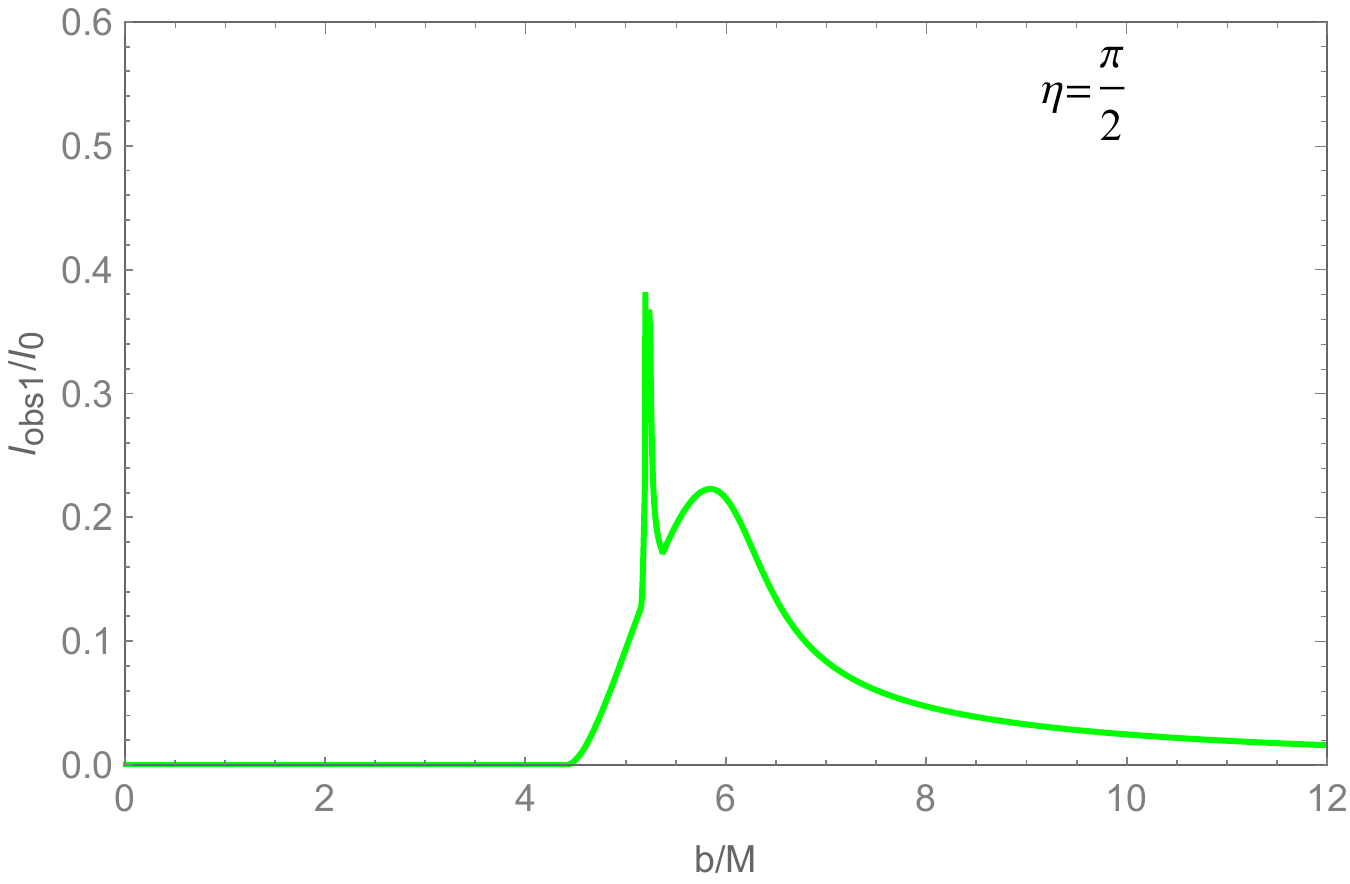}
	\end{minipage}
	\hfill
	\begin{minipage}{0.33\textwidth}
		\includegraphics[height=3.9cm,keepaspectratio]{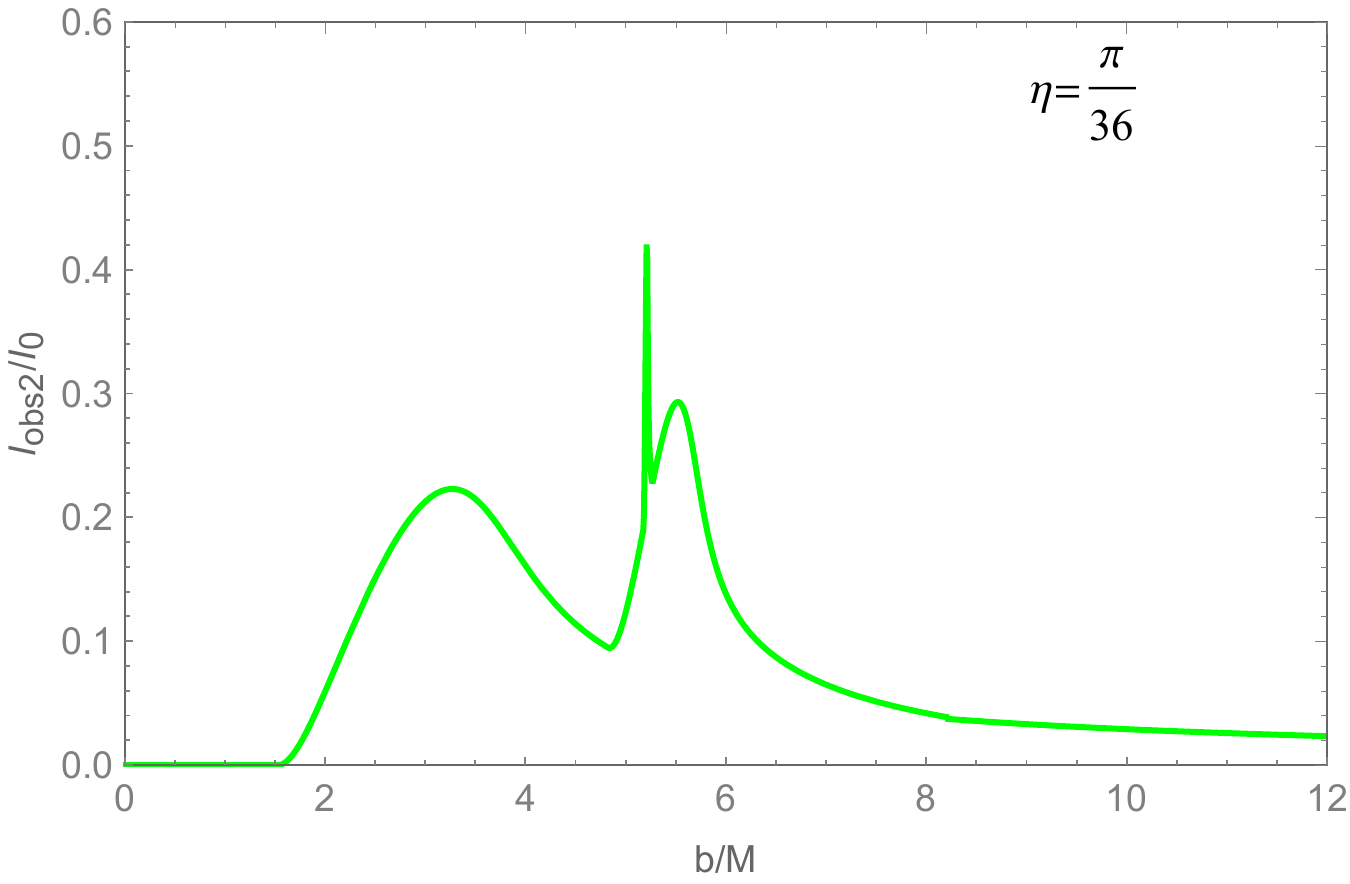}
	\end{minipage}
	\hfill
	\begin{minipage}{0.33\textwidth}
		\includegraphics[height=3.9cm]{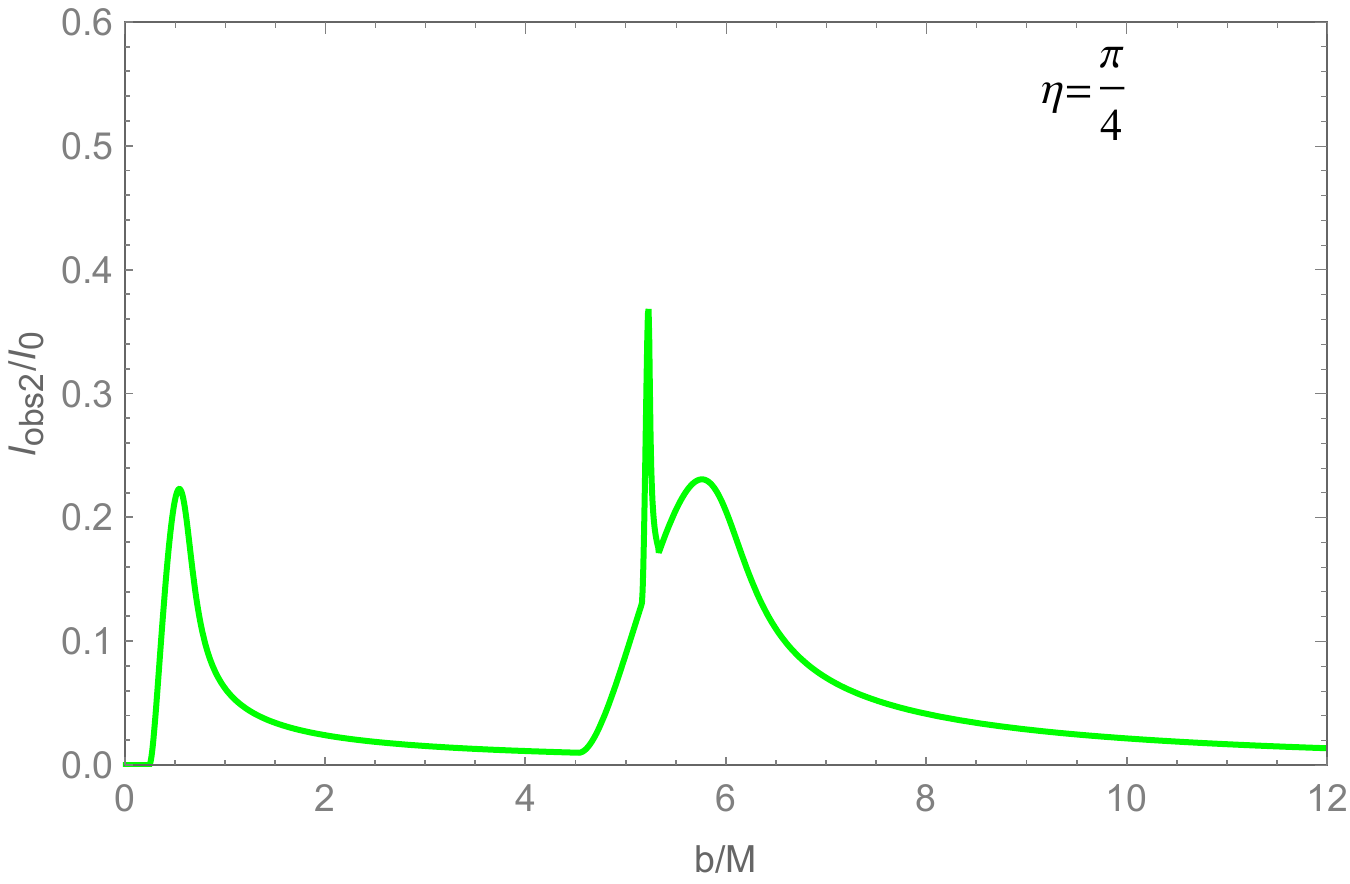}
	\end{minipage}
	\hfill
	\begin{minipage}{0.33\textwidth}
		\includegraphics[height=3.9cm,keepaspectratio]{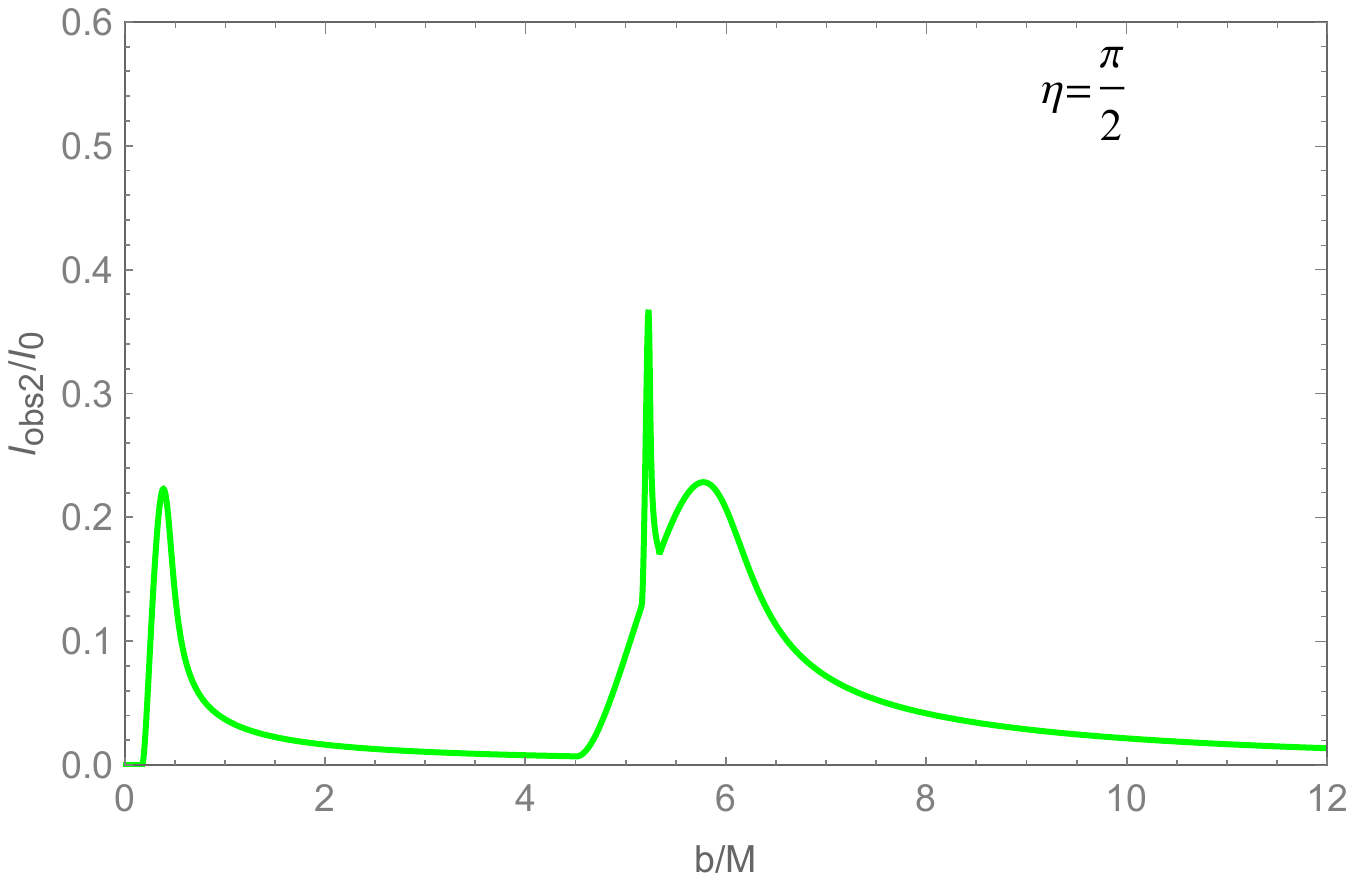}
	\end{minipage}

	\caption{For Model-III at $17\pi/36$ inclination, the counter-side and co-side intensities are shown for varying polar angles $\eta$. The upper and lower rows show $I_{\text{obs1}}$ and $I_{\text{obs2}}$ respectively.}
	\label{fig:model3_jieshou85}
\end{figure*}

In this appendix, we briefly present the received intensity profiles for selected polar angles $\eta$ in optically thin accretion disk Models II and III, shown in Figs.~\ref{fig:model2_jieshou45}--\ref{fig:model3_jieshou85}, at inclination angles of $\pi/4$ and $17\pi/36$.

%model2_45

%model2_85

%model3_45

%model3_85

%%=============================================================================================

%%=============================================================================================


\begin{thebibliography}{61}

\bibitem{Will:2014kxa}
  C.M.~Will,
  {The confrontation between general relativity and experiment}.
  Living Rev. Rel.  \textbf{17}, 4 (2014).
  {\url{https://doi.org/10.12942/lrr-2014-4}}.
  {\href{https://arxiv.org/abs/1403.7377}{{arXiv:1403.7377}}}

\bibitem{LIGOScientific:2016aoc}
  B.P.~Abbott {\em et~al.} (LIGO Scientific Collaboration and Virgo
  Collaboration),
  {Observation of Gravitational Waves from a Binary Black Hole Merger}.
  Phys. Rev. Lett.  \textbf{116}, 061102 (2016).
  {\url{https://doi.org/10.1103/PhysRevLett.116.061102}}.
  {\href{https://arxiv.org/abs/1602.03837}{{arXiv:1602.03837}}}

\bibitem{EventHorizonTelescope:2019dse}
  K.~Akiyama {\em et~al.} (Event Horizon Telescope Collaboration),
  {First M87 Event Horizon Telescope results. I. The shadow of the supermassive
  black hole}.
  Astrophys. J. Lett.  \textbf{875}, L1 (2019).
  {\url{https://doi.org/10.3847/2041-8213/ab0ec7}}.
  {\href{https://arxiv.org/abs/1906.11238}{{arXiv:1906.11238}}}

\bibitem{Kumar:2018ple}
  R.~Kumar and S.G.~Ghosh,
  {Black hole parameter estimation from its shadow}.
  Astrophys. J.  \textbf{892}, 78 (2020).
  {\url{https://doi.org/10.3847/1538-4357/ab77b0}}.
  {\href{https://arxiv.org/abs/1811.01260}{{arXiv:1811.01260}}}

\bibitem{Hioki:2022mdg}
  K.~Hioki and U.~Miyamoto,
  {Determining parameters of a spherical black hole with a thin accretion disk
  by observing its shadow}.
  Phys. Rev. D  \textbf{107}, 044042 (2023).
  {\url{https://doi.org/10.1103/PhysRevD.107.044042}}.
  {\href{https://arxiv.org/abs/2210.02164}{{arXiv:2210.02164}}}

\bibitem{Synge:1966okc}
  J.L.~Synge,
  {The escape of photons from gravitationally intense stars}.
  Mon. Not. Roy. Astron. Soc.  \textbf{131}, 463 (1966).
  {\url{https://doi.org/10.1093/mnras/131.3.463}}

\bibitem{Bardeen:1972fi}
  J.M.~Bardeen, W.H.~Press, and S.A.~Teukolsky,
  {Rotating black holes: Locally nonrotating frames, energy extraction, and
  scalar synchrotron radiation}.
  Astrophys. J.  \textbf{178}, 347 (1972).
  {\url{https://doi.org/10.1086/151796}}

\bibitem{Claudel:2000yi}
  C.-M.~Claudel, K.S.~Virbhadra, and G.F.R.~Ellis,
  {The geometry of photon surfaces}.
  J. Math. Phys.  \textbf{42}, 818 (2001).
  {\url{https://doi.org/10.1063/1.1308507}}.
  {\href{https://arxiv.org/abs/gr-qc/0005050}{{arXiv:gr-qc/0005050}}}

\bibitem{Bambi:2008jg}
  C.~Bambi and K.~Freese,
  {Apparent shape of super-spinning black holes}.
  Phys. Rev. D  \textbf{79}, 043002 (2009).
  {\url{https://doi.org/10.1103/PhysRevD.79.043002}}.
  {\href{https://arxiv.org/abs/0812.1328}{{arXiv:0812.1328}}}

\bibitem{Wei:2013kza}
  S.-W.~Wei and Y.-X.~Liu,
  {Observing the shadow of Einstein-Maxwell-Dilaton-Axion black hole}.
  J. Cosmol. Astropart. Phys.  \textbf{11}, 063 (2013).
  {\url{https://doi.org/10.1088/1475-7516/2013/11/063}}.
  {\href{https://arxiv.org/abs/1311.4251}{{arXiv:1311.4251}}}

\bibitem{Younsi:2016azx}
  Z.~Younsi, A.~Zhidenko, L.~Rezzolla, R.~Konoplya, and Y.~Mizuno,
  {New method for shadow calculations: Application to parametrized axisymmetric
  black holes}.
  Phys. Rev. D  \textbf{94}, 084025 (2016).
  {\url{https://doi.org/10.1103/PhysRevD.94.084025}}.
  {\href{https://arxiv.org/abs/1607.05767}{{arXiv:1607.05767}}}

\bibitem{Cunha:2016wzk}
  P.V.P.~Cunha, C.A.R.~Herdeiro, B.~Kleihaus, J.~Kunz, and E.~Radu,
  {Shadows of Einstein{\textendash}dilaton{\textendash}Gauss{\textendash}Bonnet
  black holes}.
  Phys. Lett. B  \textbf{768}, 373 (2017).
  {\url{https://doi.org/10.1016/j.physletb.2017.03.020}}.
  {\href{https://arxiv.org/abs/1701.00079}{{arXiv:1701.00079}}}

\bibitem{Stuchlik:2018qyz}
  Z.~Stuchl{\'\i}k, D.~Charbul{\'a}k, and J.~Schee,
  {Light escape cones in local reference frames of Kerr{\textendash}de Sitter
  black hole spacetimes and related black hole shadows}.
  Eur. Phys. J. C  \textbf{78}, 180 (2018).
  {\url{https://doi.org/10.1140/epjc/s10052-018-5578-6}}.
  {\href{https://arxiv.org/abs/1811.00072}{{arXiv:1811.00072}}}

\bibitem{Wang:2018prk}
  H.-M.~Wang, Y.-M.~Xu, and S.-W.~Wei,
  {Shadows of Kerr-like black holes in a modified gravity theory}.
  J. Cosmol. Astropart. Phys.  \textbf{03}, 046 (2019).
  {\url{https://doi.org/10.1088/1475-7516/2019/03/046}}.
  {\href{https://arxiv.org/abs/1810.12767}{{arXiv:1810.12767}}}

\bibitem{Jusufi:2019nrn}
  K.~Jusufi, M.~Jamil, P.~Salucci, T.~Zhu, and S.~Haroon,
  {Black hole surrounded by a dark matter halo in the M87 galactic center and
  its identification with shadow images}.
  Phys. Rev. D  \textbf{100}, 044012 (2019).
  {\url{https://doi.org/10.1103/PhysRevD.100.044012}}.
  {\href{https://arxiv.org/abs/1905.11803}{{arXiv:1905.11803}}}

\bibitem{Liu:2020ola}
  C.~Liu, T.~Zhu, Q.~Wu, K.~Jusufi, M.~Jamil, M.~Azreg-A{\"\i}nou, and A.~Wang,
  {Shadow and quasinormal modes of a rotating loop quantum black hole}.
  Phys. Rev. D  \textbf{101}, 084001 (2020).
  {\url{https://doi.org/10.1103/PhysRevD.101.084001}}.
  {\href{https://arxiv.org/abs/2003.00477}{{arXiv:2003.00477}}}

\bibitem{Hou:2021okc}
  Y.~Hou, M.~Guo, and B.~Chen,
  {Revisiting the shadow of braneworld black holes}.
  Phys. Rev. D  \textbf{104}, 024001 (2021).
  {\url{https://doi.org/10.1103/PhysRevD.104.024001}}.
  {\href{https://arxiv.org/abs/2103.04369}{{arXiv:2103.04369}}}

\bibitem{Ban:2024qsa}
  Z.~Ban, J.~Chen, and J.~Yang,
  {Shadows of rotating black holes in effective quantum gravity}.
  Eur. Phys. J. C  \textbf{85}, 878 (2025).
  {\url{https://doi.org/10.1140/epjc/s10052-025-14614-y}}.
  {\href{https://arxiv.org/abs/2411.09374}{{arXiv:2411.09374}}}

\bibitem{Abramowicz:2011xu}
  M.A.~Abramowicz and P.C.~Fragile,
  {Foundations of black hole accretion disk theory}.
  Living Rev. Rel.  \textbf{16}, 1 (2013).
  {\url{https://doi.org/10.12942/lrr-2013-1}}.
  {\href{https://arxiv.org/abs/1104.5499}{{arXiv:1104.5499}}}

\bibitem{Chen:2025doc}
  G.~Chen, S.~Guo, J.-S.~Li, Y.-X.~Huang, L.-F.~Li, and P.~Xu,
  {Influences of accretion flow and dilaton charge on the images of
  Einstein-Maxwell-dilation black holes}.
  Sci. China Phys. Mech. Astron.  \textbf{68}, 260413 (2025).
  {\url{https://doi.org/10.1007/s11433-024-2626-5}}.
  {\href{https://arxiv.org/abs/2502.07618}{{arXiv:2502.07618}}}

\bibitem{Zhang:2024lsf}
  Z.~Zhang, Y.~Hou, M.~Guo, and B.~Chen,
  {Imaging thick accretion disks and jets surrounding black holes}.
  J. Cosmol. Astropart. Phys.  \textbf{05}, 032 (2024).
  {\url{https://doi.org/10.1088/1475-7516/2024/05/032}}.
  {\href{https://arxiv.org/abs/2401.14794}{{arXiv:2401.14794}}}

\bibitem{Luminet:1979nyg}
  J.P.~Luminet,
  {Image of a spherical black hole with thin accretion disk}.
  Astron. Astrophys.  \textbf{75}, 228 (1979)

\bibitem{Gyulchev:2019tvk}
  G.~Gyulchev, P.~Nedkova, T.~Vetsov, and S.~Yazadjiev,
  {Image of the Janis-Newman-Winicour naked singularity with a thin accretion
  disk}.
  Phys. Rev. D  \textbf{100}, 024055 (2019).
  {\url{https://doi.org/10.1103/PhysRevD.100.024055}}.
  {\href{https://arxiv.org/abs/1905.05273}{{arXiv:1905.05273}}}

\bibitem{Guo:2022rql}
  S.~Guo, G.-R.~Li, and E.-W.~Liang,
  {Observable characteristics of the charged black hole surrounded by thin disk
  accretion in Rastall gravity}.
  Class. Quant. Grav.  \textbf{39}, 135004 (2022).
  {\url{https://doi.org/10.1088/1361-6382/ac6fa8}}.
  {\href{https://arxiv.org/abs/2205.11241}{{arXiv:2205.11241}}}

\bibitem{Huang:2023ilm}
  Y.-X.~Huang, S.~Guo, Y.-H.~Cui, Q.-Q.~Jiang, and K.~Lin,
  {Influence of accretion disk on the optical appearance of the
  Kazakov-Solodukhin black hole}.
  Phys. Rev. D  \textbf{107}, 123009 (2023).
  {\url{https://doi.org/10.1103/PhysRevD.107.123009}}.
  {\href{https://arxiv.org/abs/2311.00302}{{arXiv:2311.00302}}}

\bibitem{Guo:2023grt}
  S.~Guo, Y.-X.~Huang, Y.-H.~Cui, Y.~Han, Q.-Q.~Jiang, E.-W.~Liang, and K.~Lin,
  {Unveiling the unconventional optical signatures of regular black holes
  within accretion disk}.
  Eur. Phys. J. C  \textbf{83}, 1059 (2023).
  {\url{https://doi.org/10.1140/epjc/s10052-023-12208-0}}.
  {\href{https://arxiv.org/abs/2310.20523}{{arXiv:2310.20523}}}

\bibitem{Asukula:2023akj}
  H.~Asuk{\"u}la, M.~Hohmann, V.~Karanasou, S.~Bahamonde, C.~Pfeifer, and
  J.L.~Rosa,
  {Spherically symmetric vacuum solutions in one-parameter new general
  relativity and their phenomenology}.
  Phys. Rev. D  \textbf{109}, 064027 (2024).
  {\url{https://doi.org/10.1103/PhysRevD.109.064027}}.
  {\href{https://arxiv.org/abs/2311.17999}{{arXiv:2311.17999}}}

\bibitem{You:2024jeu}
  L.~You, Y.-H.~Feng, R.-B.~Wang, X.-R.~Hu, and J.-B.~Deng,
  {Decoding quantum gravity information with black hole accretion disk}.
  Universe  \textbf{10}, 393 (2024).
  {\url{https://doi.org/10.3390/universe10100393}}.
  {\href{https://arxiv.org/abs/2404.01418}{{arXiv:2404.01418}}}

\bibitem{Cui:2024wvz}
  Y.-H.~Cui, S.~Guo, Y.-X.~Huang, Y.~Liang, and K.~Lin,
  {Optical appearance of numerical black hole solutions in higher derivative
  gravity}.
  Eur. Phys. J. C  \textbf{84}, 772 (2024).
  {\url{https://doi.org/10.1140/epjc/s10052-024-13153-2}}.
  {\href{https://arxiv.org/abs/2408.03387}{{arXiv:2408.03387}}}

\bibitem{You:2024uql}
  L.~You, R.-b.~Wang, S.-J.~Ma, J.-B.~Deng, and X.-R.~Hu,
  {Optical properties of Euler-Heisenberg black hole in the Cold Dark Matter
  Halo}.
  {\href{https://arxiv.org/abs/2403.12840}{{arXiv:2403.12840}}}

\bibitem{Gralla:2019xty}
  S.E.~Gralla, D.E.~Holz, and R.M.~Wald,
  {Black hole shadows, photon rings, and lensing rings}.
  Phys. Rev. D  \textbf{100}, 024018 (2019).
  {\url{https://doi.org/10.1103/PhysRevD.100.024018}}.
  {\href{https://arxiv.org/abs/1906.00873}{{arXiv:1906.00873}}}

\bibitem{Peng:2020wun}
  J.~Peng, M.~Guo, and X.-H.~Feng,
  {Influence of quantum correction on black hole shadows, photon rings, and
  lensing rings}.
  Chin. Phys. C  \textbf{45}, 085103 (2021).
  {\url{https://doi.org/10.1088/1674-1137/ac06bb}}.
  {\href{https://arxiv.org/abs/2008.00657}{{arXiv:2008.00657}}}

\bibitem{Zeng:2020vsj}
  X.-X.~Zeng and H.-Q.~Zhang,
  {Influence of quintessence dark energy on the shadow of black hole}.
  Eur. Phys. J. C  \textbf{80}, 1058 (2020).
  {\url{https://doi.org/10.1140/epjc/s10052-020-08656-7}}.
  {\href{https://arxiv.org/abs/2007.06333}{{arXiv:2007.06333}}}

\bibitem{Zeng:2020dco}
  X.-X.~Zeng, H.-Q.~Zhang, and H.~Zhang,
  {Shadows and photon spheres with spherical accretions in the four-dimensional
  Gauss{\textendash}Bonnet black hole}.
  Eur. Phys. J. C  \textbf{80}, 872 (2020).
  {\url{https://doi.org/10.1140/epjc/s10052-020-08449-y}}.
  {\href{https://arxiv.org/abs/2004.12074}{{arXiv:2004.12074}}}

\bibitem{Cardoso:2021sip}
  V.~Cardoso, F.~Duque, and A.~Foschi,
  {Light ring and the appearance of matter accreted by black holes}.
  Phys. Rev. D  \textbf{103}, 104044 (2021).
  {\url{https://doi.org/10.1103/PhysRevD.103.104044}}.
  {\href{https://arxiv.org/abs/2102.07784}{{arXiv:2102.07784}}}

\bibitem{Gan:2021xdl}
  Q.~Gan, P.~Wang, H.~Wu, and H.~Yang,
  {Photon ring and observational appearance of a hairy black hole}.
  Phys. Rev. D  \textbf{104}, 044049 (2021).
  {\url{https://doi.org/10.1103/PhysRevD.104.044049}}.
  {\href{https://arxiv.org/abs/2105.11770}{{arXiv:2105.11770}}}

\bibitem{Uniyal:2022vdu}
  A.~Uniyal, R.C.~Pantig, and A.~{\"O}vg{\"u}n,
  {Probing a non-linear electrodynamics black hole with thin accretion disk,
  shadow, and deflection angle with M87* and Sgr A* from EHT}.
  Phys. Dark Univ.  \textbf{40}, 101178 (2023).
  {\url{https://doi.org/10.1016/j.dark.2023.101178}}.
  {\href{https://arxiv.org/abs/2205.11072}{{arXiv:2205.11072}}}

\bibitem{Wang:2022yvi}
  H.-M.~Wang, Z.-C.~Lin, and S.-W.~Wei,
  {Optical appearance of Einstein-{\AE}ther black hole surrounded by thin
  disk}.
  Nucl. Phys. B  \textbf{985}, 116026 (2022).
  {\url{https://doi.org/10.1016/j.nuclphysb.2022.116026}}.
  {\href{https://arxiv.org/abs/2205.13174}{{arXiv:2205.13174}}}

\bibitem{Zeng:2021dlj}
  X.-X.~Zeng, G.-P.~Li, and K.-J.~He,
  {The shadows and observational appearance of a noncommutative black hole
  surrounded by various profiles of accretions}.
  Nucl. Phys. B  \textbf{974}, 115639 (2022).
  {\url{https://doi.org/10.1016/j.nuclphysb.2021.115639}}.
  {\href{https://arxiv.org/abs/2106.14478}{{arXiv:2106.14478}}}

\bibitem{Guerrero:2021ues}
  M.~Guerrero, G.J.~Olmo, D.~Rubiera-Garcia, and D.S.-C.~G{\'o}mez,
  {Shadows and optical appearance of black bounces illuminated by a thin
  accretion disk}.
  J. Cosmol. Astropart. Phys.  \textbf{08}, 036 (2021).
  {\url{https://doi.org/10.1088/1475-7516/2021/08/036}}.
  {\href{https://arxiv.org/abs/2105.15073}{{arXiv:2105.15073}}}

\bibitem{Hou:2022eev}
  Y.~Hou, Z.~Zhang, H.~Yan, M.~Guo, and B.~Chen,
  {Image of a Kerr-Melvin black hole with a thin accretion disk}.
  Phys. Rev. D  \textbf{106}, 064058 (2022).
  {\url{https://doi.org/10.1103/PhysRevD.106.064058}}.
  {\href{https://arxiv.org/abs/2206.13744}{{arXiv:2206.13744}}}

\bibitem{Zeng:2022pvb}
  X.-X.~Zeng, K.-J.~He, G.-P.~Li, E.-W.~Liang, and S.~Guo,
  {QED and accretion flow models effect on optical appearance of
  Euler{\textendash}Heisenberg black holes}.
  Eur. Phys. J. C  \textbf{82}, 764 (2022).
  {\url{https://doi.org/10.1140/epjc/s10052-022-10733-y}}.
  {\href{https://arxiv.org/abs/2209.05938}{{arXiv:2209.05938}}}

\bibitem{Rosa:2022tfv}
  J.L.~Rosa and D.~Rubiera-Garcia,
  {Shadows of boson and Proca stars with thin accretion disks}.
  Phys. Rev. D  \textbf{106}, 084004 (2022).
  {\url{https://doi.org/10.1103/PhysRevD.106.084004}}.
  {\href{https://arxiv.org/abs/2204.12949}{{arXiv:2204.12949}}}

\bibitem{Yang:2022btw}
  J.~Yang, C.~Zhang, and Y.~Ma,
  {Shadow and stability of quantum-corrected black holes}.
  Eur. Phys. J. C  \textbf{83}, 619 (2023).
  {\url{https://doi.org/10.1140/epjc/s10052-023-11800-8}}.
  {\href{https://arxiv.org/abs/2211.04263}{{arXiv:2211.04263}}}

\bibitem{Zhang:2023okw}
  C.~Zhang, Y.~Ma, and J.~Yang,
  {Black hole image encoding quantum gravity information}.
  Phys. Rev. D  \textbf{108}, 104004 (2023).
  {\url{https://doi.org/10.1103/PhysRevD.108.104004}}.
  {\href{https://arxiv.org/abs/2302.02800}{{arXiv:2302.02800}}}

\bibitem{Wang:2023vcv}
  X.-J.~Wang, X.-M.~Kuang, Y.~Meng, B.~Wang, and J.-P.~Wu,
  {Rings and images of Horndeski hairy black hole illuminated by various thin
  accretions}.
  Phys. Rev. D  \textbf{107}, 124052 (2023).
  {\url{https://doi.org/10.1103/PhysRevD.107.124052}}.
  {\href{https://arxiv.org/abs/2304.10015}{{arXiv:2304.10015}}}

\bibitem{Sui:2023tje}
  T.-T.~Sui, Z.-L.~Wang, and W.-D.~Guo,
  {The effect of scalar hair on the charged black hole with the images from
  accretions disk}.
  Eur. Phys. J. C  \textbf{84}, 441 (2024).
  {\url{https://doi.org/10.1140/epjc/s10052-024-12807-5}}.
  {\href{https://arxiv.org/abs/2311.10946}{{arXiv:2311.10946}}}

\bibitem{Gao:2023mjb}
  X.-J.~Gao, T.-T.~Sui, X.-X.~Zeng, Y.-S.~An, and Y.-P.~Hu,
  {Investigating shadow images and rings of the charged Horndeski black hole
  illuminated by various thin accretions}.
  Eur. Phys. J. C  \textbf{83}, 1052 (2023).
  {\url{https://doi.org/10.1140/epjc/s10052-023-12231-1}}.
  {\href{https://arxiv.org/abs/2311.11780}{{arXiv:2311.11780}}}

\bibitem{Meng:2024puu}
  Y.~Meng, X.-M.~Kuang, X.-J.~Wang, B.~Wang, and J.-P.~Wu,
  {Images of hairy Reissner{\textendash}Nordstr{\"o}m black hole illuminated by
  static accretions}.
  Eur. Phys. J. C  \textbf{84}, 305 (2024).
  {\url{https://doi.org/10.1140/epjc/s10052-024-12686-w}}.
  {\href{https://arxiv.org/abs/2401.05634}{{arXiv:2401.05634}}}

\bibitem{Darvishi:2024ndu}
  M.~Darvishi, M.~Heydari-Fard, and M.~Mohseni,
  {On optical appearance of Einstein-Maxwell-{\AE}ther black holes surrounded
  by various accretions}.
  Chin. J. Phys.  \textbf{93}, 632 (2025).
  {\url{https://doi.org/10.1016/j.cjph.2024.12.018}}.
  {\href{https://arxiv.org/abs/2405.13079}{{arXiv:2405.13079}}}

\bibitem{Zare:2024dtf}
  S.~Zare, L.M.~Nieto, X.-H.~Feng, S.-H.~Dong, and H.~Hassanabadi,
  {Shadows, rings and optical appearance of a magnetically charged regular
  black hole illuminated by various accretion disks}.
  J. Cosmol. Astropart. Phys.  \textbf{08}, 041 (2024).
  {\url{https://doi.org/10.1088/1475-7516/2024/08/041}}.
  {\href{https://arxiv.org/abs/2406.07300}{{arXiv:2406.07300}}}

\bibitem{Chen:2025ifv}
  J.~Chen and J.~Yang,
  {Shadows and optical appearance of quantum-corrected black holes illuminated
  by static thin accretions}.
  Eur. Phys. J. C  \textbf{85}, 512 (2025).
  {\url{https://doi.org/10.1140/epjc/s10052-025-14230-w}}.
  {\href{https://arxiv.org/abs/2503.06215}{{arXiv:2503.06215}}}

\bibitem{Rani:2025esb}
  S.~Rani, A.~Jawad, M.~Heydari-Fard, and U.~Zafar,
  {Thermodynamic and shadow analysis of Dehnen type dark matter Halo corrected
  Schwarzschild black hole surrounded by thin disk}.
  Eur. Phys. J. C  \textbf{85}, 677 (2025).
  {\url{https://doi.org/10.1140/epjc/s10052-025-14388-3}}

\bibitem{Luo:2025xjb}
  Z.~Luo, M.~Tang, and Z.~Xu,
  {Shadows and observational images of a Schwarzschild-like black hole
  surrounded by a Dehnen-type dark matter halo}.
  J. Cosmol. Astropart. Phys.  \textbf{10}, 065 (2025).
  {\url{https://doi.org/10.1088/1475-7516/2025/10/065}}.
  {\href{https://arxiv.org/abs/2505.20115}{{arXiv:2505.20115}}}

\bibitem{Li:2025ftb}
  X.-Q.~Li, Y.~Kim, B.-H.~Lee, H.-P.~Yan, and X.-J.~Yue,
  {Black holes immersed in modified Chaplygin-like dark fluid and cloud of
  strings: Geodesics, shadows, and images}.
  Chin. Phys. C  \textbf{49}, 105104 (2025).
  {\url{https://doi.org/10.1088/1674-1137/add9fa}}.
  {\href{https://arxiv.org/abs/2505.12077}{{arXiv:2505.12077}}}

\bibitem{CraigWalker:2018vam}
  R.~Craig~Walker, P.E.~Hardee, F.B.~Davies, C.~Ly, and W.~Junor,
  {The structure and dynamics of the subparsec jet in M87 based on 50 VLBA
  observations over 17 years at 43 GHz}.
  Astrophys. J.  \textbf{855}, 128 (2018).
  {\url{https://doi.org/10.3847/1538-4357/aaafcc}}.
  {\href{https://arxiv.org/abs/1802.06166}{{arXiv:1802.06166}}}

\bibitem{Wald:1984rg}
  R.M.~Wald,
  \emph{{General Relativity}}
  (Chicago University Press, Chicago, 1984).
  {\url{https://doi.org/10.7208/chicago/9780226870373.001.0001}}

\bibitem{Liang:2023ahd}
  C.~Liang and B.~Zhou,
  \emph{{Differential Geometry and General Relativity: Volume 1}}
  (Springer, Singapore, 2023).
  {\url{https://doi.org/10.1007/978-981-99-0022-0}}

\bibitem{Igata:2025glk}
  T.~Igata, M.~Omamiuda, and Y.~Takamori,
  {Gravitational lensing and accretion disk imaging of a Buchdahl dense core}.
  Phys. Rev. D  \textbf{111}, 084062 (2025).
  {\url{https://doi.org/10.1103/PhysRevD.111.084062}}.
  {\href{https://arxiv.org/abs/2502.11755}{{arXiv:2502.11755}}}

\bibitem{Yang:2024ulu}
  X.~Yang,
  {Observational appearance of the spherically symmetric black hole in PFDM}.
  Phys. Dark Univ.  \textbf{44}, 101467 (2024).
  {\url{https://doi.org/10.1016/j.dark.2024.101467}}

\bibitem{Shakura:1972te}
  N.I.~Shakura and R.A.~Sunyaev,
  {Black holes in binary systems. Observational appearance}.
  Astron. Astrophys.  \textbf{24}, 337 (1973)

\end{thebibliography}
\end{document}